\documentclass[useAMS,usenatbib]{mn2e}
\usepackage{txfonts}
\usepackage{color, supertabular, multicol}

\usepackage{graphicx}
\usepackage{amssymb}
\usepackage{threeparttable}

\newcommand{\oiii}{[O {\sc iii}]}

\newcommand{\feii}{Fe {\sc ii}}
\newcommand{\soiii}{\ensuremath{\rm [O\,{\scriptscriptstyle III}]}}
\mathchardef\mhyphen="2D

\title[Similarity of type 1 and type 2 quasar nebulae]{Similarity of ionized gas nebulae around unobscured and obscured quasars} 

\author[G. Liu et al.]{Guilin Liu$^1$\thanks{E-mail: liu@pha.jhu.edu}, 
Nadia L. Zakamska$^1$\thanks{E-mail: zakamska@pha.jhu.edu} and 
Jenny E. Greene$^2$\thanks{E-mail: jgreene@astro.princeton.edu}
\\
$^{1}$Center for Astrophysical Sciences, Department of Physics \& Astronomy, Johns Hopkins University, 
3400 N. Charles St., Baltimore, MD 21218, USA \\
$^{2}$Department of Astrophysical Sciences, Princeton University, Princeton, NJ 08544, USA 
}

\begin{document}

\date{Published in MNRAS, 2014, Volume 442, Issue 2, p.1303--1318}

\pagerange{\pageref{firstpage}--\pageref{lastpage}} \pubyear{2013}

\maketitle

\label{firstpage}

\begin{abstract}


Quasar feedback is suspected to play a key role in the evolution of massive galaxies, by removing or reheating gas in quasar host galaxies and thus limiting the amount of star formation. In this paper we continue our investigation of quasar-driven winds on galaxy-wide scales. We conduct Gemini Integral Field Unit spectroscopy of a sample of luminous unobscured (type 1) quasars, to determine the morphology and kinematics of ionized gas around these objects, predominantly via observations of the \oiii$\lambda$5007\AA\ emission line. We find that ionized gas nebulae extend out to $\sim$13 kpc from the quasar, that they are smooth and round, and that their kinematics are inconsistent with gas in dynamical equilibrium with the host galaxy. The observed morphological and kinematic properties are strikingly similar to those of ionized gas around obscured (type 2) quasars with matched \oiii\ luminosity, with marginal evidence that nebulae around unobscured quasars are slightly more compact. Therefore in samples of obscured and unobscured quasars carefully matched in \oiii\ luminosity we find support for the standard geometry-based unification model of active galactic nuclei, in that the intrinsic properties of the quasars, of their hosts and of their ionized gas appear to be very similar. Given the apparent ubiquity of extended ionized regions, we are forced to conclude that either the quasar is at least partially illuminating pre-existing gas or that both samples of quasars are seen during advanced stages of quasar feedback. In the latter case, we may be biased by our \oiii-based selection against quasars in the early ``blow-out'' phase, for example due to dust obscuration.

\end{abstract}


\section{Introduction}
\label{sec:intro}

The discovery of a tight relationship between the masses of black
holes in nearby galaxies and the velocities and masses of their
stellar populations \citep[e.g.,][]{mago98,gebh00} strongly suggests 
that the active phase of black hole evolution has profound effects on the
formation of massive galaxies. If the energy output of the black hole can 
somehow couple to the surrounding gas -- for example by blowing large-scale 
winds -- then the observed correlations can be reproduced \citep{hopk06}. 
The physical properties of such winds, their launching mechanisms, their 
impact on galaxies, and their incidence in various types of active galaxies 
are currently the subject of intensive observational and theoretical research 
efforts \citep[e.g.,][]{crot06, nesv08, moe09, zubo12, wagn13}.  

In the past several years, we have undertaken an observational
campaign to map out the kinematics of the ionized gas around luminous
obscured quasars \citep{zaka03, reye08} using Magellan, Gemini and
other facilities in search of signatures of quasar-driven winds
\citep{gree09, gree11, gree12, liu13a, liu13b, hain13, hain14}. 
In our observations, we are focusing on the most powerful quasars,
where feedback effects are expected to be strongest, and we take the 
observational advantages provided by circumnuclear obscuration to 
maximize sensitivity to faint extended emission associated with quasar
feedback. Our goal is to determine whether quasars can launch powerful 
winds via radiation pressure \citep{murr95, prog00} in the most 
common radio-quiet mode of accretion activity, without the aid of the 
relativistic jets known to drive gas outflows in some objects \citep{nesv08}. 

In December 2010, we conducted a Gemini-North Multi-Object 
Spectrograph (GMOS-N) Integral Field Unit (IFU) campaign which targeted
a sample of eleven obscured ``type 2'' luminous radio-quiet quasars at $z\sim0.5$. 
In the papers describing our results \citep{liu13a,liu13b}, we present the
analysis of the extents, morphologies and gas kinematics of the narrow 
emission line regions of these objects, predominantly via the \oiii$\lambda$5007\AA\ 
line (hereafter \oiii). We detect extended emission line nebulae in every case, 
extending out to 15--40 kpc from the center of the galaxy. Compared to ionized gas 
nebulae around radio-loud objects at low and high redshifts, our targets show 
morphologies that are more regular and smooth and less elongated \citep{liu13a}. 
The nebulae in our sample display well-organized velocity fields with velocity 
dispersions $\sim 10^3$ km s$^{-1}$ over the entire face of the nebulae \citep{liu13b}. 
The most likely explanation for these observations is that the quasars in our sample 
have ionized gas winds with large covering factors and propagation velocities 
($\sim 800$ km s$^{-1}$) that likely exceed the escape velocities of their host 
galaxies; we estimate the kinetic energies of the outflowing ionized gas to be 
well in excess of $10^{44}$ erg s$^{-1}$. Our analysis demonstrates that obscured 
radio-quiet quasars can drive gas outflows of similar scale, luminosity and 
velocity as those seen in some powerful radio galaxies \citep{nesv08}. 

The standard geometric unification model postulates that type 1 and type 2 quasars differ only by the orientation of the observer's line of sight relative to distribution of the obscuring material \citep{anto93}. In this case, the observer's line of sight is blocked in type 2 objects, but photons from the quasar can escape along other directions, scatter off the interstellar material in the host galaxy and reach the observer. We find strong support for this picture in luminous type 2 quasars: both extended scattering regions \citep{zaka06} and scattered broad emission lines \citep{zaka05} are seen in the data. Thus the objects in our sample are unambiguously seen as type 1 quasars along the unobscured directions. If type 1 and type 2 objects differ only by the orientation of the circumnuclear obscuration relative to the observer's line of sight, then we expect to see very similar distributions of ionized gas around quasars of both types (modulo perhaps some minor differences if the ionized gas is distributed in bicones which would be viewed closer to or farther from the axis in type 1 and type 2 sources, respectively). 

However, there may be more to the story. Theoretical models of galaxy formation have long utilized evolutionary scenarios in which galaxy mergers induce both star formation and nuclear activity. This active phase, characterized by largely obscured (type 2) accretion and star formation, is then terminated by a quasar-driven wind which clears out the surrounding gas and triggers a transition of the active black hole into an unobscured (type 1) quasar phase \citep[e.g.,][]{sand96,hopk06,hopk10}. A variety of observational studies support a transition of this type, finding that type 2 quasar host galaxies exhibit more energetically significant star formation than those of type 1's \citep[e.g.,][]{zaka08,lacy07}. In this evolutionary paradigm of quasar obscuration, one might expect to find more prominent winds in the type 2 objects (more characteristic of the ``blow-out'' phase) than in the type 1's. 

In this paper we present an observational test of evolutionary models and geometric models of quasar obscuration. We investigate whether the properties of the ionized gas nebulae around luminous quasars are associated in any way with the circumnuclear obscuration that determines the optical type of the active nucleus. We select a sample of 12 type 1 quasars that are well-matched in redshift and \oiii\ luminosity to our previously studied sample of luminous obscured quasars \citep{liu13a, liu13b} and conduct observations of the ionized gas around these objects in a manner identical to the one we employed in our previous work. In Section \ref{sec:data} we describe sample selection, observations, data reduction and calibrations. In Section \ref{sec:maps}, we present maps of the ionized gas emission, and in Section \ref{sec:compare} we compare properties of nebulae in obscured and unobscured quasars. In Section \ref{sec:discuss} we discuss the implications of our findings, discuss morphologies of the nebulae, and we summarize in Section \ref{sec:summary}. As in \citet{liu13a,liu13b}, we adopt a $h$=0.71, $\Omega_m$=0.27, $\Omega_{\Lambda}$=0.73 cosmology throughout this paper; objects are identified as SDSS Jhhmmss.ss+ddmmss.s in the tables and are shortened to SDSS Jhhmm+ddmm elsewhere; and the rest-frame wavelengths of the emission lines are given in air.

\begin{table*}
\caption{Properties of our Gemini unobscured quasar sample.}
\setlength{\tabcolsep}{1.4mm}
\label{tab1}
\begin{center}
\begin{tabular}{ccccrccccccc}
\hline
\multicolumn{1}{c}{Object name} &
\multicolumn{1}{c}{$f_{\rm 1.4~GHz}$} &
\multicolumn{1}{c}{$M_B$} &
\multicolumn{1}{c}{$R$} &
\multicolumn{1}{c}{$z$} &
\multicolumn{1}{c}{PA} &
\multicolumn{1}{c}{$t_{\rm exp}$} &
\multicolumn{1}{c}{$\lambda_{\rm c}$} &
\multicolumn{1}{c}{Seeing} &
\multicolumn{1}{c}{$\log L_{\soiii}$} &
\multicolumn{1}{c}{$\eta$} &
\multicolumn{1}{c}{$\nu L_{\rm \nu,\;8\;\mu m}$} \\
\multicolumn{1}{c}{(1)} &
\multicolumn{1}{c}{(2)} &
\multicolumn{1}{c}{(3)} &
\multicolumn{1}{c}{(4)} &
\multicolumn{1}{c}{(5)} &
\multicolumn{1}{c}{(6)} &
\multicolumn{1}{c}{(7)} &
\multicolumn{1}{c}{(8)} &
\multicolumn{1}{c}{(9)} &
\multicolumn{1}{c}{(10)} &
\multicolumn{1}{c}{(11)} &
\multicolumn{1}{c}{(12)} 
\\
\hline

SDSS J023342.57$-$074325.8 & $<$1.1 & $-$24.8 &  $<$0.4 & 0.4538 & 300 & 1620$\times$2 & 800 & 0.42 & 43.13 & 2.94$\pm$0.03 & 45.05 \\
SDSS J030422.39$+$002231.8 & $<$0.8 & $-$26.9 & $<$$-$$0.3$ & 0.6385 &   0 & 1620$\times$2 & 800 & 0.47 & 42.82 & 3.89$\pm$0.37 & 45.91 \\
SDSS J031154.51$-$070741.9 & $<$1.1 & $-$25.4 &  $<$0.5 & 0.6330 & 180 & 1620$\times$2 & 800 & 0.60 & 42.94 & 3.93$\pm$0.14 & 45.84 \\
SDSS J041210.17$-$051109.1 &  ~~3.2 & $-$26.0 &  ~~~0.5 & 0.5492 & 185 & 1620$\times$2 & 760 & 0.39 & 43.50 & 2.58$\pm$0.07 & 45.79 \\
SDSS J075352.98$+$315341.6 & $<$1.0 & $-$24.5 &  $<$0.5 & 0.4938 & 180 & 1620$\times$2 & 760 & 0.51 & 42.62 & 2.57$\pm$0.05 & 44.63 \\
SDSS J080954.38$+$074355.1 & $<$1.0 & $-$26.6 &  $<$0.0 & 0.6527 & 270 & 1620$\times$2 & 800 & 0.48 & 43.25 & 3.88$\pm$0.07 & 45.66 \\
SDSS J084702.55$+$294011.0 & $<$1.0 & $-$25.0 &  $<$0.5 & 0.5662 & 270 & 1620$\times$2 & 760 & 0.60 & 42.71 & 3.12$\pm$0.11 & 44.98 \\
SDSS J090902.21$+$345926.5 & $<$1.0 & $-$25.5 &  $<$0.3 & 0.5749 & 210 & 1620$\times$2 & 800 & 0.63 & 43.12 & 4.22$\pm$0.07 & 45.62 \\
SDSS J092423.42$+$064250.6 & $<$1.1 & $-$26.1 &  $<$0.1 & 0.5884 & 280 & 1620$\times$2 & 800 & 0.48 & 42.95 & 4.54$\pm$0.22 & 45.55 \\
SDSS J093532.45$+$534836.5 & $<$1.0 & $-$25.3 &  $<$0.6 & 0.6864 & 270 & 1620$\times$2 & 760 & 0.70 & 43.20 & 3.73$\pm$0.03 & 45.26 \\
SDSS J114417.78$+$104345.9 & $<$1.0 & $-$25.3 &  $<$0.5 & 0.6785 &   0 & 1620$\times$2 & 760 & 0.54 & 43.30 & 4.68$\pm$0.20 & 45.24 \\
SDSS J221452.10$+$211505.1 & $<$2.5 & $-$25.1 &  $<$0.6 & 0.4752 & 284 & 1620$\times$2 & 800 & 0.45 & 42.78 & 3.58$\pm$0.10 & 45.08 \\

\hline
\end{tabular}
\tablenotes{{\bf Notes.} -- 
(1) Object name. 
(2) Observed radio flux at 1.4 GHz in mJy, taken from the FIRST survey \citep{beck95,whit97} and the NVSS survey
\citep[][for SDSS J0412$-$0511 and SDSS J2214$+$2115]{cond98}. None of the sources with FIRST coverage are detected, so we report here the upper limits for point sources \citep[5$\sigma$+0.25 mJy,][]{coll05}. 
SDSS J2214$+$2115 does not have coverage in FIRST and is not included in the NVSS point-source catalog; thus we place an upper limit of $f_{\rm1.4~GHz}\lesssim 2.5$ mJy.
(3) Absolute magnitude in the rest-frame $B$ band, derived by convolving the SDSS spectra to the transmission curve of the Johnson B filter and converting the flux density averaged over the bandwidth ($\Delta \lambda=940\;\AA$) to the AB magnitude at the effective wavelength midpoint $\lambda_{\rm eff}=4450\;\AA$ \citep{binn98}.}
(4) Radio-to-optical flux ratio, defined as $R=\log\,(F_{\rm 1.4~GHz}/F_B)$, where $F_{\rm 1.4~GHz}$ and $F_B$ are the rest-frame 1.4 GHz and $B$-band fluxes, respectively. $K$-correction for the radio flux is calculated following \citet{zaka04}, assuming a power-law spectrum of $F_{\nu}\propto \nu^{-0.5}$. All of our objects satisfy the conventional criterion that requires radio-quiet/weak objects to have $R<1$ \citep{kell89}.
(5) Redshift, from \citet{shen11}. 
(6) Position angle of the shorter axis of the field of view, in degrees east of north. 
(7) Exposure time in seconds and number of exposures.
(8) Central wavelength of the grating R400 used for the object in nm.
(9) Full width at half maximum (FWHM) of the seeing at the observing site, in arcseconds.
(10) Total luminosity of the \oiii$\lambda$5007\AA\ line (logarithmic scale, in erg s$^{-1}$), derived from our data calibrated against SDSS spectra. The SDSS DR10 spectra of the sample objects are accurate within $\lesssim$5\%. We estimate that subtracting continuum and \feii\ introduces $\sim$10\% uncertainty and the procedure of flux calibration against SDSS spectra introduces another $\sim$10\%. Thus the uncertainty of $\log L_{\soiii}$ is about 15\%.
(11) Absolute value of the best-fit power-law exponent of the outer part of the \oiii\ profile along the major axes (i.e., $I_{R,\soiii}\propto R^{-\eta}$, see Figure~\ref{fig:psf}).
(12) Luminosity at rest-frame 8 \micron\ (logarithmic scale, in erg s$^{-1}$), interpolated from WISE photometry.
\end{center}
\end{table*}

\begin{table*}
\caption{Properties of our Gemini unobscured quasar sample (continued).\label{tab2}}
\setlength{\tabcolsep}{1.4mm}
\begin{center}
\begin{tabular}{crcrcrrrcrcrrrrrr}
\hline
\multicolumn{1}{c}{Object name} &
\multicolumn{1}{c}{$R_{5\sigma}^{\rm cont}$} &
\multicolumn{1}{c}{$\epsilon_{5\sigma}^{\rm cont}$} &
\multicolumn{1}{c}{$R_{5\sigma}$} &
\multicolumn{1}{c}{$\epsilon_{5\sigma}$} &
\multicolumn{1}{c}{$R_{\rm eff}^{\rm cont}$} &
\multicolumn{1}{c}{$R_{\rm eff}$} &
\multicolumn{1}{c}{$R_{\rm obs}$} &
\multicolumn{1}{c}{$\epsilon_{\rm obs}$} &
\multicolumn{1}{c}{$R_{\rm int}$} &
\multicolumn{1}{c}{$\epsilon_{\rm int}$} &
\multicolumn{1}{c}{$\Delta v_{\rm max}$} &
\multicolumn{1}{c}{$\langle W_{80} \rangle$} &
\multicolumn{1}{c}{$W_{\rm 80,max}$} &
\multicolumn{1}{c}{$\nabla W_{80}$} &
\multicolumn{1}{c}{$\langle A\rangle$} &
\multicolumn{1}{c}{$\langle K\rangle$} \\
\multicolumn{1}{c}{(1)} &
\multicolumn{1}{c}{(2)} &
\multicolumn{1}{c}{(3)} &
\multicolumn{1}{c}{(4)} &
\multicolumn{1}{c}{(5)} &
\multicolumn{1}{c}{(6)} &
\multicolumn{1}{c}{(7)} &
\multicolumn{1}{c}{(8)} &
\multicolumn{1}{c}{(9)} &
\multicolumn{1}{c}{(10)} &
\multicolumn{1}{c}{(11)} &
\multicolumn{1}{c}{(12)} &
\multicolumn{1}{c}{(13)} &
\multicolumn{1}{c}{(14)} &
\multicolumn{1}{c}{(15)} &
\multicolumn{1}{c}{(16)} &
\multicolumn{1}{c}{(17)} \\
\hline

SDSS J0233$-$0743 &  8.5 & 0.15 & 16.6 & 0.35 & 2.7 & 2.8 & 12.4 & 0.35 & 10.7 & 0.44 & 164 &  540 &  544 & $-$4.8 & $-$0.05 & 1.22 \\
SDSS J0304$+$0022 & 12.3 & 0.02 &  9.8 & 0.15 & 2.9 & 2.8 &  8.2 & 0.12 &  7.9 & 0.19 & 576 & 1630 & 2002 & $-$9.9 & $-$0.12 & 0.87 \\
SDSS J0311$-$0707 & 13.2 & 0.14 & 11.0 & 0.14 & 3.6 & 3.3 &  8.7 & 0.10 &  7.9 & 0.09 & 286 & 1353 & 1260 & $-$5.3 & $-$0.26 & 1.97 \\
SDSS J0412$-$0511 &  9.9 & 0.04 & 13.7 & 0.18 & 2.4 & 2.3 & 13.1 & 0.18 & 11.1 & 0.14 & 266 & 1248 & 1476 & $-$4.8 &    0.04 & 2.19 \\
SDSS J0753$+$3153 &  7.8 & 0.10 &  8.7 & 0.07 & 2.6 & 2.6 &  6.3 & 0.06 &  5.2 & 0.03 & 165 &  333 &  578 &    0.7 &    0.04 & 1.18 \\
SDSS J0809$+$0743 & 12.4 & 0.03 & 12.5 & 0.10 & 3.6 & 3.6 & 11.2 & 0.14 & 10.2 & 0.14 & 192 &  879 &  962 & $-$5.2 & $-$0.20 & 1.42 \\
SDSS J0847$+$2940 &  8.3 & 0.05 & 11.9 & 0.21 & 4.0 & 3.8 &  8.6 & 0.16 &  7.4 & 0.09 & 115 &  473 &  678 & $-$5.2 & $-$0.10 & 1.73 \\
SDSS J0909$+$3459 & 11.7 & 0.15 & 13.3 & 0.12 & 4.4 & 3.9 & 11.4 & 0.13 &  9.7 & 0.12 & 101 &  610 & 1075 & $-$3.1 &    0.05 & 1.52 \\
SDSS J0924$+$0642 & 11.5 & 0.05 &  9.1 & 0.02 & 2.6 & 2.7 &  7.9 & 0.07 &  7.7 & 0.12 &  83 &  914 & 1089 & $-$0.8 & $-$0.13 & 1.15 \\
SDSS J0935$+$5348 & 11.5 & 0.04 & 13.6 & 0.12 & 4.6 & 4.5 & 11.7 & 0.14 & 11.0 & 0.13 & 249 &  750 &  780 & $-$2.5 &    0.02 & 1.26 \\
SDSS J1144$+$1043 & 10.3 & 0.10 & 14.2 & 0.11 & 4.6 & 4.7 & 13.0 & 0.11 & 12.4 & 0.11 & 305 &  715 &  958 & $-$4.1 & $-$0.04 & 1.38 \\
SDSS J2214$+$2115 &  9.7 & 0.04 & 10.8 & 0.05 & 2.6 & 2.5 &  7.1 & 0.04 &  5.8 & 0.04 & 238 &  688 &  773 & $-$7.1 & $-$0.26 & 1.66 \\

\hline
\end{tabular}
\tablenotes{{\bf Notes.} -- 
(1) Object name. 
(2, 3) Semi-major axis (in kpc) and ellipticity ($\epsilon\equiv1-b/a$) of the best-fit ellipse which encloses pixels 
with S/N$\geqslant$5 in the continuum map.
(4, 5) Semi-major axis (in kpc) and ellipticity of the best-fit ellipse which encloses pixels 
with S/N$\geqslant$5 in the \oiii$\lambda$5007\AA\ line map.
(6, 7) Half-light isophotal radius (effective radius) of the continuum and \oiii$\lambda$5007\AA\ line emission 
(semi-major axis, in kpc).
(8, 9) Isophotal radius (semi-major axis, in kpc) and ellipticity at the observed limiting surface brightness of 
$10^{-16}$ erg s$^{-1}$ cm$^{-2}$ arcsec$^{-2}$.
(10, 11) Isophotal radius (semi-major axis, in kpc) and ellipticity at the intrinsic limiting surface brightness 
(corrected for cosmological dimming) of $10^{-15}$ erg s$^{-1}$ cm$^{-2}$ arcsec$^{-2}$.
(12) Maximum median velocity range across the kinematic maps of the nebulae (see Figure \ref{fig:Vmed}), in km s$^{-1}$. For each object, the 5\% tails on either side of the distribution of median velocities are excluded for determination of $\Delta v_{\rm max}$ to minimize the effect of the noise.
(13, 16, 17) Widths $W_{80}$ (km s$^{-1}$), asymmetries $A$ and kurtosis $K$ values of the integrated \oiii$\lambda$5007\AA\ velocity profiles, measured from the SDSS fiber spectrum.
(14) Maximum $W_{80}$ and most negative $v_{02}$ values in their respective spatially-resolved maps (km s$^{-1}$). Like for $\Delta v_{\rm max}$, the 5\% tails on either side of their respective distributions are excluded.
(15) Observed percentage change of $W_{80}$ per unit distance from the brightness center, in units of \% kpc$^{-1}$.
   It is defined as $\nabla W_{80}\equiv\delta W_{80}/R_5$, where $R_5$ is the maximum radius for the region where the
   peak of the \oiii$\lambda$5007\AA\ line is detected with S/N$\geqslant$5, and
   $\delta W_{80}=100\times(W_{\rm 80,R=0}-W_{80,R_{5}})/W_{\rm 80,R=0}$.
}
\end{center}
\end{table*}

\section{Data and measurements}
\label{sec:data}

\subsection{Sample selection}
\label{sec:selection}

\begin{figure*}
\includegraphics[scale=0.5,trim=0cm 0mm 0mm 0mm]{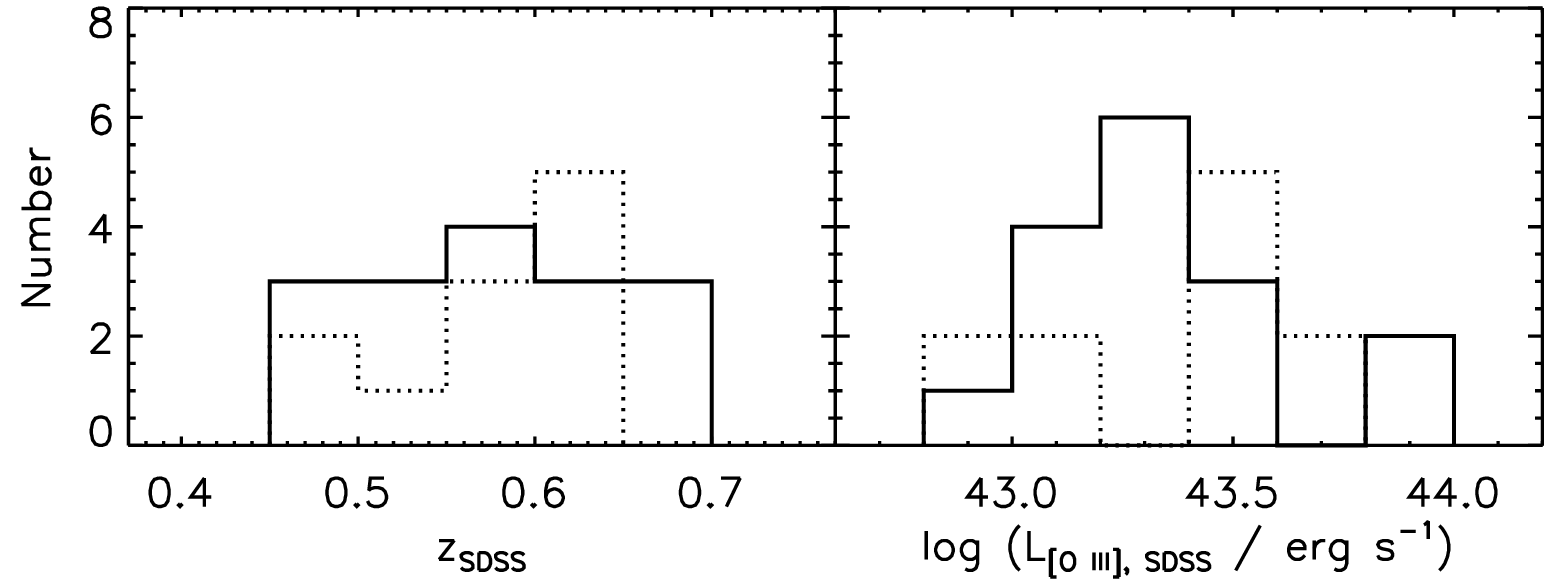}
\caption{Distribution of the redshifts and \oiii\
luminosities of the proposed type 1 quasars (solid lines). The sample 
is selected from the Sloan Digital Sky Survey (SDSS) quasar catalog \citep{shen11}
to match both quantities to the type 2 quasar sample that has 
been observed in our previous Gemini GMOS program (dotted lines).}
\label{fig:match}
\end{figure*}

In our previous Gemini IFU campaign completed in December 2010, we mapped ionized gas nebulae around eleven type 2 radio-quiet quasars selected from \citep{reye08}. These objects were selected to be as luminous as possible in the \oiii\ line ($L_{\soiii}\geq10^{42.8}$ erg s$^{-1}$, corresponding to estimated intrinsic luminosities of the active nucleus of $M_B<-26.2$ mag) while being at low enough redshift ($0.4<z<0.6$) to maximize the spatial information. In this work, we study a sample of twelve type 1 radio-quiet/weak quasars selected from \citep{shen11} according to the following criteria:

\begin{enumerate}

\item We match the \oiii$\lambda$5007\AA\ line luminosities of the
selected unobscured quasars to the 11 obscured quasars we studied in
\citet{liu13a,liu13b}. The selected quasars have \oiii\ line luminosities 
$L_{\soiii}>10^{42.9}$ erg s$^{-1}$ (Figure \ref{fig:match}).

\item As we do for \citet{liu13a,liu13b}, we select the objects at redshifts 
$z=0.4$--0.6 (Figure \ref{fig:match}).

\item All quasars are selected to be radio-quiet/weak. Conservatively, we require
their radio flux at 1.4 GHz from FIRST and NVSS \citep{beck95, whit97, cond98} to be 
$f_{\rm 1.4\;GHz}<10$ mJy, a criterion identical to the one applied for the obscured sample.

\item We focus on objects with high \oiii\ equivalent widths, which also tend to have low \feii\ luminosity \citep{boro92}. The reasons are two-fold. First, the relatively low levels of continuum and \feii\ emission reduce contamination from the bright point-like quasar making it easier to perform measurements of faint extended emission. Second, we thus maximize our chances of resolving the extended \oiii\ emission in light of the recent suggestion that the most extended and luminous ionized gas nebulae are found around quasars with weak emission from \feii\ lines and complexes \citep{mats12}. As a result, considerable \feii\ emission only exists in two of our targets (SDSS J0304+0022 and SDSS J0924+0642). In general, \oiii\ selection results in a bias toward objects with radio jets \citep{boro02}, but because we also impose a stringent limit on the radio emission, such objects are unlikely to dominate our sample. The origin of radio emission in radio-weak objects is a topic of active debate \citep{cond13,mull13,zaka14}. For our purposes the important aspects of the selection are that (1) our sources are not dominated by powerful relativistic jets, and (2) the type 1 and type 2 sources are selected to have similar radio properties to enable a direct comparison.

\end{enumerate}

\begin{figure*}
\includegraphics[scale=0.6,trim=0cm 0mm 0mm 0mm]{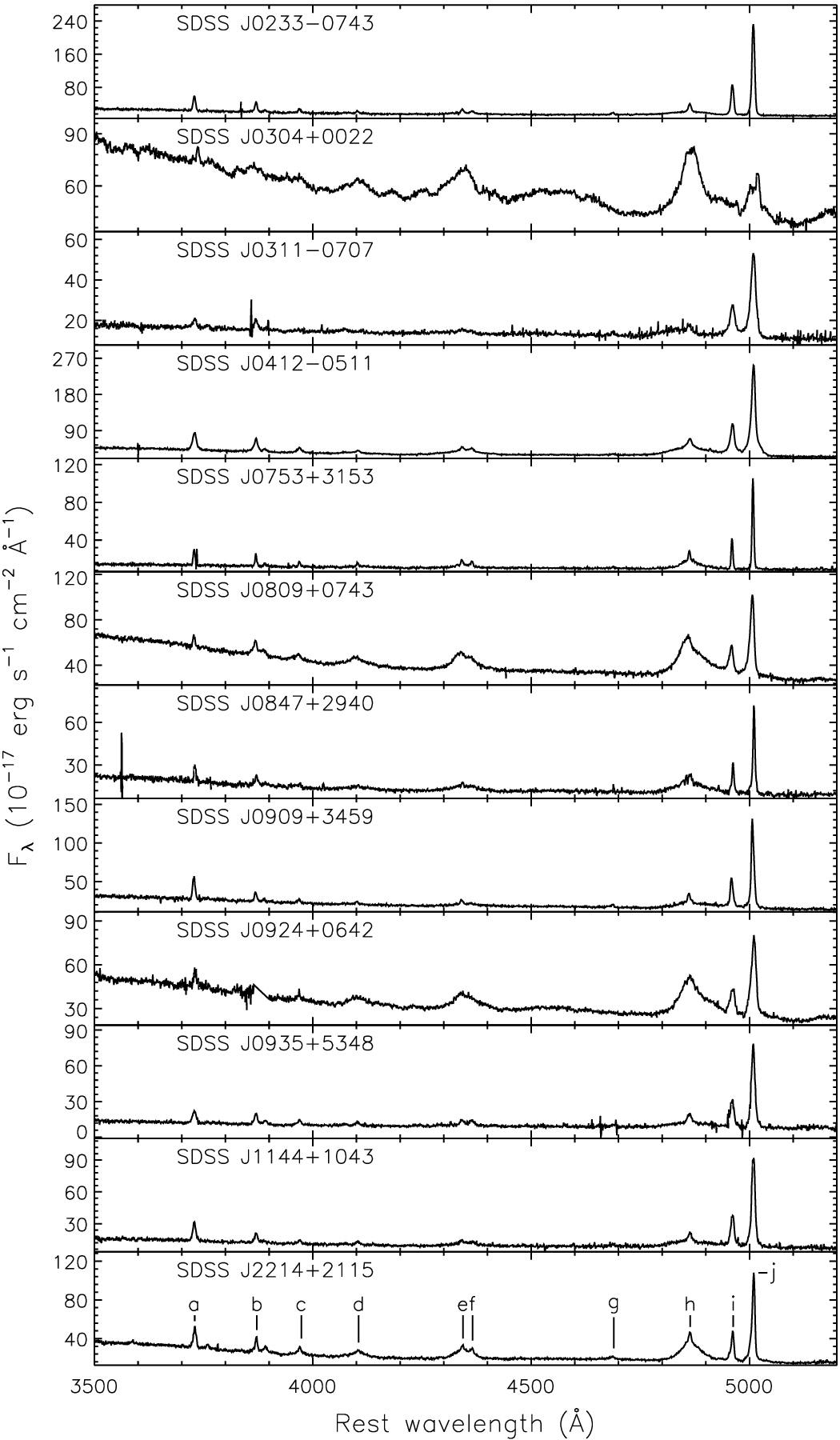}
\includegraphics[scale=0.6,trim=0cm 0mm 0mm 0mm]{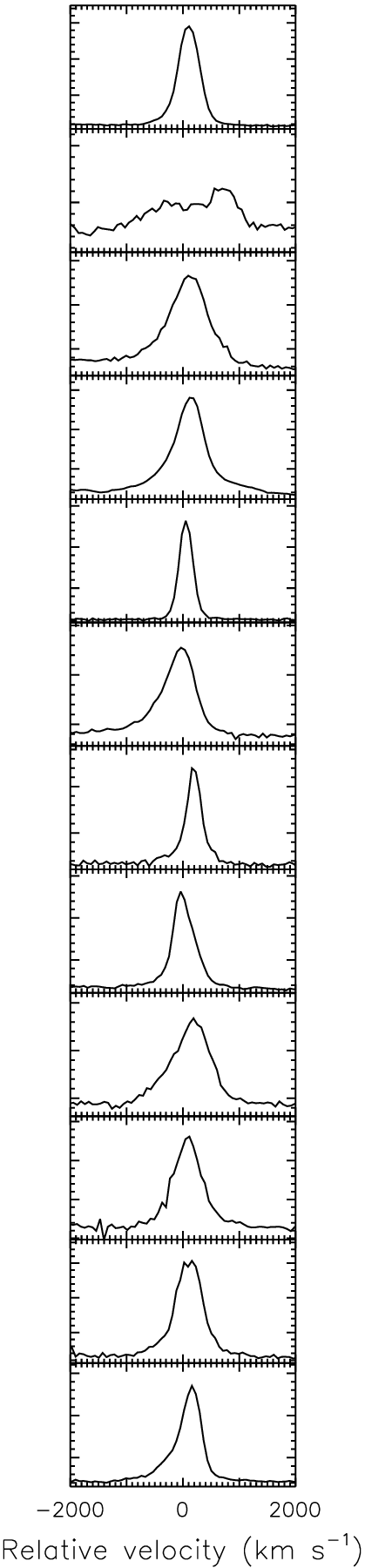}
\caption{Spectra of the twelve proposed targets
in their respective rest frames, from Data Release 8 of the
Sloan Digital Sky Survey. We concentrate on quasars that have high equivalent widths of \oiii\ and low equivalent width of \feii\ \citep{boro92}. The marked emission lines are: 
(a) [O {\sc ii}] 3727\AA; (b) [Ne {\sc iii}] 3869\AA; 
(c) [Ne {\sc iii}] 3871\AA; (d) H$\delta$ 4102\AA; 
(e) H$\gamma$ 4341\AA; (f) \oiii\ 4363\AA; (g) He {\sc ii} 4686\AA;
(h) H$\beta$ 4861\AA; (i) \oiii\ 4959\AA; (j) \oiii\ 5007\AA. 
The velocity substructures in the [O {\sc iii}] 5007\AA\ 
line is shown in the right panels, where de-redshifting is
performed using their SDSS redshifts (Table \ref{tab1}).} 
\label{fig:sdss}
\end{figure*}

We show the SDSS spectra of our chosen targets in Figure \ref{fig:sdss}. Although all spatial information 
is lost, these spectra (collected within a 3\arcsec\ fiber) cover a much broader wavelength range than our
Gemini data.

\subsection{Observations and data reduction} 

We observed twelve radio-quiet/weak unobscured quasars (Table~\ref{tab1}) with GMOS-N IFU \citep{alli02} between 2012 September and 2013 January (program ID: GN-2012B-Q-29, PI: G. Liu) to determine the spatial distribution of their emission. We use the two-slit mode that covers a 5\arcsec$\times$7\arcsec\ field of view, translating to a physical scale of 30$\times$42 kpc$^2$ at $z=0.5$, the typical redshift of our objects. The science field of view is sampled by 1000 contiguous 0.2\arcsec-diameter hexagonal lenslets, and simultaneous sky observations are obtained by 500 lenslets located $\sim$1\arcmin\ away. The seeing at the time of our observations was between 0.4\arcsec\ and 0.7\arcsec\ (2.4--4.2 kpc at $z\sim 0.5$). 

All objects were observed in the {\it i}-band (7060--8500 \AA) so as to cover the rest frame wavelengths $4100<\lambda<5200$ \AA\ and thus the \oiii-H$\beta$ region. To ensure that none of the important emission lines in this region is severely affected by the slit gaps, we tune the central wavelength to either 760 or 800 nm, according to their respective redshifts. The grating R400-G5305 we used has a spectral resolution of $R$=1918. As the width of our observed \oiii\ line is always well above the instrumental resolution \citep[full width at half maximum of unresolved sky lines is FWHM$=137\pm$12 km s$^{-1}$;][]{liu13b}, the velocity structure of our objects is well resolved. The basic information for our quasar sample and the Gemini campaign is summarized in Table \ref{tab1}.

For each object in each band, we took two exposures of 1620 sec each 
without spatial offset. We perform the data reduction in the same way as for the obscured quasars \citep{liu13a}. We use the Gemini package 
for IRAF\footnote{The Image Reduction and Analysis Facility (IRAF) is 
distributed by the National Optical Astronomy Observatories which is operated 
by the Association of Universities for Research in Astronomy, Inc. under 
cooperative agreement with the National Science Foundation.}, following the 
standard procedure for GMOS IFU described in the tasks {\sl gmosinfoifu} and 
{\sl gmosexamples}\footnote{http://www.gemini.edu/sciops/data/IRAFdoc/gmosinfoifu.html},
except that (a) we use an overscan instead of a bias image throughout the
data reduction, and adjust the relevant parameters so that the bias correction
is applied only once on each image, and (b) we set the parameter ``weights''
of {\sl gfreduce} to ``none'' (in contrast to ``variance'' as suggested by
the standard example) to avoid significantly increased noise in some parts
of the extracted spectrum. The  final product of the data reduced from each exposure is a data cube with
0.1\arcsec\ spatial pixels (``spaxels''). The two frames are finally combined 
using the tasks {\sl imcombine} by taking the mean spectra in each spaxel.

\subsection{Flux Calibration and Continuum / \feii\ Subtraction} 
\label{sec:calib}

We flux-calibrate our data using the spectra of our science targets from 
the SDSS Data Release 10 \citep{ahn13}\footnote{http://www.sdss3.org/dr10}. 
SDSS spectra are collected by fibers with a 3\arcsec~diameter at a typical seeing of 
$\sim$2\arcsec, and SDSS spectro-photometric calibrations are likely good
to 5\% or better \citep{adel08}. Since SDSS fiber fluxes are calibrated using 
point spread function (PSF) magnitudes, SDSS spectrophotometry is corrected for 
fiber losses.

In view of the limited wavelength coverage of our IFU data, we choose the
rest-frame wavelength range of 4980\AA\ to 5050\AA\ which is covered with 
good sensitivity for all objects to perform the flux calibration.
To simulate the SDSS fiber observations, we convolve the IFU image at each 
wavelength with a Gaussian kernel whose Full Width at Half Maximum (FWHM) 
satisfies $\rm FWHM^2+seeing^2={2\arcsec}^2$ to mimic the SDSS observing 
conditions and then we extract the spectrum using a 3\arcsec-diameter 
circular aperture. We then collapse the spectrum between rest-frame 4980 
and 5050\AA\ and compare the resultant flux to that of the SDSS spectrum
after converting SDSS vacuum wavelengths to air.
The calibration of the SDSS data includes a PSF correction to recover the 
flux outside the fibers assuming a 2\arcsec\ seeing, which needs to be removed 
for our purpose. In the last step of our calibration, we downgrade the image 
of our standard star to 2\arcsec\ resolution and find that a 3\arcsec\ circular 
aperture centered on the star encloses $\sim$80\% of the total flux. This
factor is taken into account for the final calibration of the IFU data 
against the SDSS spectra. 

\begin{figure*}
\centering
\includegraphics[scale=0.55,trim=20mm 0mm 10mm 0mm]{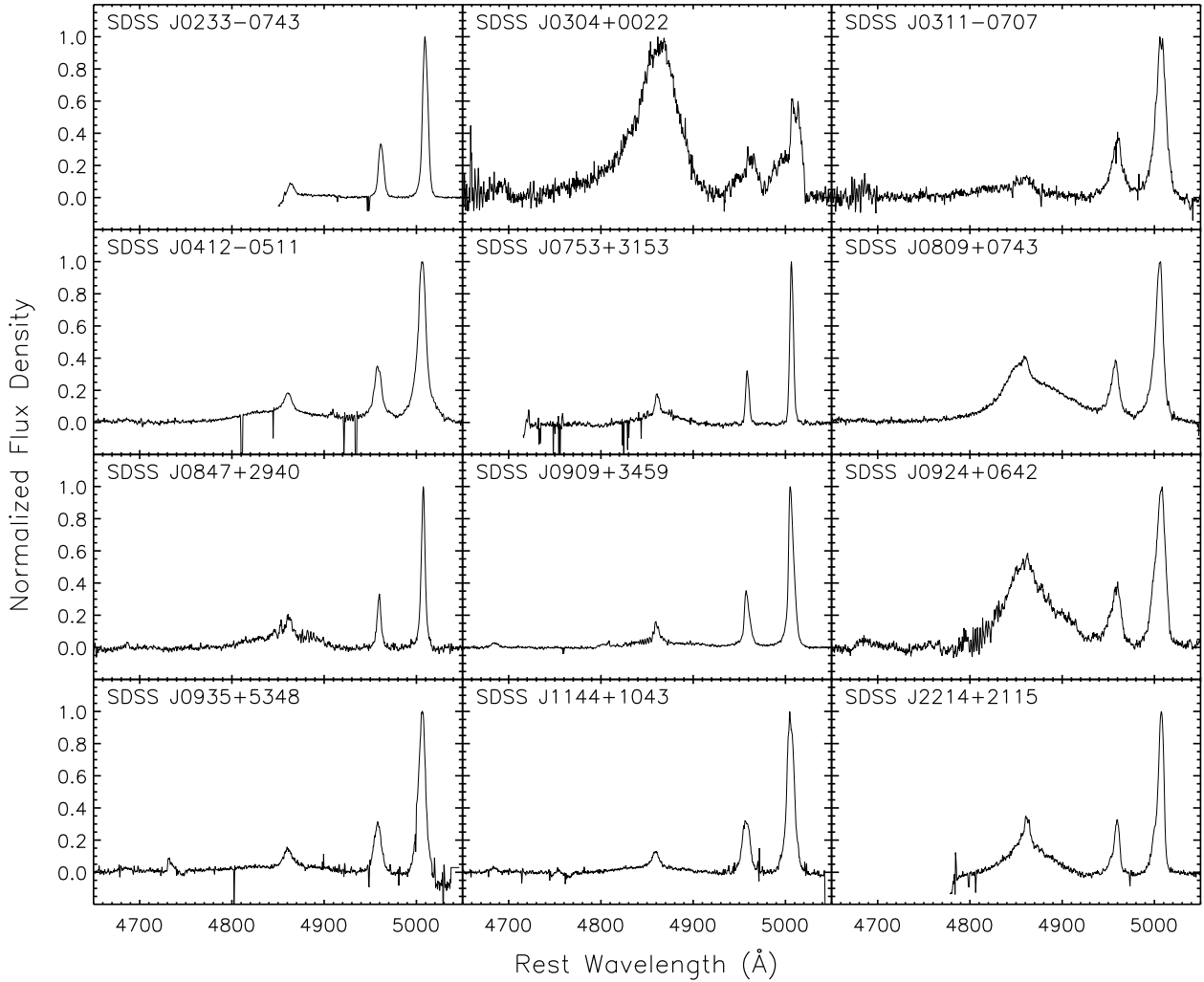}
\caption{Gemini spectra of our quasar sample after continuum and \feii\
subtraction, spatially collapsed within a circular annulus between 0.5\arcsec\ 
and 2\arcsec\ radii. Flux densities are normalized by the peak flux density
in the \oiii-H$\beta$ wavelength range. In the integrated spectrum of SDSS 
J0304$+$0022 the \oiii$\lambda$4959\AA\ is not detected, but it is recovered 
after continuum and \feii\ subtraction.}
\label{fig:outer_spec}
\end{figure*}

In order to remove the contamination from \feii\ emission, we fit the continuum in the vicinity of the \oiii-H$\beta$ region with the sum of a polynomial and the \feii\ template from \citet{boro92}. As part of the fit, we smooth the \feii\ template using a Gaussian  kernel, whose width is a free fitting parameter, to take into account that the velocity dispersion of \feii\ emission varies from object to object. The polynomial is set  to be quadratic for all targets with the exception of SDSS~J0924+0642, for which cubic functions are necessary to produce reasonable fits. The Gemini spectra after continuum and \feii\ subtraction, coadded spatially within a circular annulus between 0.5\arcsec\ and 2\arcsec\ radii, are shown in Figure \ref{fig:outer_spec}. 

\feii\ contamination to our \oiii\ analysis is insignificant or negligible in most (10 out of 12) of our targets, because the sample is pre-selected to have low \feii\ equivalent width. Considerable \feii\ contamination is present only in SDSS J0304+0022 and SDSS J0924+0642. Specifically, in Figure \ref{fig:sdss}, the strong \feii\ contamination in the \oiii-H$\beta$ wavelength region essentially swamps the \oiii$\lambda$4959\AA\ emission, whereas after \feii\ and continuum subtraction this feature is revealed with the same velocity structure as the \oiii$\lambda$5007\AA\ line (Figure \ref{fig:outer_spec}). 

\section{\oiii\ line fitting and surface brightness maps}
\label{sec:maps}

For obscured quasars, we created the \oiii\ maps by directly collapsing the datacube over the \oiii\ wavelength range \citep{liu13a}. This was possible because the direct emission from the quasar itself is blocked by circumnuclear obscuration. In our sample of unobscured quasars, the analysis is more complicated because the overall emission is dominated by the directly observable continuum and the broad emission lines that originate near the supermassive black hole, rather than in the much more extended host galaxy. Even after we perform continuum and \feii\ subtraction from the overall spectrum of each spaxel, \oiii\ maps of type 1 quasars are contaminated by the residual continuum. 

\begin{figure*}
\begin{flushleft}
\includegraphics[scale=0.36,trim=0cm 0mm 30mm 0mm,clip=clip]{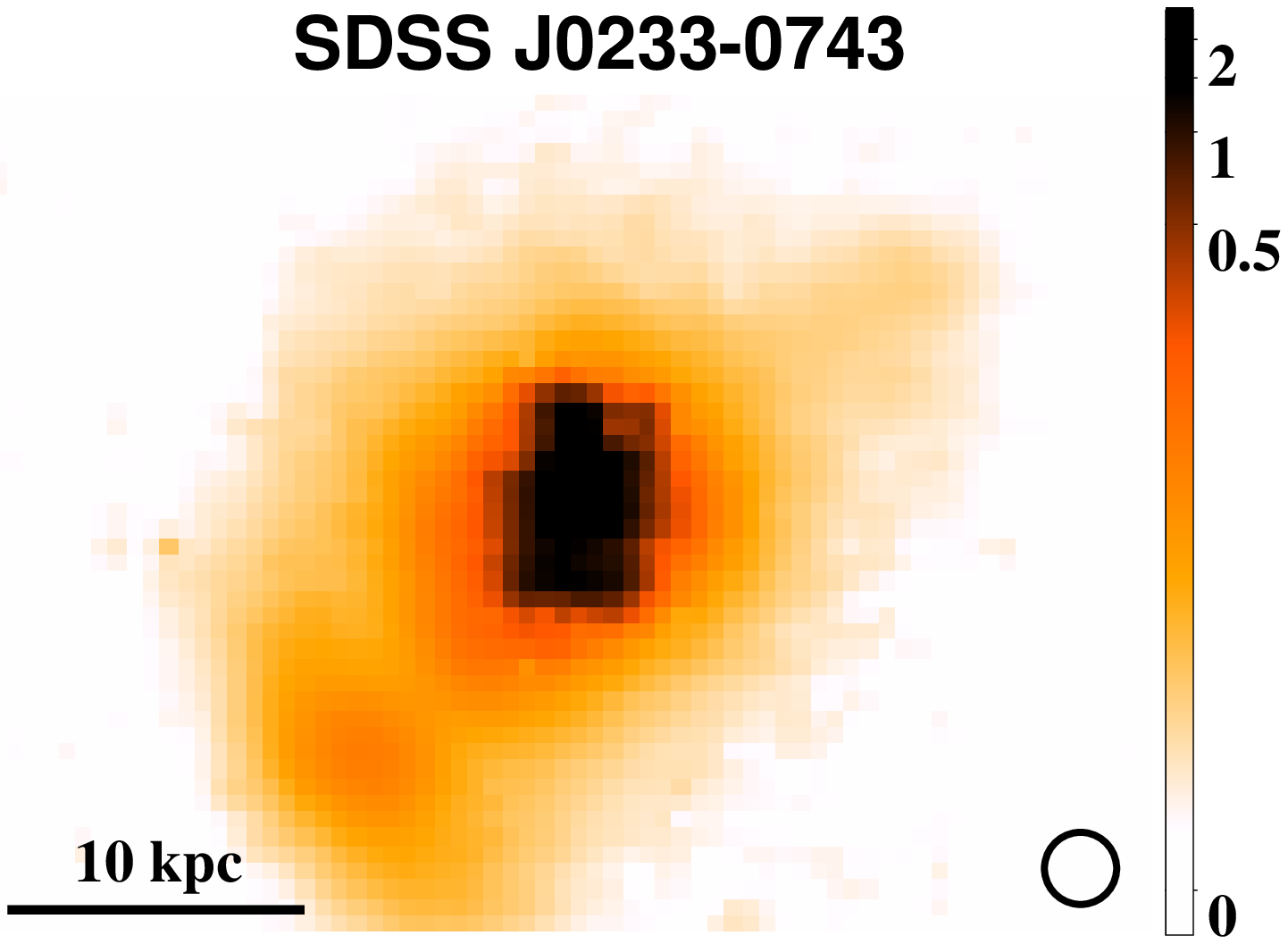}%
\includegraphics[scale=0.36,trim=0cm 0mm 30mm 0mm,clip=clip]{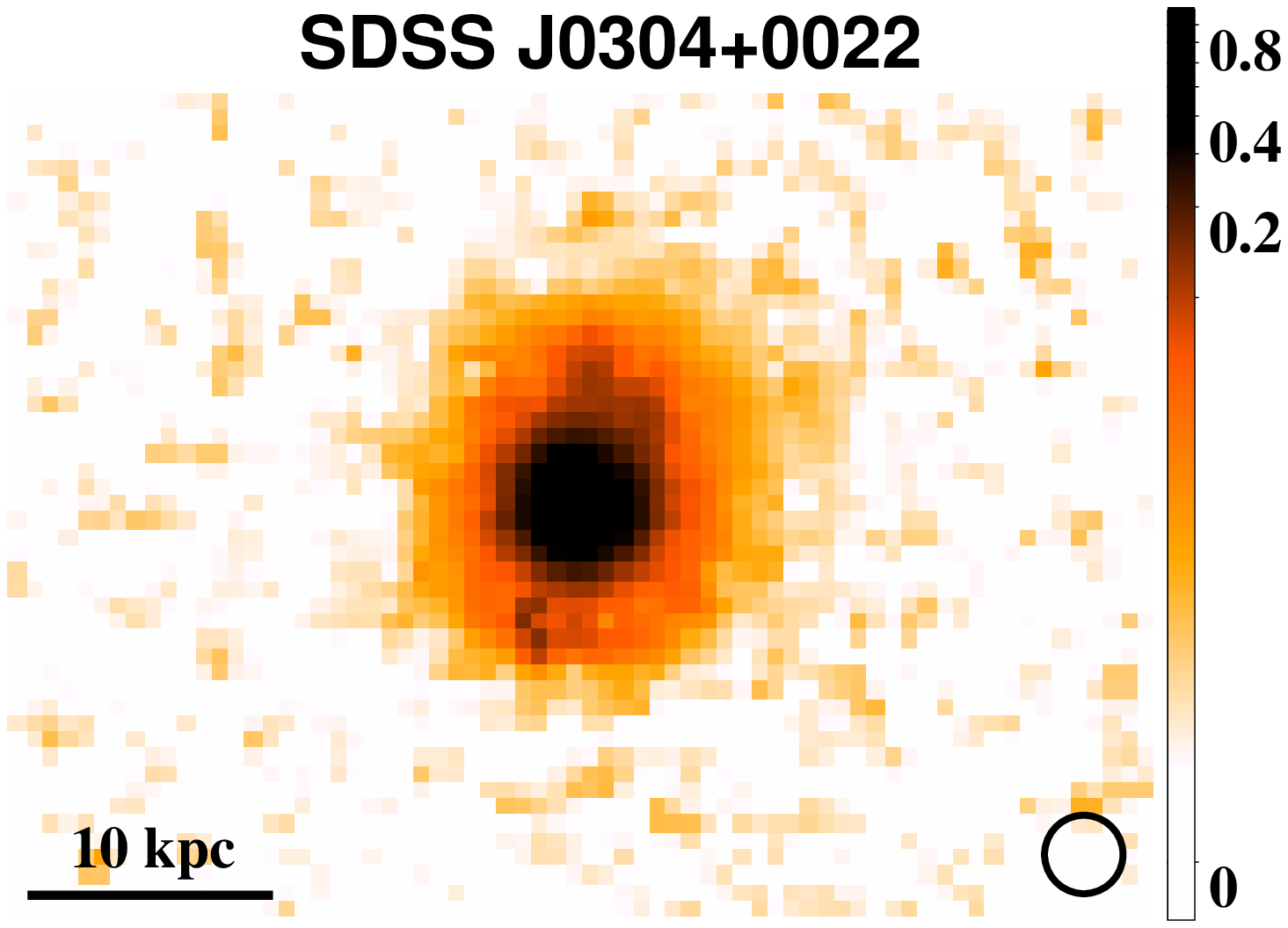}%
\includegraphics[scale=0.36,trim=0cm 0mm 30mm 0mm,clip=clip]{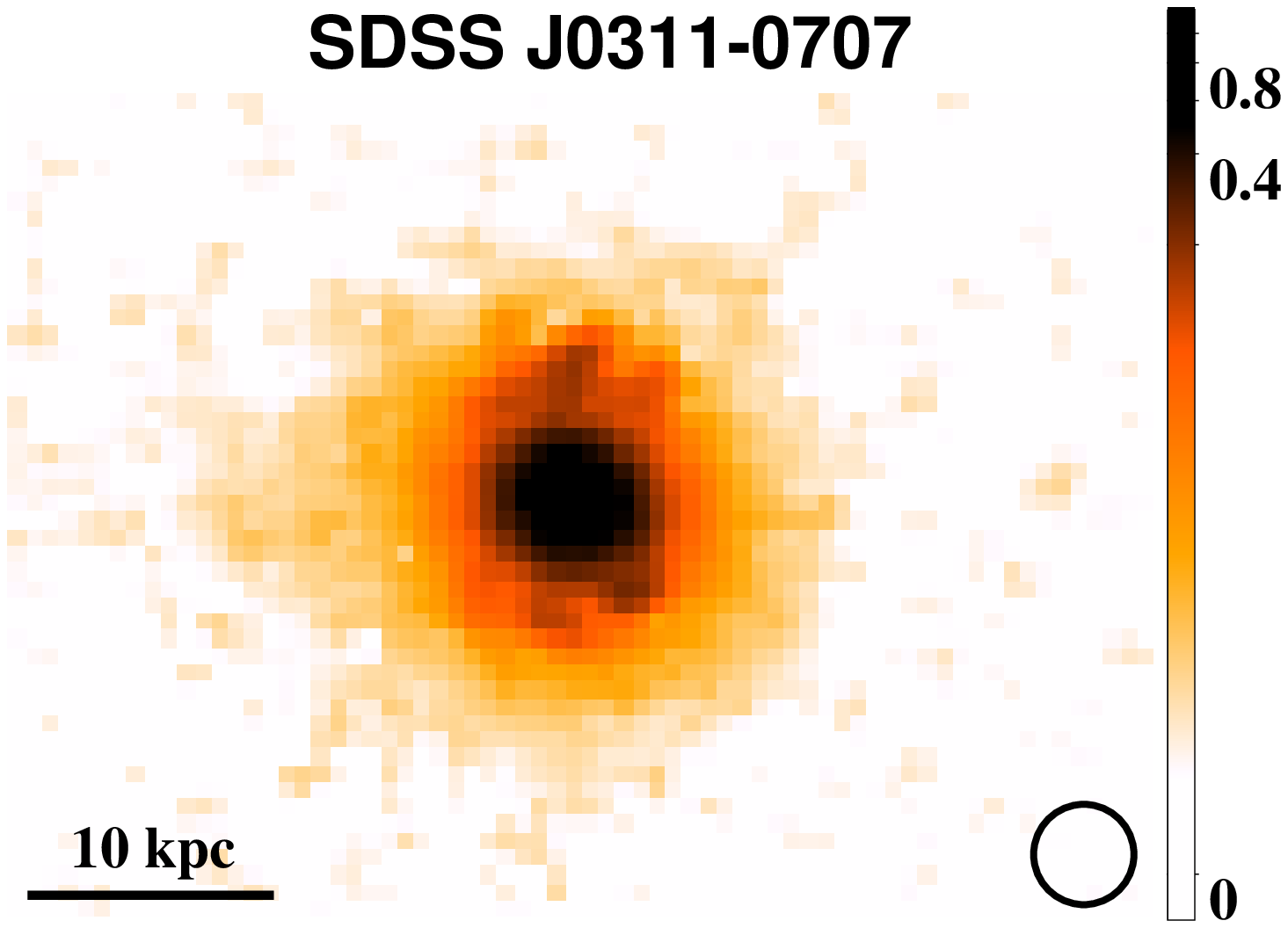}\\
\includegraphics[scale=0.36,trim=0cm 0mm 30mm 0mm,clip=clip]{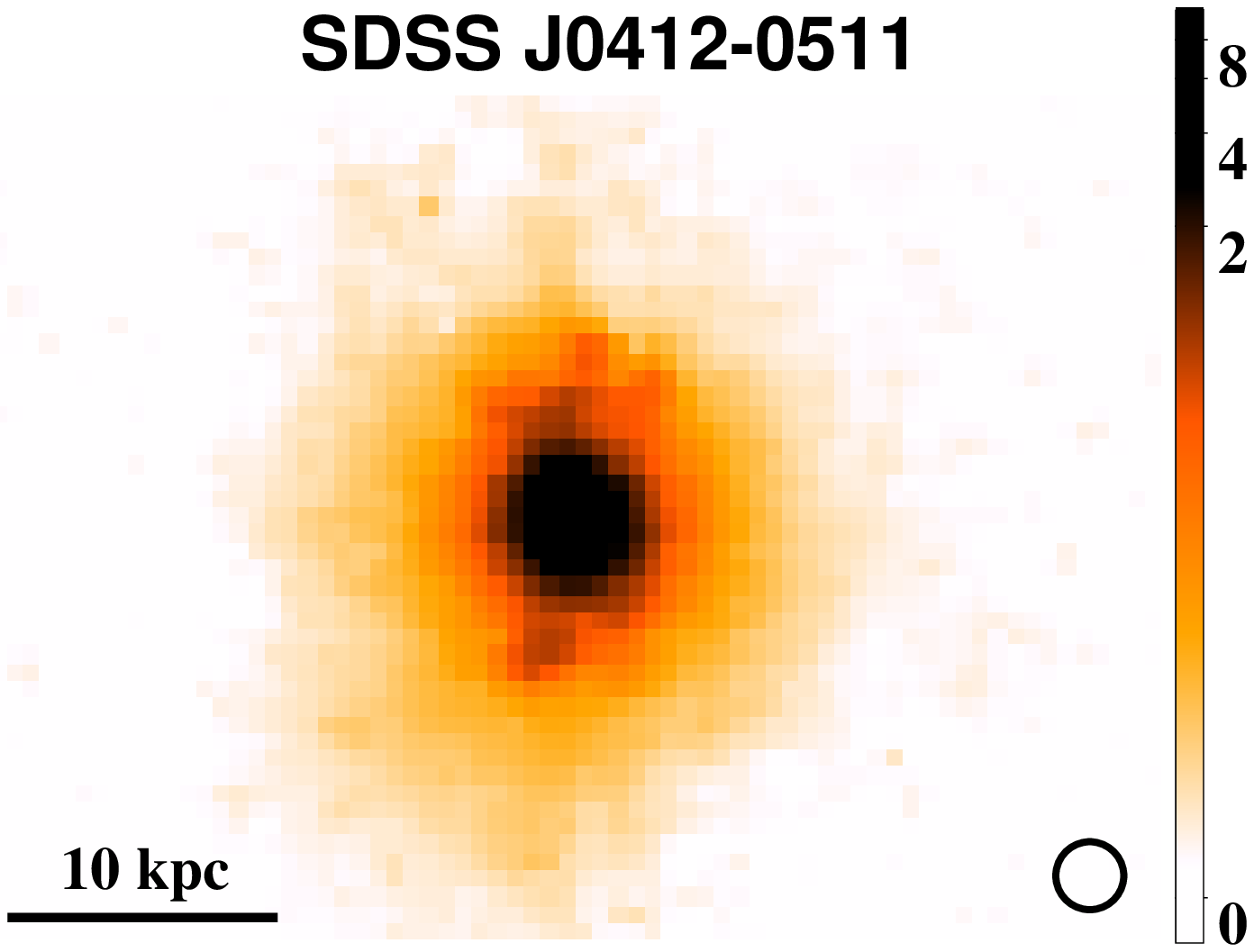}%
\includegraphics[scale=0.36,trim=0cm 0mm 30mm 0mm,clip=clip]{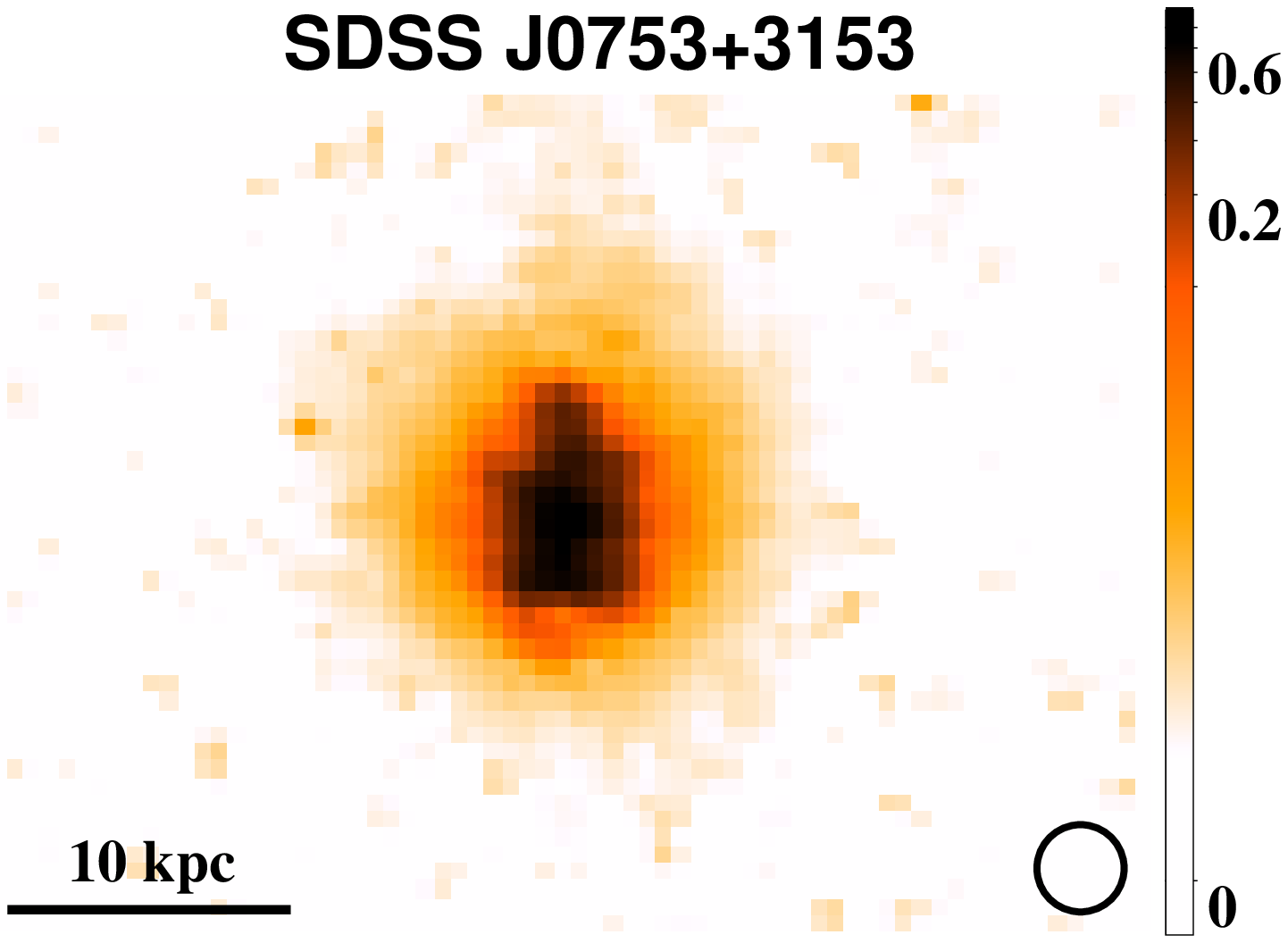}%
\includegraphics[scale=0.36,trim=0cm 0mm 30mm 0mm,clip=clip]{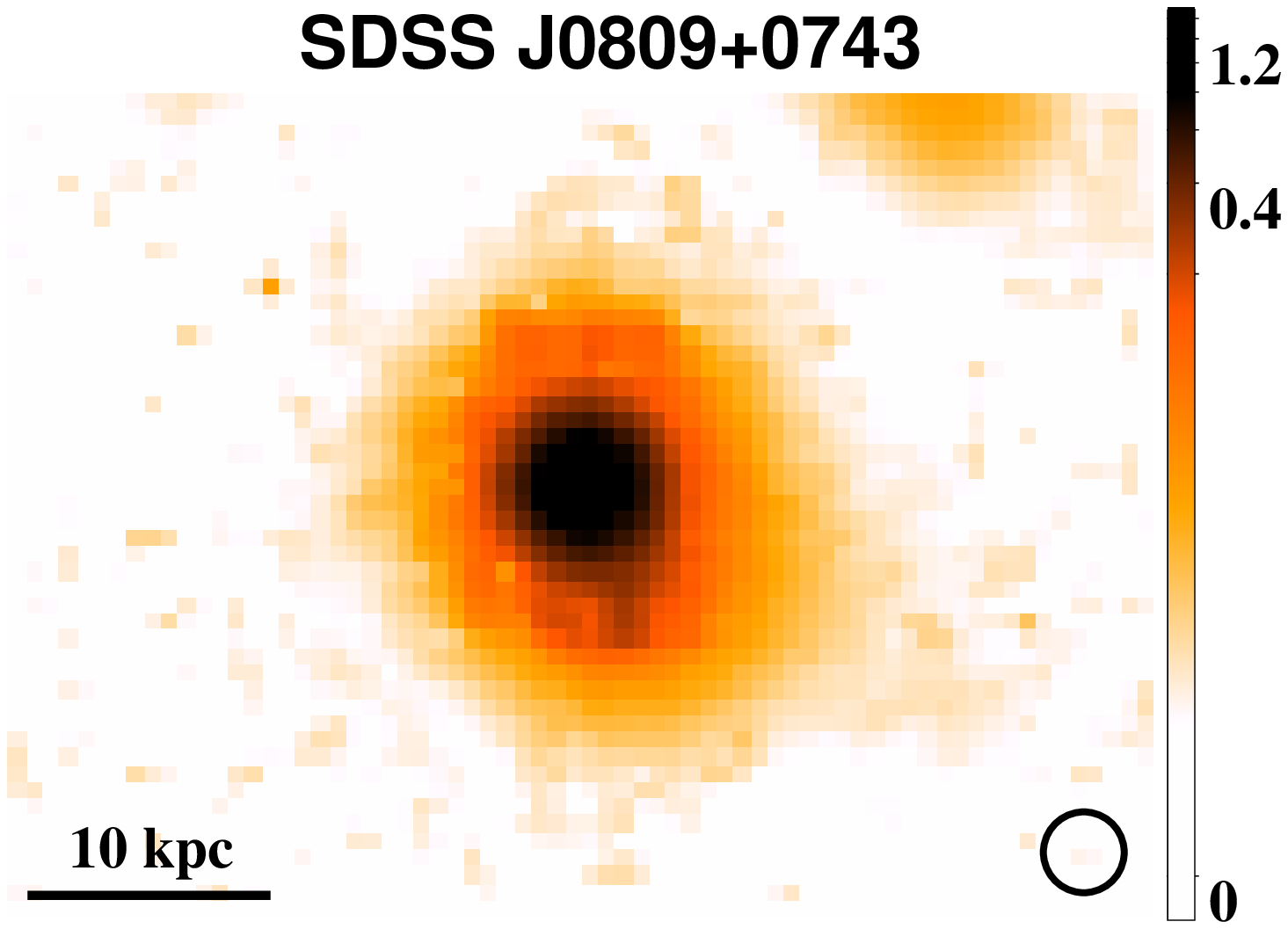}\\
\includegraphics[scale=0.36,trim=0cm 0mm 30mm 0mm,clip=clip]{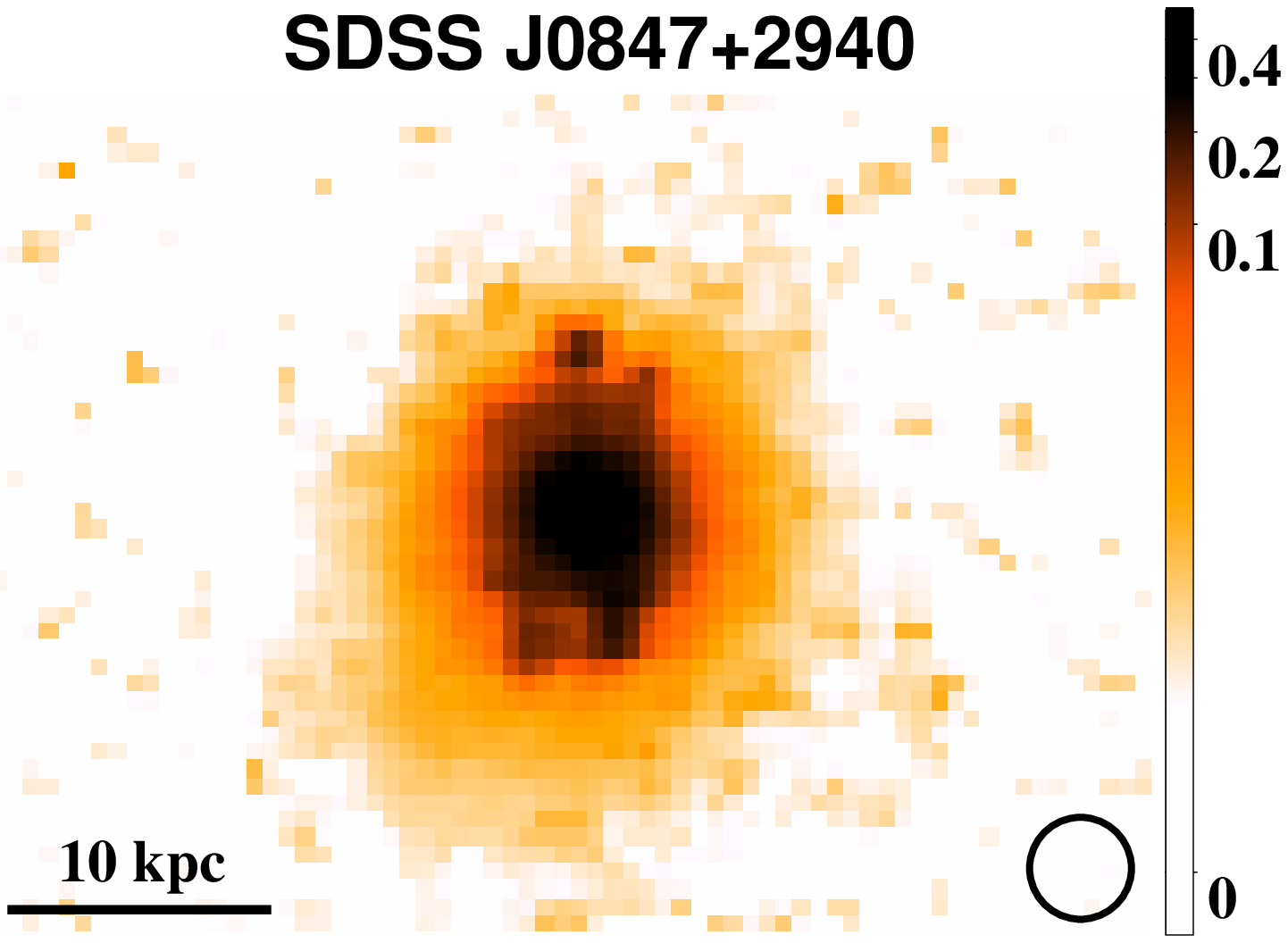}%
\includegraphics[scale=0.36,trim=0cm 0mm 30mm 0mm,clip=clip]{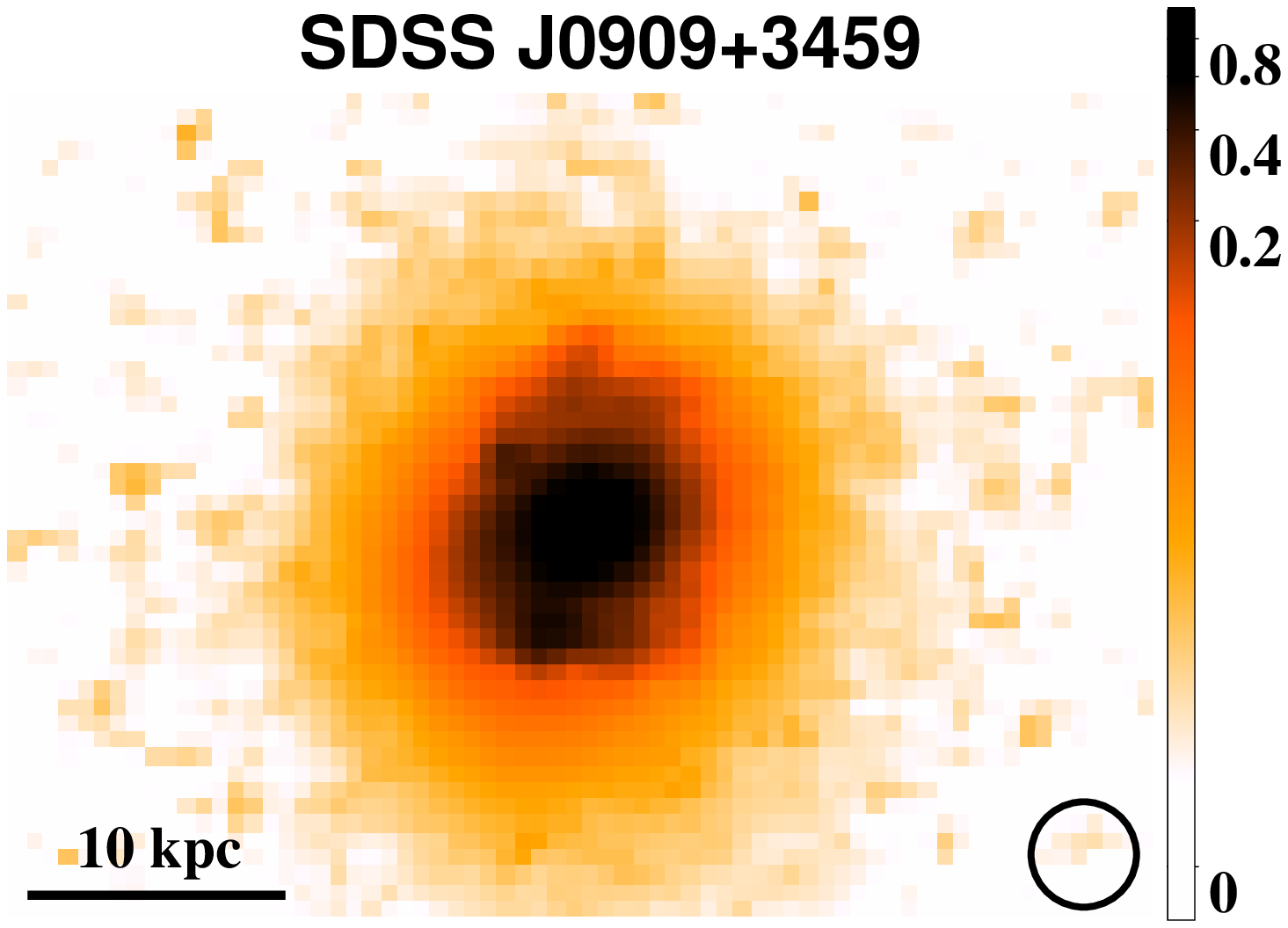}%
\includegraphics[scale=0.36,trim=0cm 0mm 30mm 0mm,clip=clip]{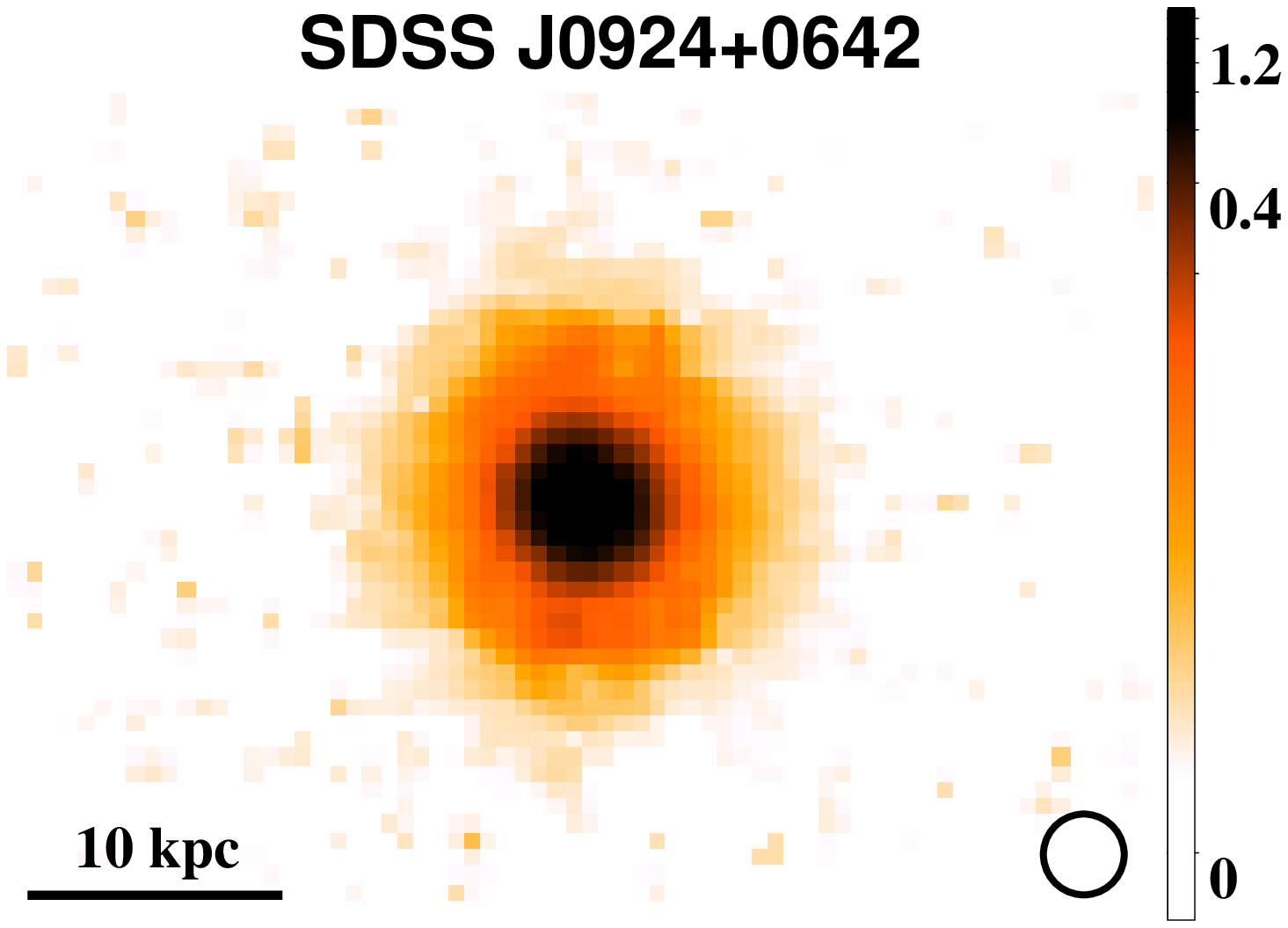}\\
\includegraphics[scale=0.36,trim=0cm 0mm 30mm 0mm,clip=clip]{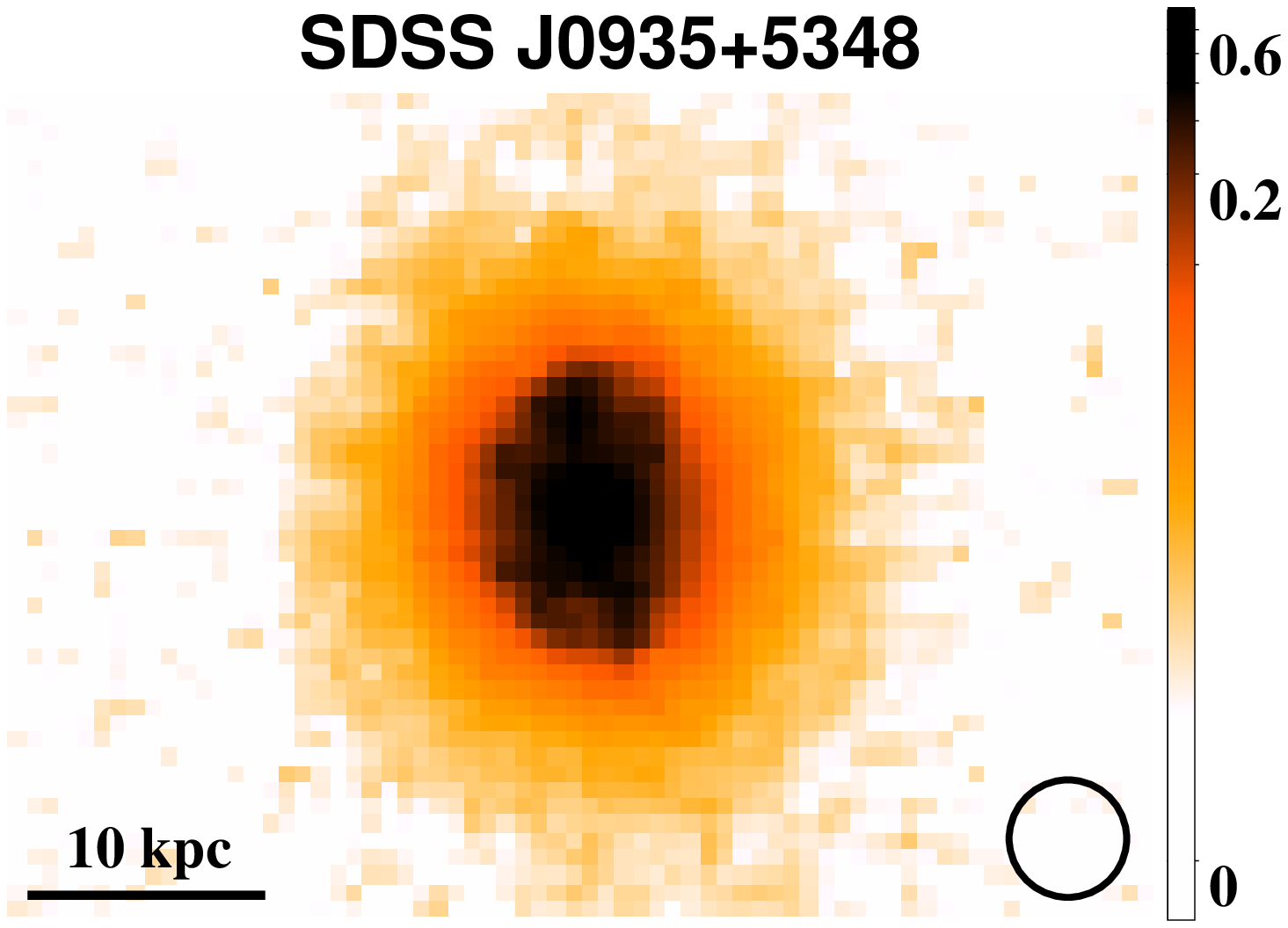}%
\includegraphics[scale=0.36,trim=0cm 0mm 30mm 0mm,clip=clip]{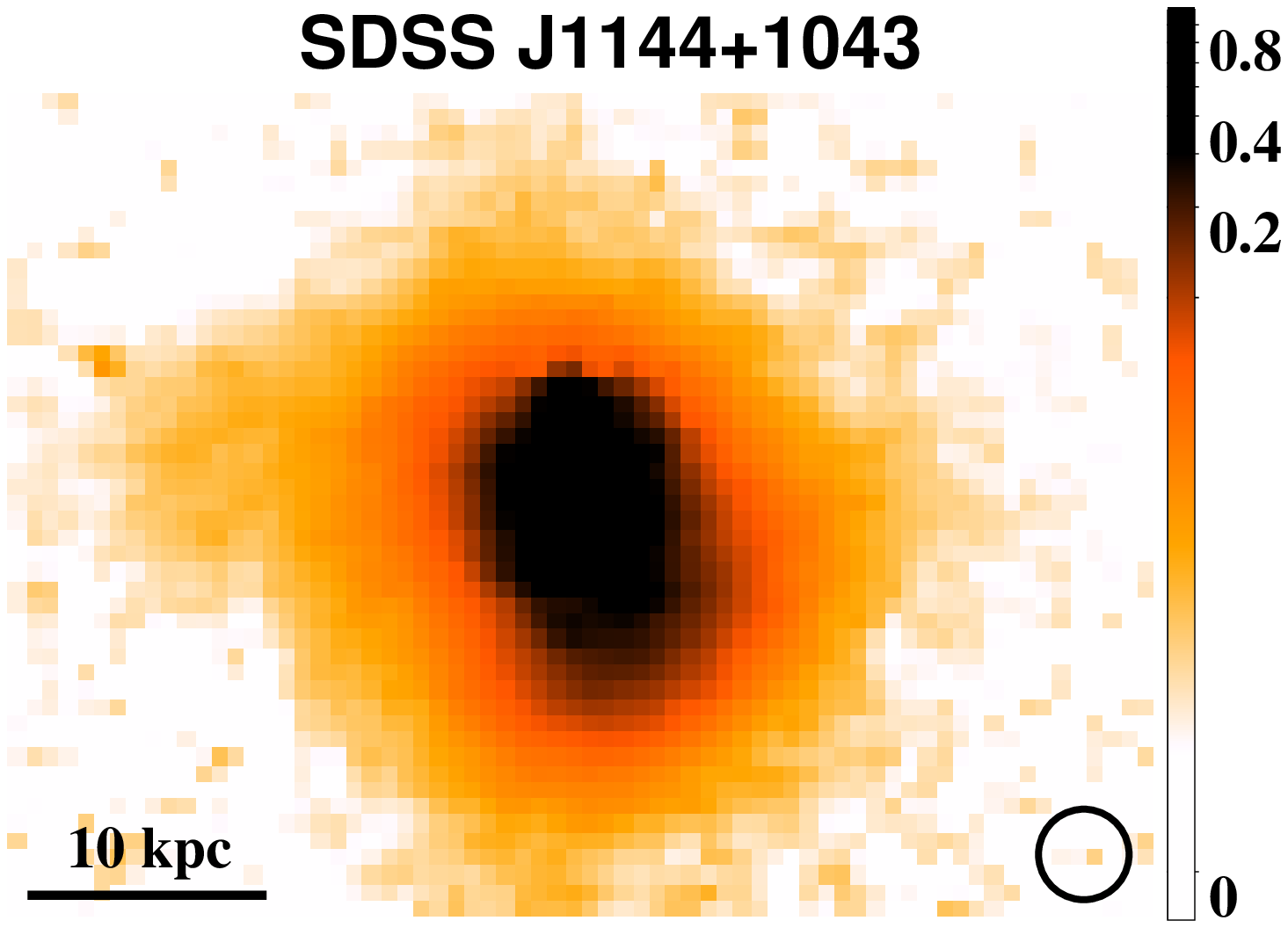}%
\includegraphics[scale=0.36,trim=0cm 0mm 30mm 0mm,clip=clip]{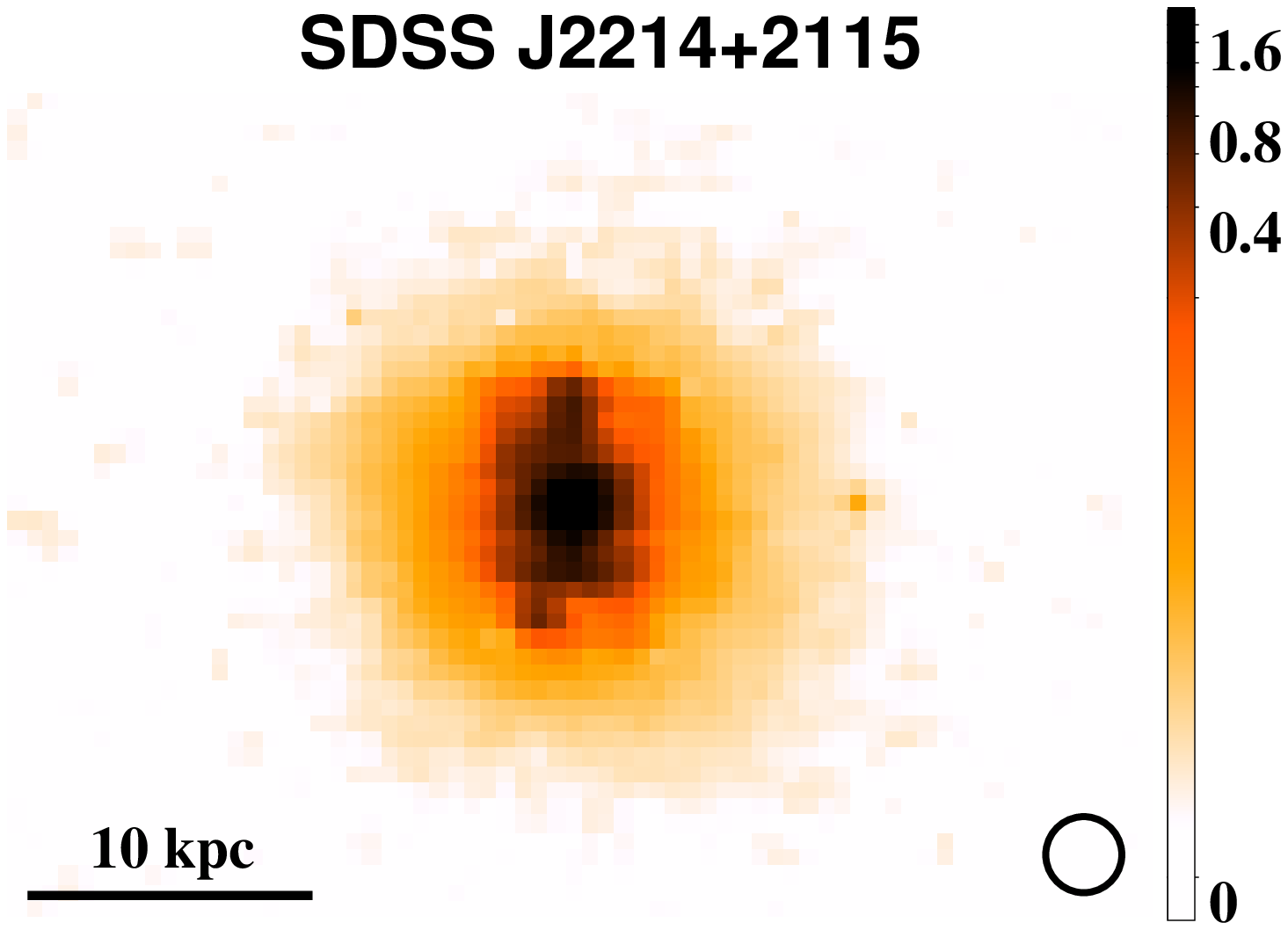}
\caption{Intensity maps of the \oiii\ line in the 12 unobscured quasars in our sample, shown on a logarithmic scale. The tick marks of the color bars are in units of $10^{-14}$ erg$^{-1}$ s$^{-1}$ cm$^{-2}$ arcsec$^{-2}$. In each spaxel, the \oiii\ flux is calculated from a multi-Gaussian fit to the spectral profile of the line. The seeing at the observing site is depicted by the open circle whose diameter is the FWHM of the PSF. SDSS J0924$+$0642 is the only object that is likely unresolved.}
\label{fig:flx}
\end{flushleft}
\end{figure*}

We can minimize these residuals by taking advantage of the spectroscopic information to better isolate the \oiii\ emission through line fitting. In each spaxel, we perform linear continuum + $N$-component Gaussian fits to the \oiii\ line profiles from the flux-calibrated data cubes following the strategy described in \citet[][Section 2.2]{liu13b}. As in the case of obscured quasars, we find that a combination of $N=1$--3 Gaussians is sufficient for every source in our sample so that the reduced $\chi^2$ is $<2$ except for sporadic problematic spaxels. The intensity of the \oiii\ line in each spaxel is then computed from the multi-Gaussian fit, not from the observed profile. The final \oiii\ surface brightness maps are shown in Figure \ref{fig:flx}. The surface brightness sensitivity (rms noise) of these maps is in the range $\sigma =$ (0.5--1.5) $\times 10^{-17}$ erg s$^{-1}$ cm$^{-2}$ arcsec$^{-2}$.

\begin{figure*}
\includegraphics[scale=0.6,trim=0cm 0mm 0mm 0mm]{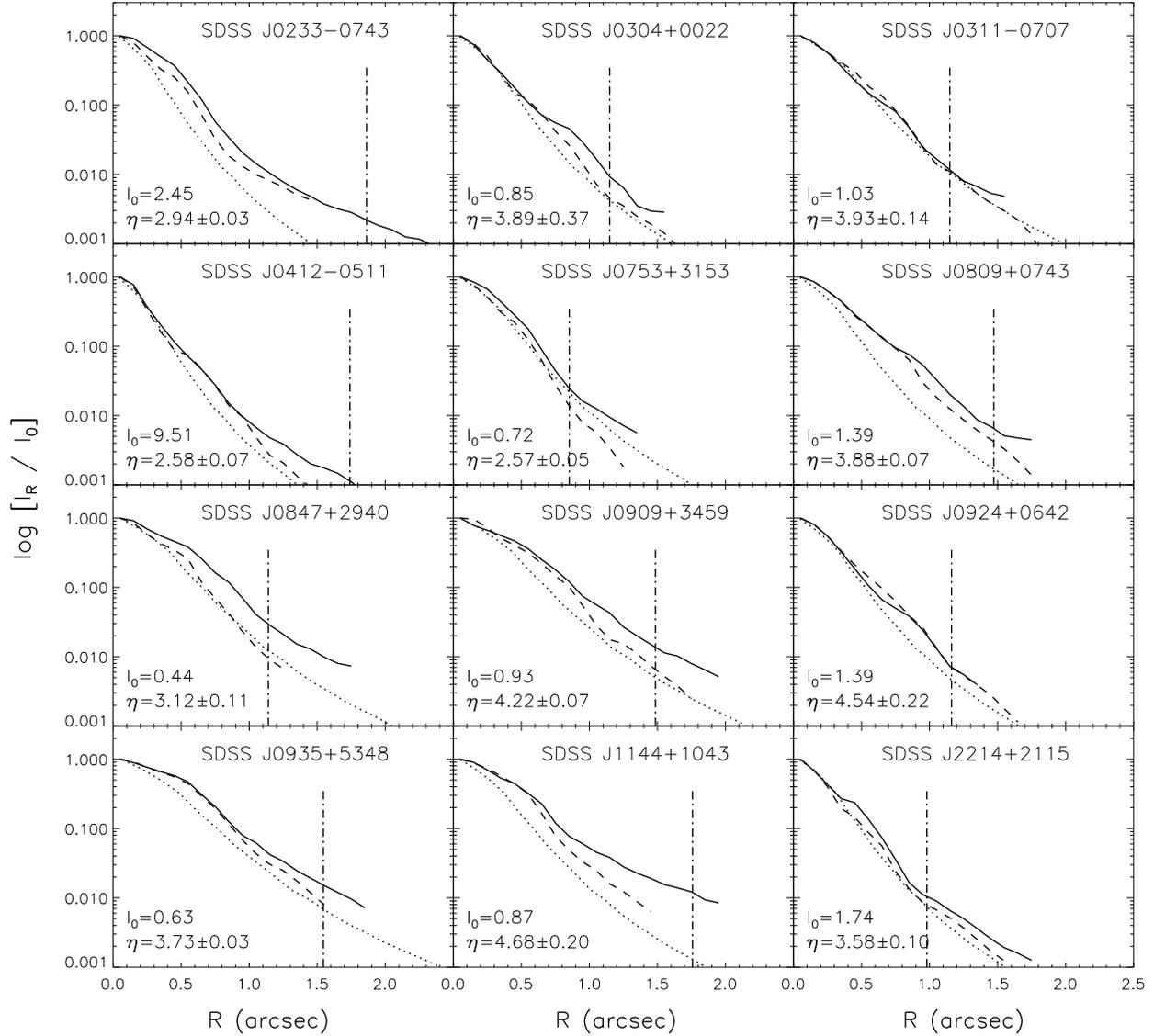}
\caption{Radial profiles of the surface brightness of \oiii\ emission (solid lines), of quasar continuum emission (dashed lines), and of the PSF (dotted lines). \oiii\ and quasar continuum profiles are clipped at 5$\sigma$. The peak \oiii\ surface brightness ($I_0$) marked on each panel is in units of $10^{-14}$ erg s$^{-1}$ cm$^{-2}$ arcsec$^{-2}$. Also given in each panel is the absolute value of the best-fitting exponent of power-law fits ($\eta$) to the outer regions of the nebulae along the semi-major axes (i.e. $I_{R,\soiii} \propto R^{-\eta}$). Vertical dash-dot lines mark the isophotal radii at a surface brightness level of $10^{-15}/(1+z)^4$ erg s$^{-1}$ cm$^{-2}$ arcsec$^{-1}$ (i.e. $R_{\rm int}$ listed in Table \ref{tab2}). }
\label{fig:psf}
\end{figure*}

In Figure \ref{fig:psf} we show a comparison of the surface brightness profiles of the \oiii\ emission and quasar continuum with the PSFs. All profiles are extracted using simple circular annuli. To determine the PSF for each observation, we directly measure the FWHM of the seeing from a sample of field stars in the acquision images taken right before the science exposure. We then use the radial profile of the standard star observed with the IFU, but rescale it in the spatial direction to reproduce the correct FWHM of the science observation. 

By comparing the radial profiles of the PSFs and the \oiii\ emission, we find that the majority (8/12) of the target \oiii\ nebulae are clearly more extended than the PSF. Among the remaining targets with \oiii\ profiles approximating the PSF, 3 quasars (SDSS J0311$-$0707, SDSS J0753$+$3153 and SDSS J2214$+$2115) are marginally resolved, because of unambiguous changes in radial velocity ($>$160 km s$^{-1}$) across their extents (Figure \ref{fig:Vmed}) or variations in their velocity dispersions (Figure \ref{fig:W80}). Kinematic differences across the nebulae give a strong indication that these sources are extended. The remaining object SDSS J0924$+$0642 is not resolved. Its spatial profile is consistent with the PSF, and its velocity field, although well organized, has a measured maximum velocity difference of only $\sim80$ km s$^{-1}$, and its velocity dispersion is almost constant in all parts of the nebula. 


In order to characterize the physical extents of the nebulae around obscured quasars, we previously defined four different size measures and discussed their advantages 
and disadvantages \citep{liu13a}: $R_{5\sigma}$, the semi-major axis of the best-fitting ellipse enclosing S/N$\geqslant$5 spaxels; $R_{\rm eff}$, the half-light radius;
$R_{\rm obs}$, the isophotal radius at a surface brightness of 
$10^{-16}$ erg s$^{-1}$ cm$^{-1}$ arcsec$^{-2}$; $R_{\rm int}$, the isophotal 
radius at a surface brightness of $10^{-15}/(1+z)^4$ erg s$^{-1}$ cm$^{-1}$ 
arcsec$^{-2}$ corrected for cosmological dimming.
In Table \ref{tab2} we report all of these quantities for the \oiii\
line and the first two size measures for the continuum as well. As we pointed out in \citet{liu13a}, the most physically motivated measure for the \oiii\ extent is $R_{\rm int}$, which is independent of redshift and the depth of the data and is thus most suitable for comparing sizes of nebulae from different observations. 
For the 4 quasars that are marginally or unresolved, we report the measured sizes as upper limits.

We also create continuum images of our quasars by collapsing the spectrum over a wavelength interval free of line signatures. Depending on the quasar redshift and wavelength coverage of the IFU data, the median wavelength range we use is 4700 to 5200 \AA\ excluding line emission features. The spatial profiles of the continuum emisson are depicted by dashed lines in Figure \ref{fig:psf}. The continuum is more compact than \oiii\ emission, as is expected since it is dominated by the emission of the point-like quasar. The continuum emission is approximately consistent with the PSF in about 5 objects (noted in Table \ref{tab1}), but is resolved or marginally resolved in the other objects. The resolved blue continuum may be due to quasar light scattered off of the interstellar matter of the host galaxy \citep{borg08} or due to star formation in the quasar host \citep{leta07,silv09}. 

\section{Ionized gas nebulae in unobscured and obscured quasars}
\label{sec:compare}

Our observations of the type 1 quasars are well matched to those of the type 2 quasar sample we conducted previously in redshift, \oiii\ luminosity, and parameters of observations and data reduction. The major difference between the two samples is that the issues related to the PSF of the unobscured quasar need to be addressed carefully for type 1 objects. Because most \oiii\ nebulae around type 1 quasars are well-resolved, we are in a good position to directly compare the ionized gas distribution around the two populations. 

\subsection{Physical extents and morphology}

Unobscured quasars, like their obscured peers, are surrounded by ionized gas nebulae that extend over a spatial scale comparable to the typical size of a galaxy in every case. 


{\it Sizes.} Taking data from Table \ref{tab2}, we first compare the sizes of the \oiii\ nebulae using the semi-major axes of the elliptical isophotes fitted at the 5--$\sigma$ surface brightness limit. In the type 2 sample, SDSS J0841$+$2042 and SDSS J1039$+$4512 can be categorized as marginally resolved by our standard. Excluding marginally or unresolved targets from both samples, we find the median and the standard deviation of the detected \oiii\ nebulae to be $\langle R_{5\sigma}^{\rm T1}\rangle=13.3\pm2.0$ for the unobscured sample and $\langle R_{5\sigma}^{\rm T2}\rangle=14.1\pm3.6$ for the obscured quasars, and $\langle R_{\rm int}^{\rm T1}\rangle=10.7\pm1.7$ for the unobscured objects and $\langle R_{\rm int}^{\rm T2}\rangle=12.9\pm3.4$ for the obscured ones when $R_{\rm int}$ (sizes of the nebulae at a cosmologically corrected fixed surface brightness limit) is considered. 
To compare the survival distributions of the two samples with censored data, we perform the logrank test using the {\em twosampt} task from the {\sc iraf stsdas} package, finding the probabilities that they are drawn from the same parent distribution are 0.57 and 0.39 for $R_{5\sigma}$ and $R_{\rm int}$, respectively.
We therefore conclude that no significant difference is seen between the two samples in their sizes, though the nebula of the unobscured quasars appear slightly more compact than those of the obscured sample, and the fraction of marginally or unresolved objects is slightly higher (type 1: 4/12, type 2: 2/11). 

{\it Morphology.} In addition to the similar spatial extents, the
unobscured quasar nebulae show regular morphology and are nearly perfectly 
round, which is also very similar to the obscured sample. 
With marginally or unresolved objects excluded from both type 1 and type
2 samples, the median ellipticity measured 
at a surface brightness of $10^{-15}/(1+z)^4$ erg s$^{-1}$ cm$^{-2}$ arcsec$^{-2}$ is 
$\langle \epsilon_{\rm int}^{\rm T1}\rangle=0.14\pm0.11$ for the unobscured objects and 
$\langle \epsilon_{\rm int}^{\rm T2}\rangle=0.14\pm0.11$. 
The logrank test gives a probability of 1.00 that both samples follow the same distribution.
Therefore, we do not observe significant morphological difference between the two samples.

{\it Radial profiles.} The radial distribution of \oiii\ emission
from the unobscured quasars also follow loci very similar to those of the 
obscured sample. When the outer ($R \gtrsim 1$\arcsec) part of the \oiii\ surface brightness 
profile is fit by a power law $I_{R,\soiii}\propto R^{-\eta}$ and the
4 marginally or unresolved quasars are excluded, we find that
the best-fit exponent $\eta$ ranges from 2.58 to 4.68 (Table \ref{tab1}) 
and has a median $\langle \eta^{\rm T1}\rangle=3.88\pm0.70$, well 
consistent with the results from the obscured quasar sample which has 
$\eta=3.00$--5.70 and a median value of $\langle \eta^{\rm T2}\rangle=3.53\pm0.87$.
The logrank test shows that the probabilities that the two samples are drawn 
from the same parent distribution is 0.28.

Hence, we conclude that no significant difference in sizes, morphology or radial 
distribution is seen between the gas nebulae of unobscured and obscured 
quasar samples.

\subsection{Kinematic properties}

\begin{figure*}
\begin{flushleft}
\includegraphics[scale=0.33,trim=0cm 0mm 10mm 0mm]{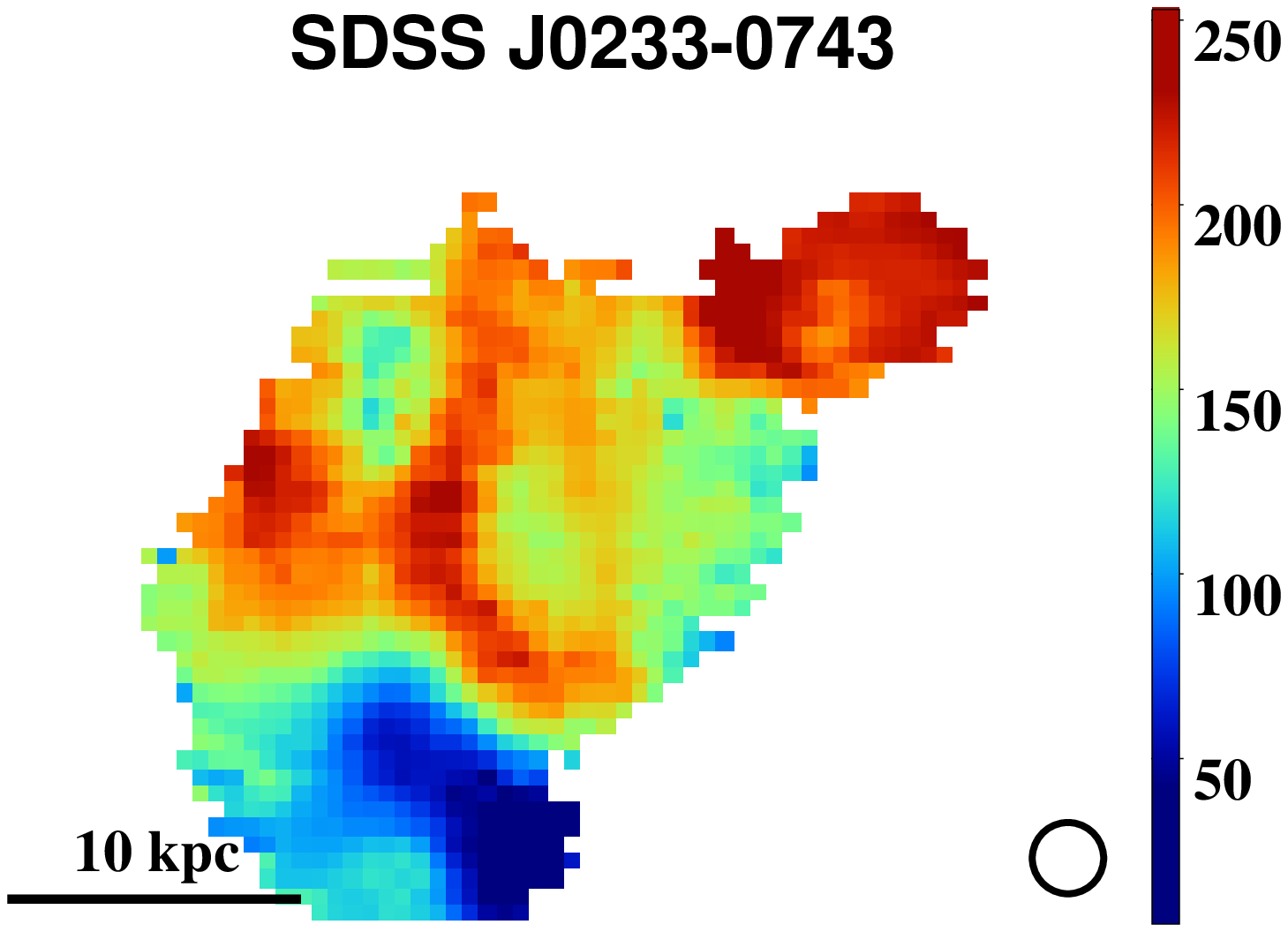}%
\includegraphics[scale=0.33,trim=0cm 0mm 10mm 0mm]{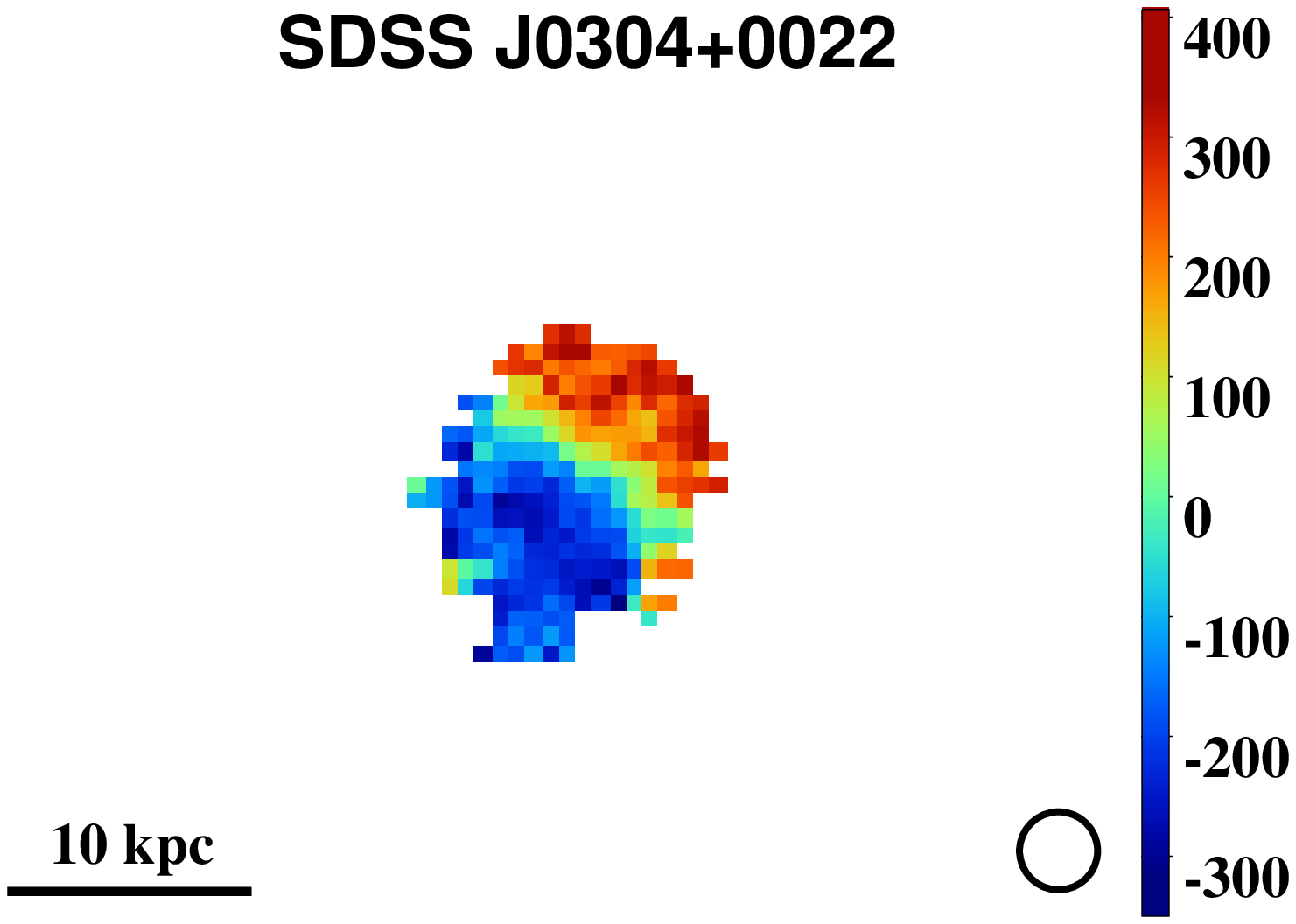}%
\includegraphics[scale=0.33,trim=0cm 0mm 10mm 0mm]{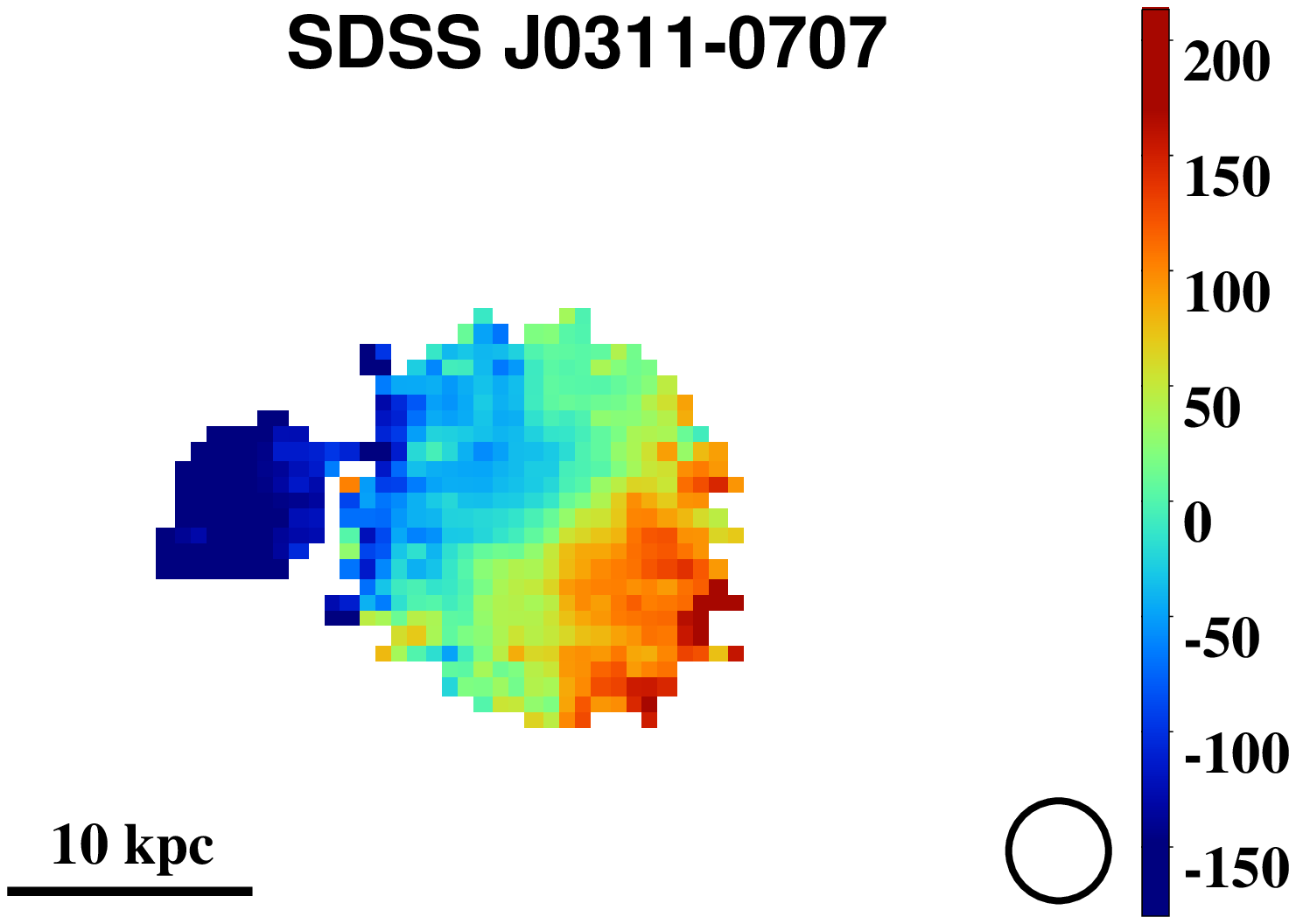}\\
\includegraphics[scale=0.33,trim=0cm 0mm 10mm 0mm]{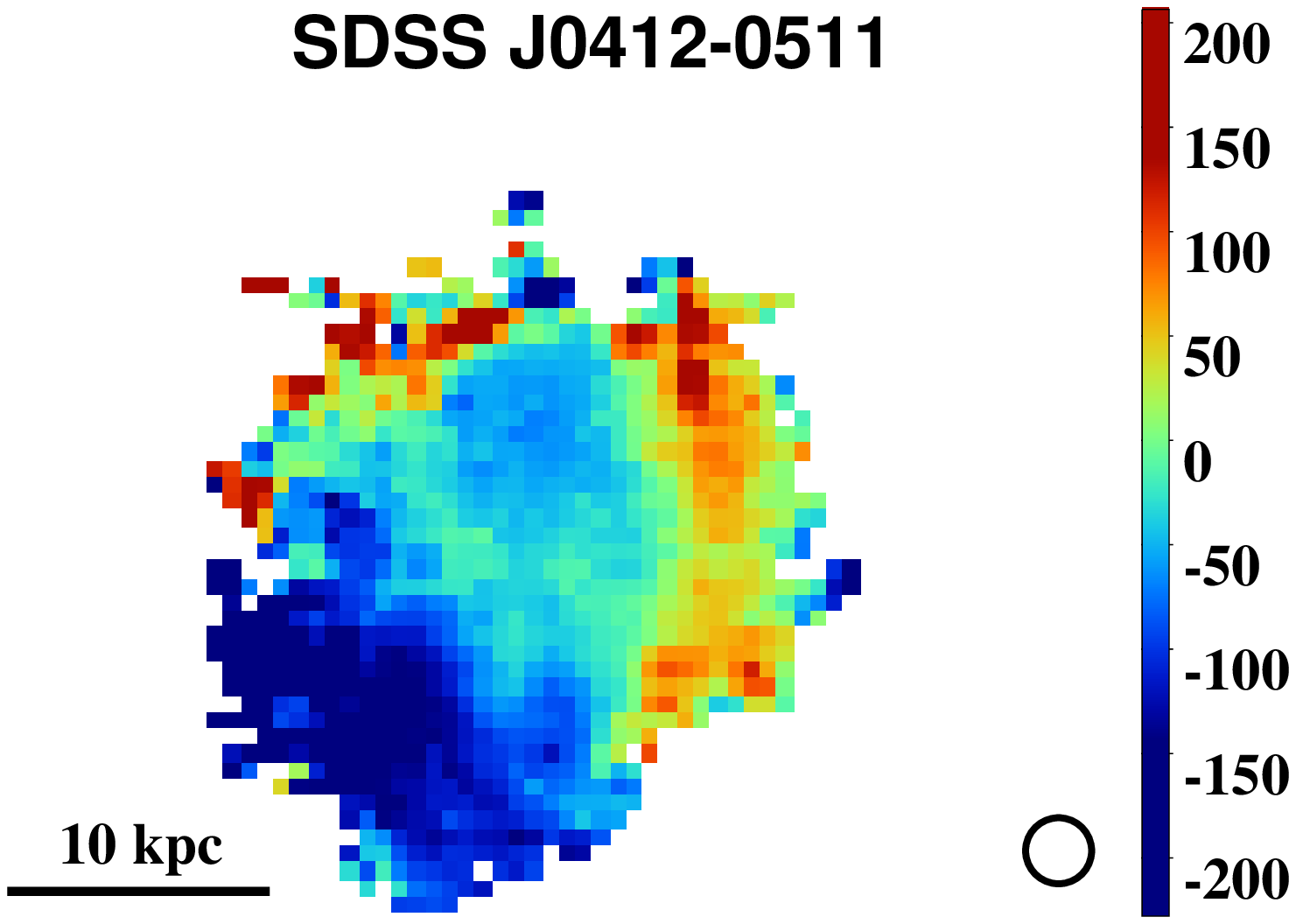}%
\includegraphics[scale=0.33,trim=0cm 0mm 10mm 0mm]{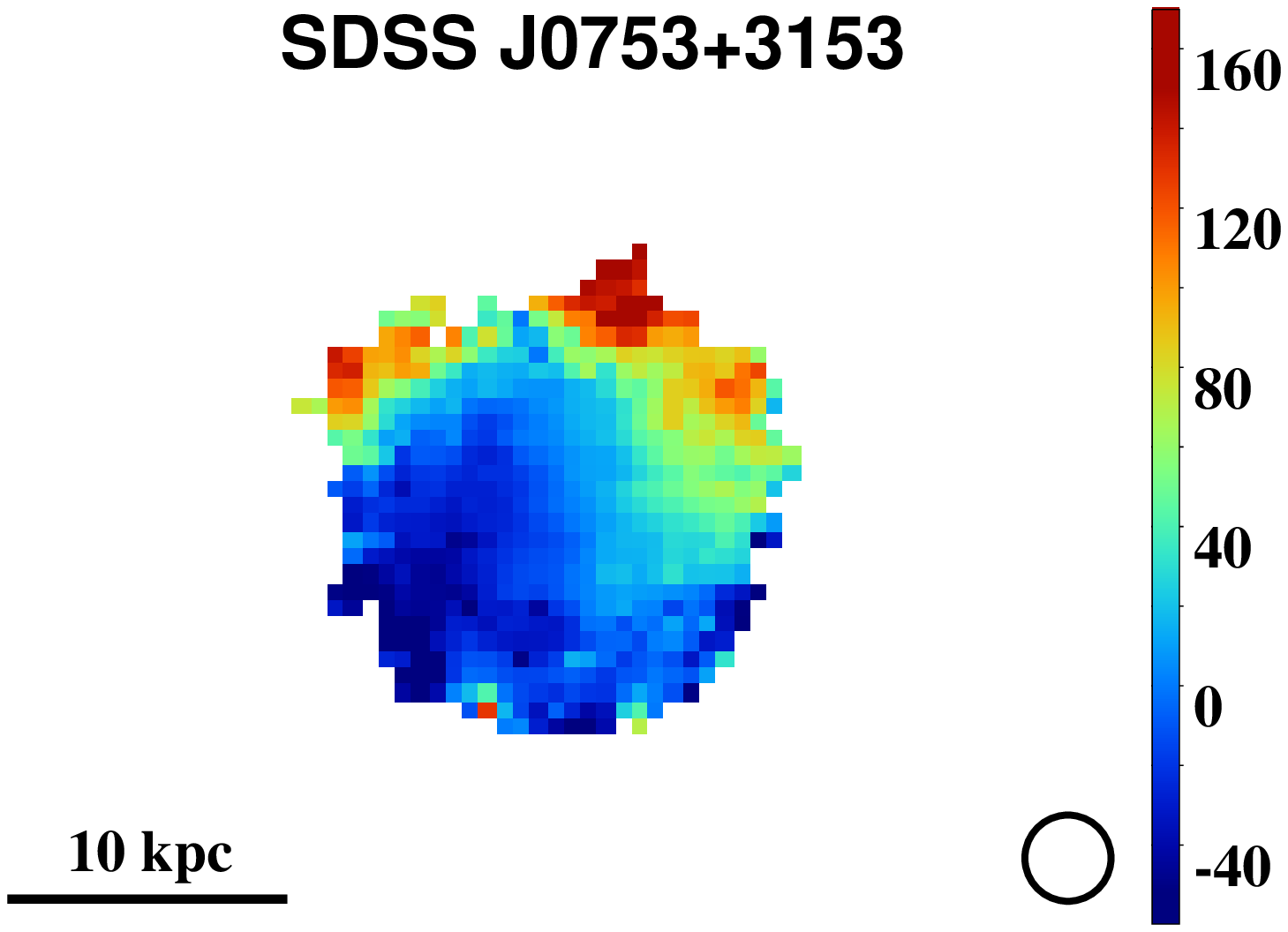}%
\includegraphics[scale=0.33,trim=0cm 0mm 10mm 0mm]{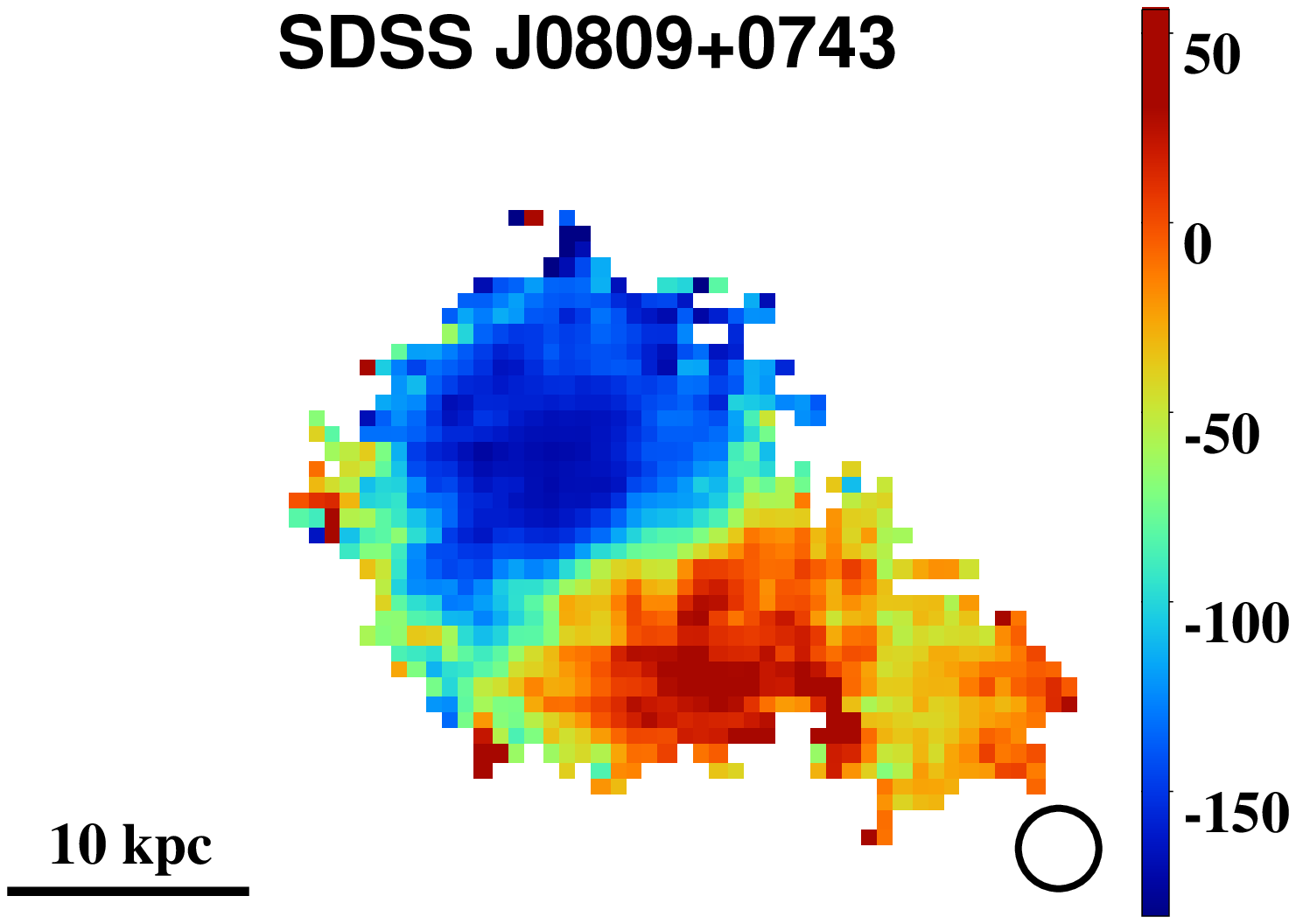}\\
\includegraphics[scale=0.33,trim=0cm 0mm 10mm 0mm]{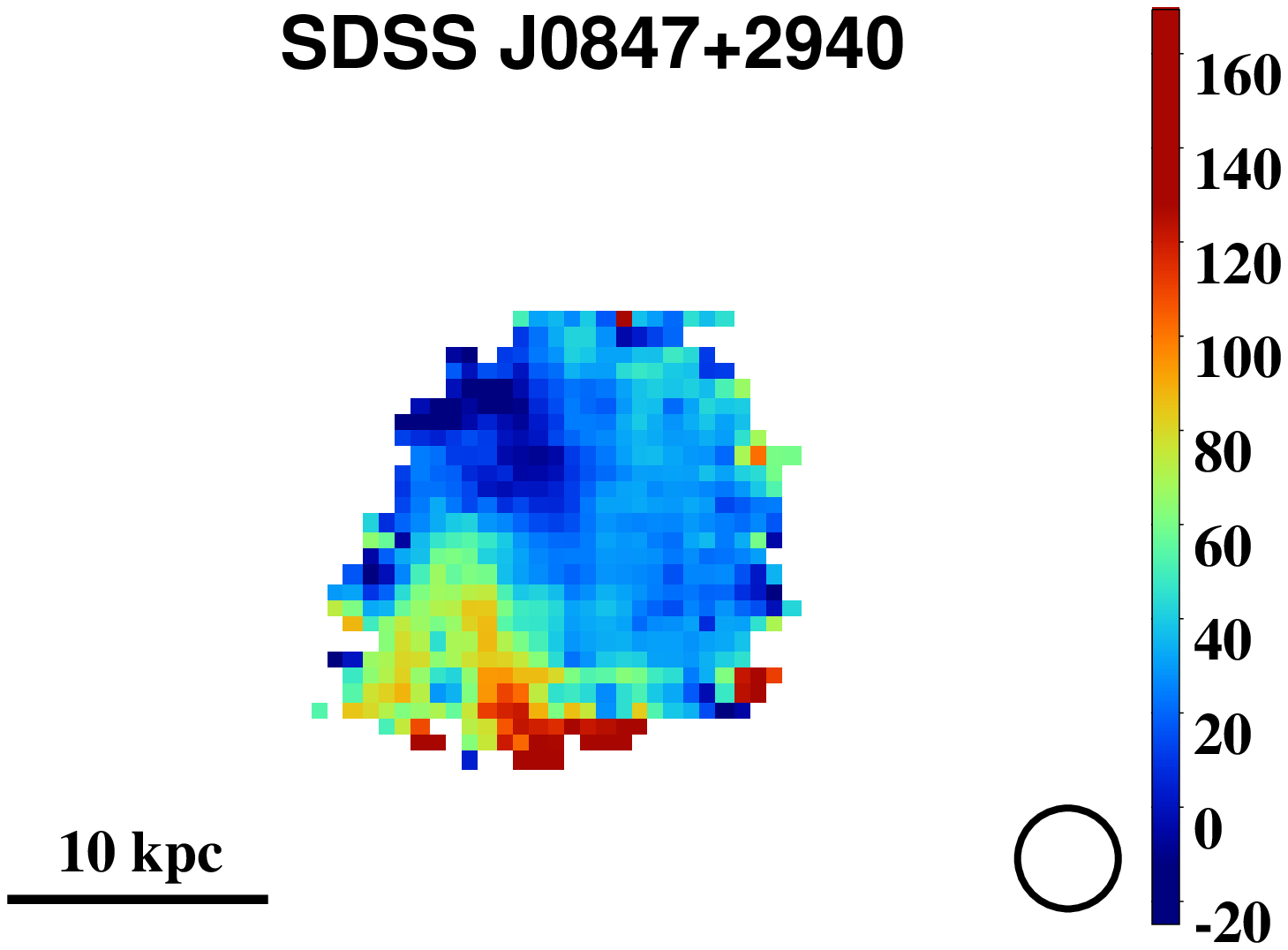}%
\includegraphics[scale=0.33,trim=0cm 0mm 10mm 0mm]{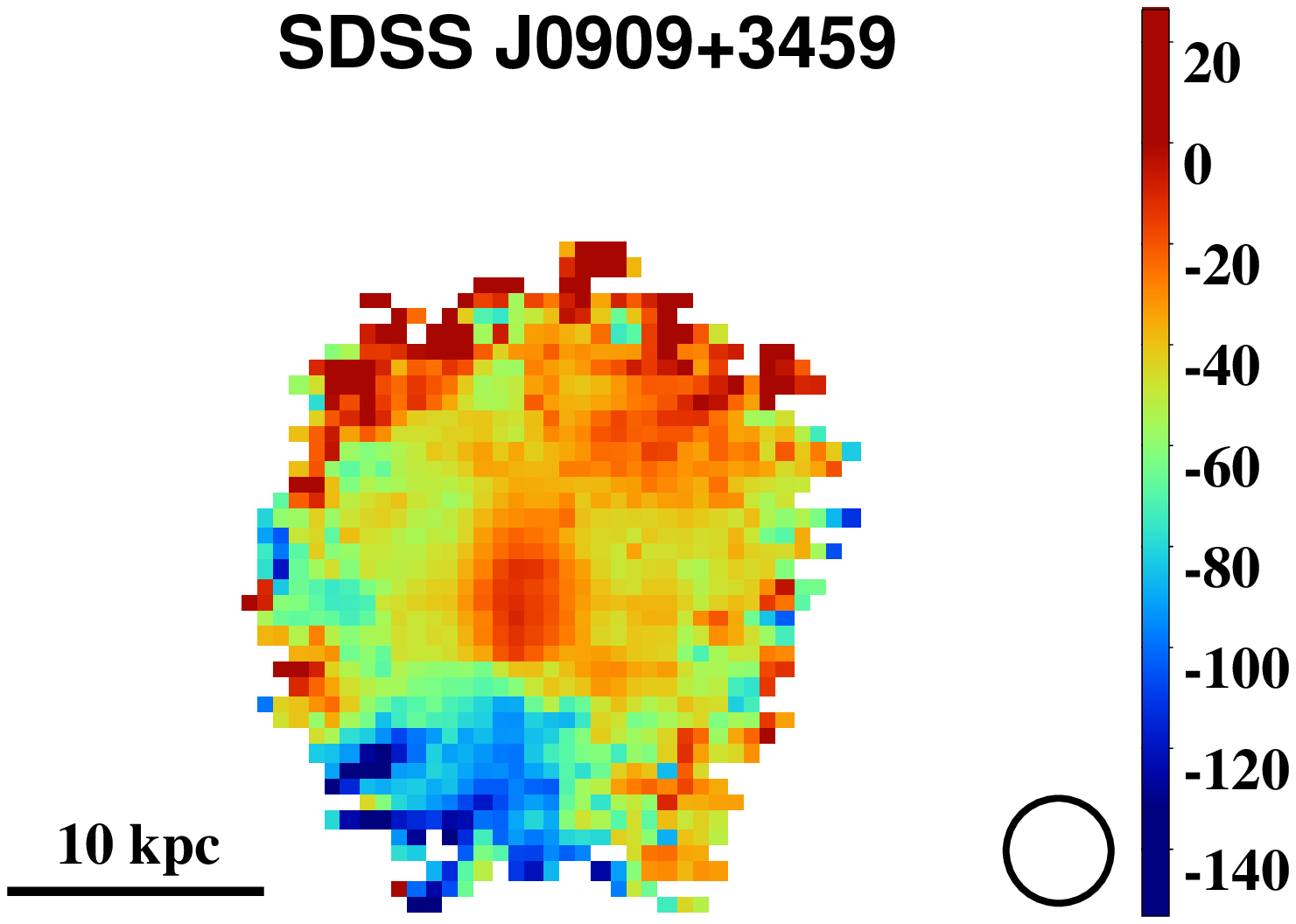}%
\includegraphics[scale=0.33,trim=0cm 0mm 10mm 0mm]{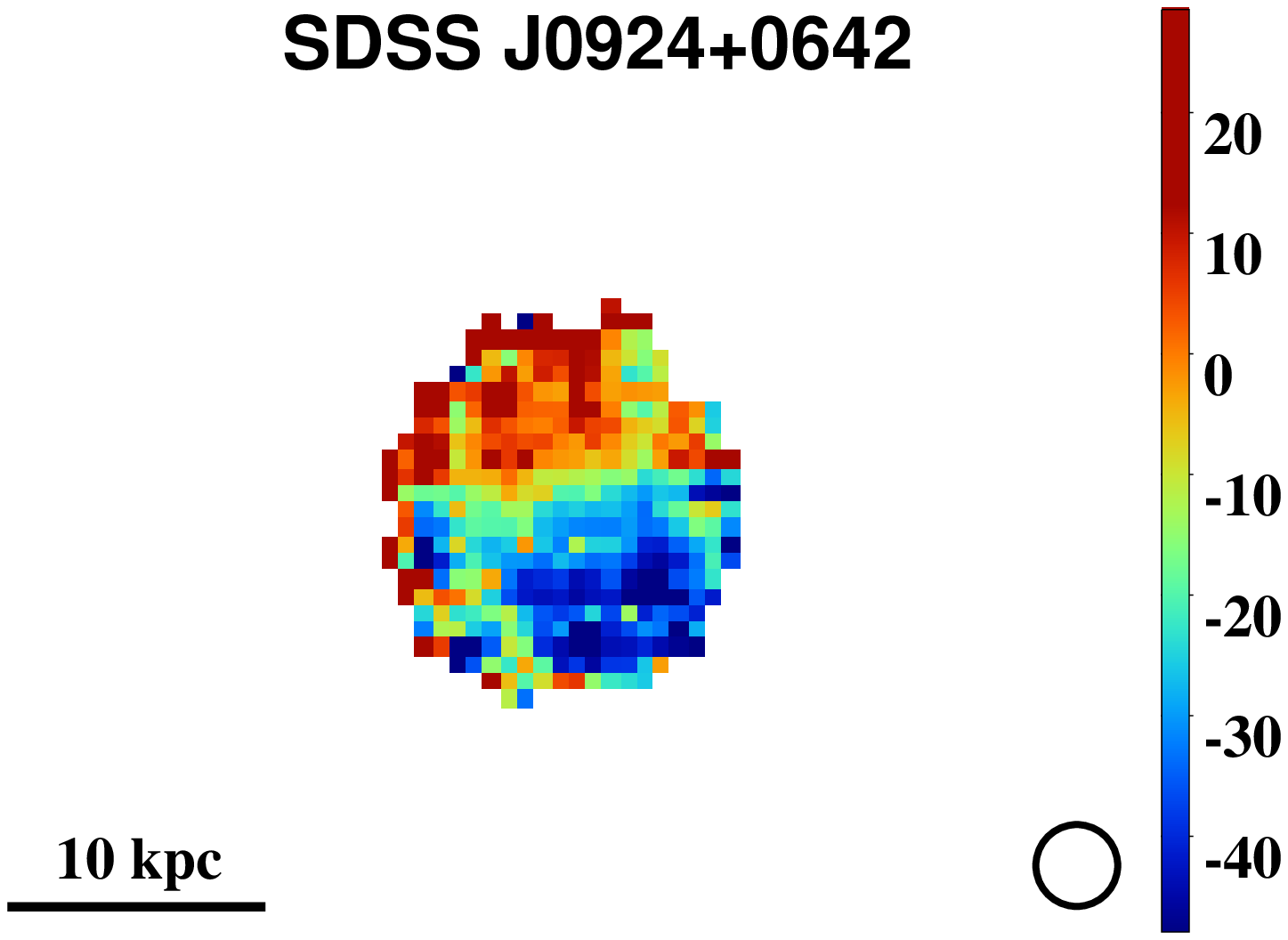}\\
\includegraphics[scale=0.33,trim=0cm 0mm 10mm 0mm]{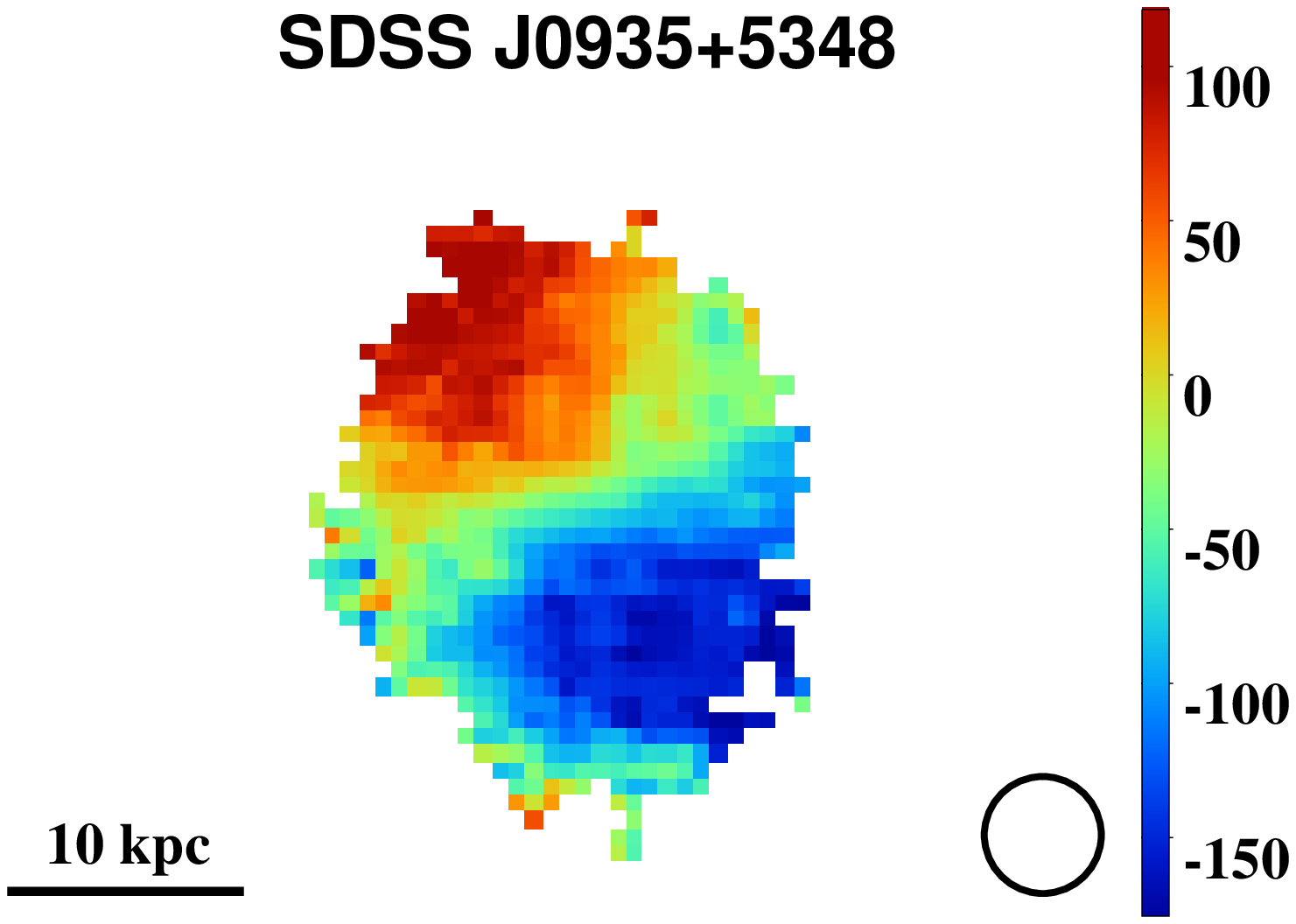}%
\includegraphics[scale=0.33,trim=0cm 0mm 10mm 0mm]{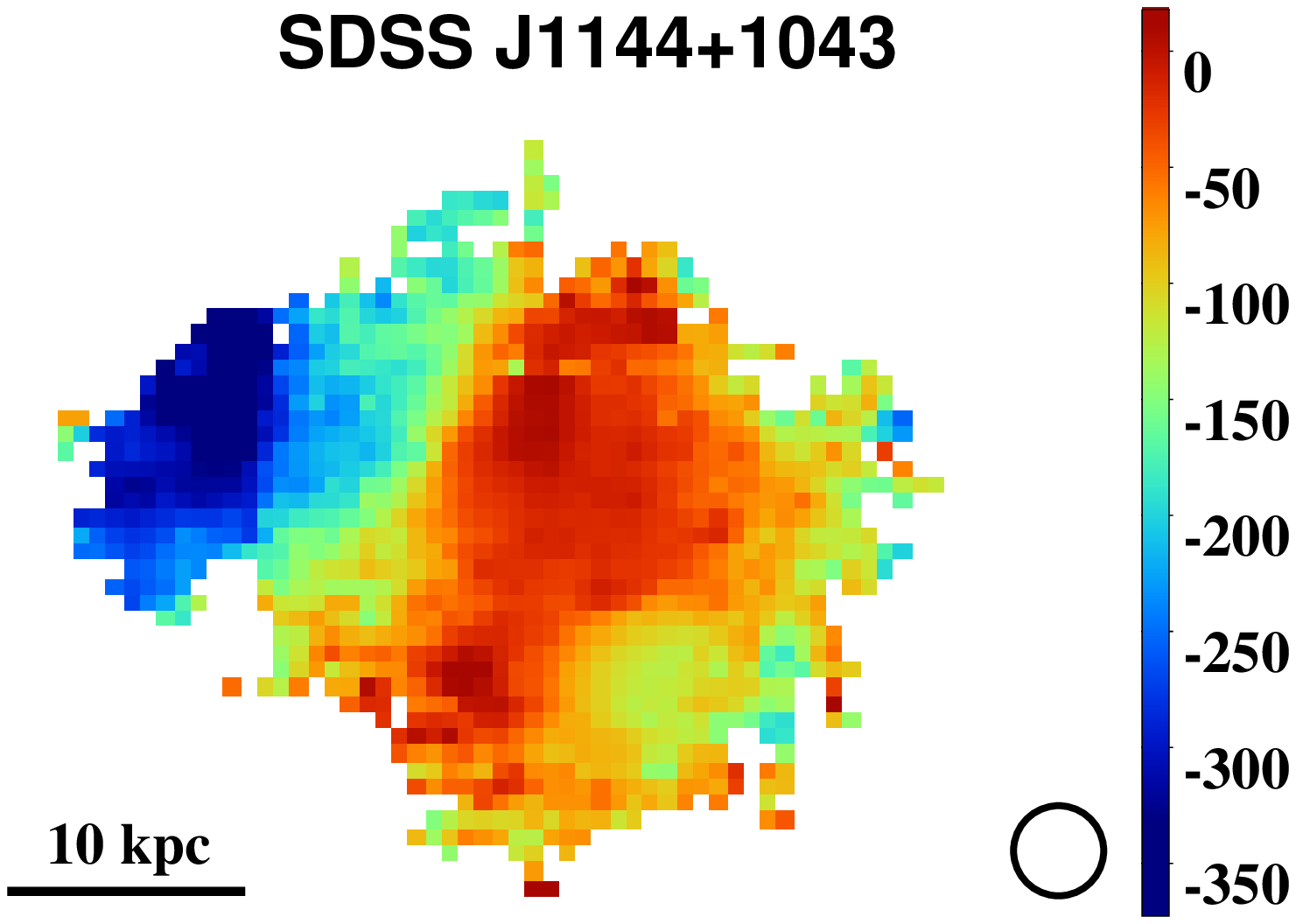}%
\includegraphics[scale=0.33,trim=0cm 0mm 10mm 0mm]{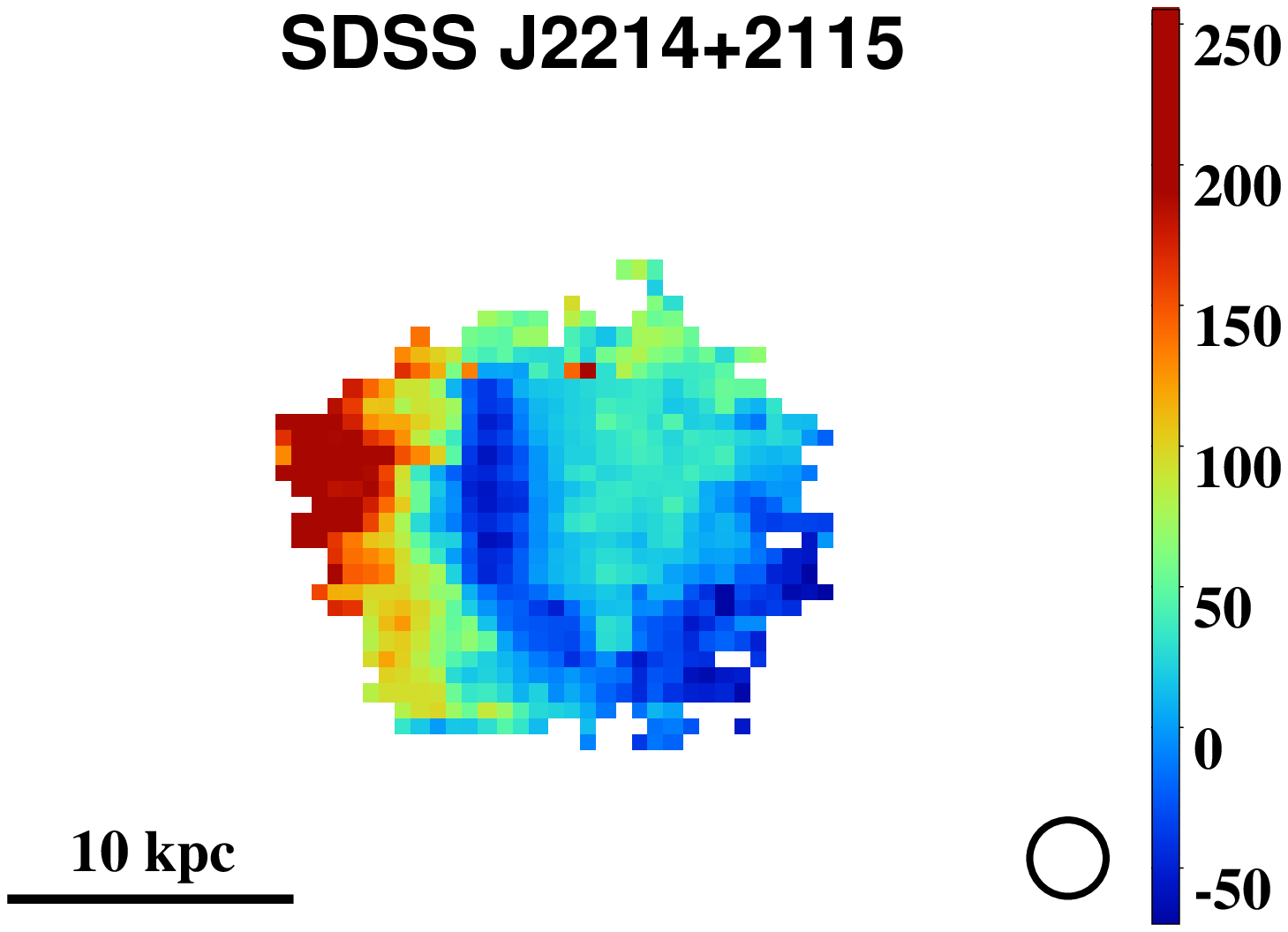}
\caption{The line-of-sight velocity maps of our quasars, in km s$^{-1}$. In every spaxel, we use multi-Gaussian fits to the \oiii\ emission line to determine the median line-of-sight velocity. These values are determined relative to the overall quasar redshift (as measured by the SDSS pipeline by fitting a quasar template to the overall fiber spectrum) and may be different by a few tens of km s$^{-1}$ from the redshift of the host galaxy \citep{liu13b}. Only spaxels where the peak of the \oiii\ $\lambda$5007\AA\ line is detected with S/N $>5$ are plotted.}
\label{fig:Vmed}
\end{flushleft} 
\end{figure*}

\begin{figure*}
\includegraphics[scale=0.33,trim=0cm 0mm 10mm 0mm]{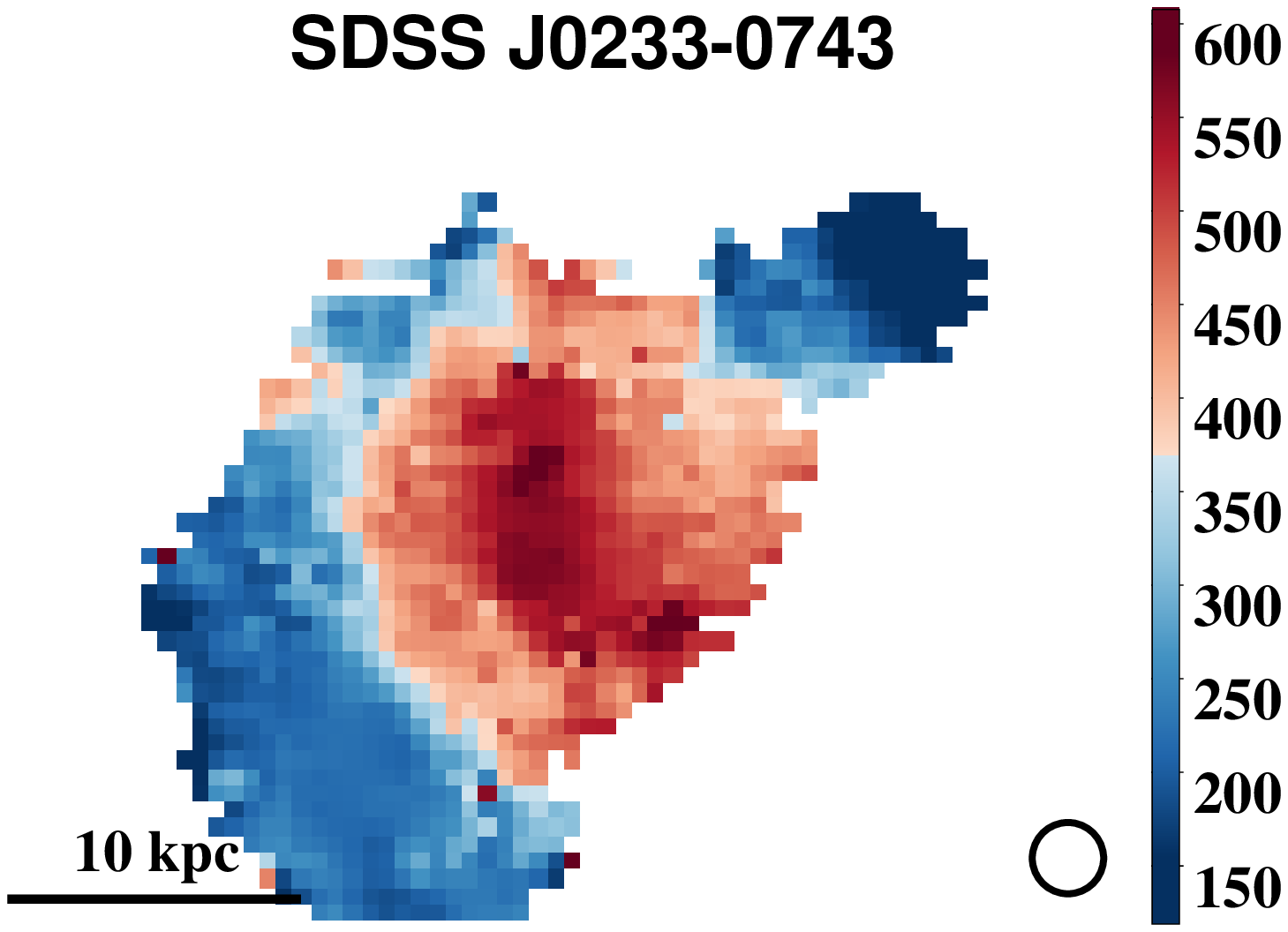}%
\includegraphics[scale=0.33,trim=0cm 0mm 10mm 0mm]{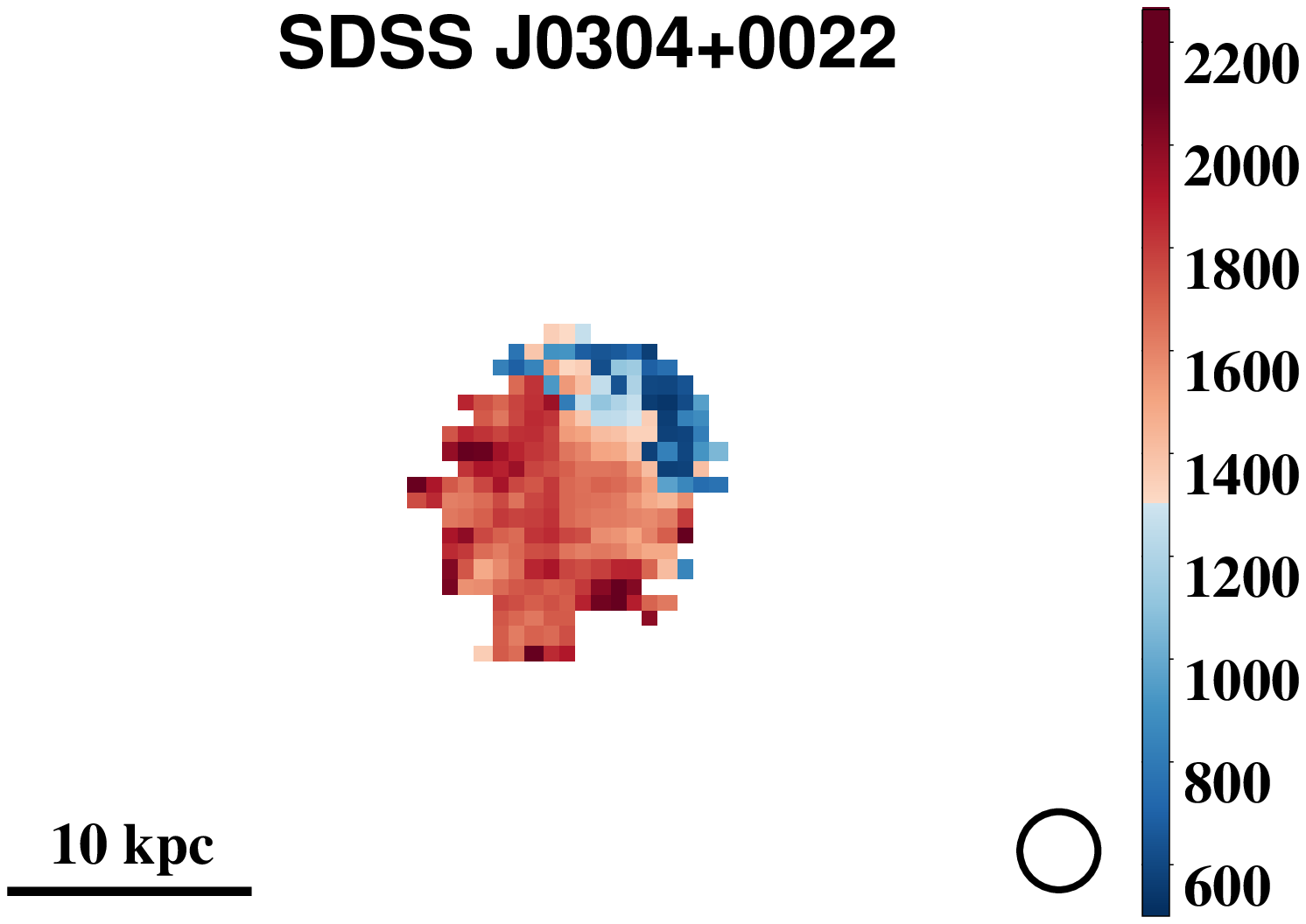}%
\includegraphics[scale=0.33,trim=0cm 0mm 10mm 0mm]{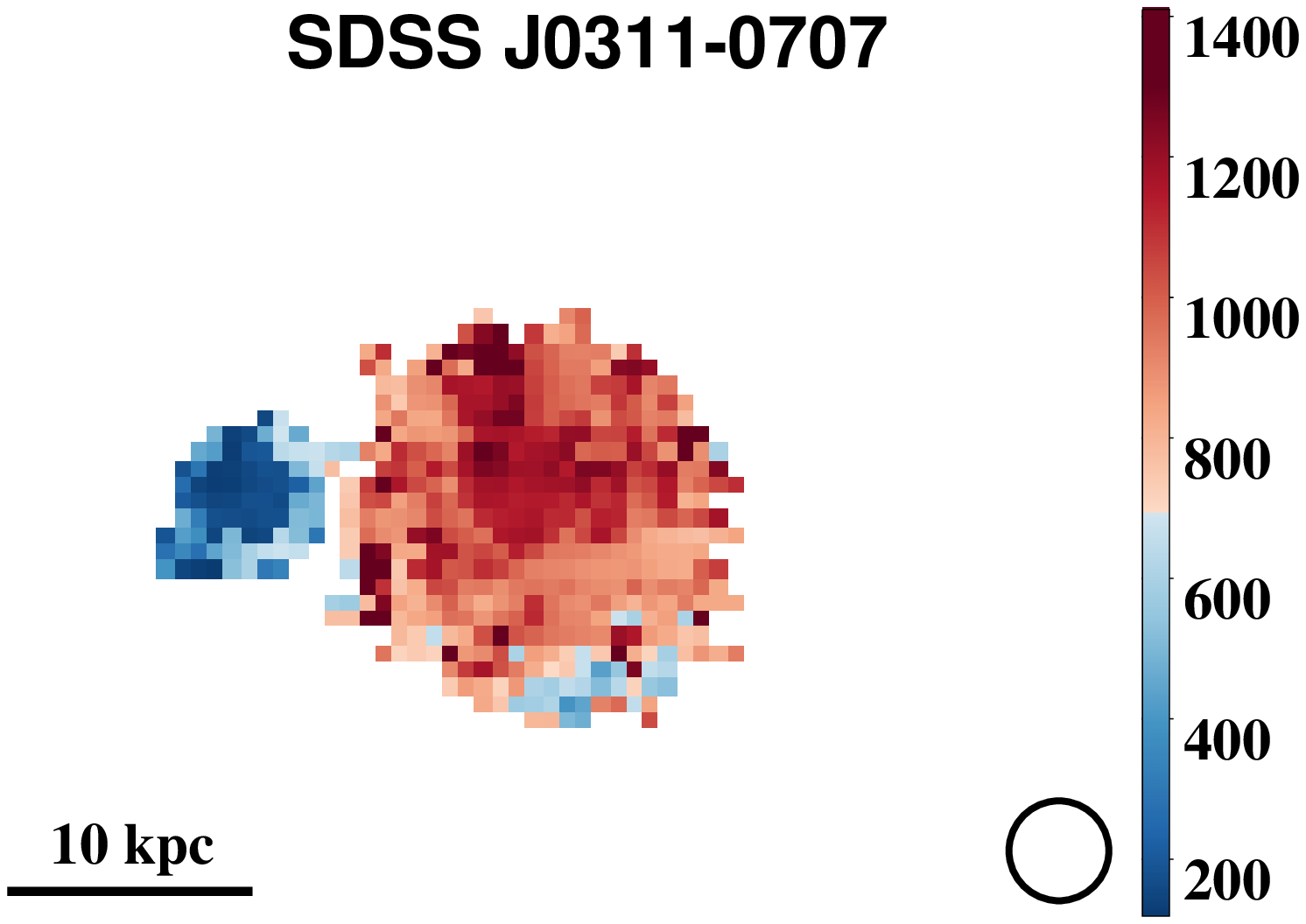}\\
\includegraphics[scale=0.33,trim=0cm 0mm 10mm 0mm]{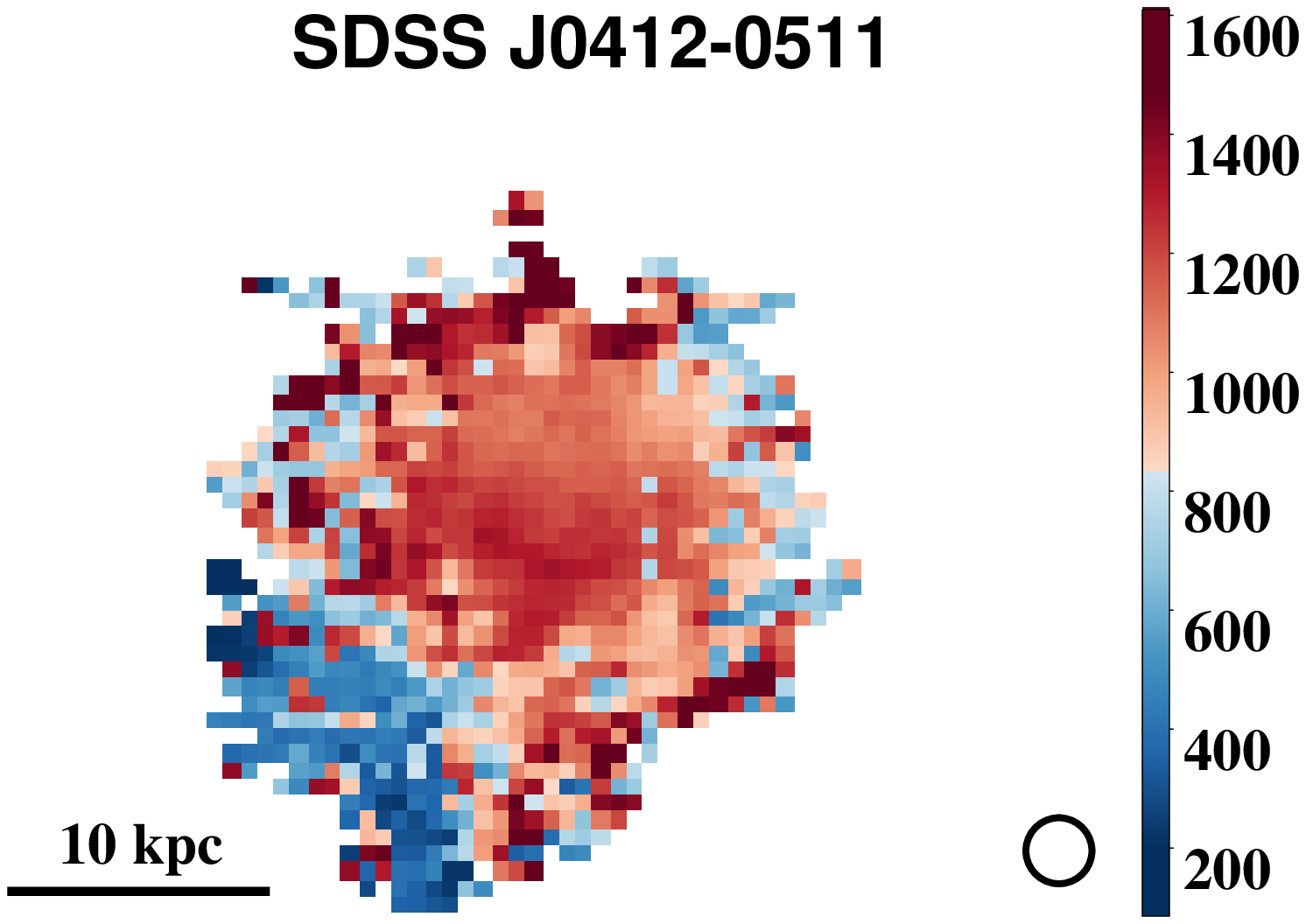}%
\includegraphics[scale=0.33,trim=0cm 0mm 10mm 0mm]{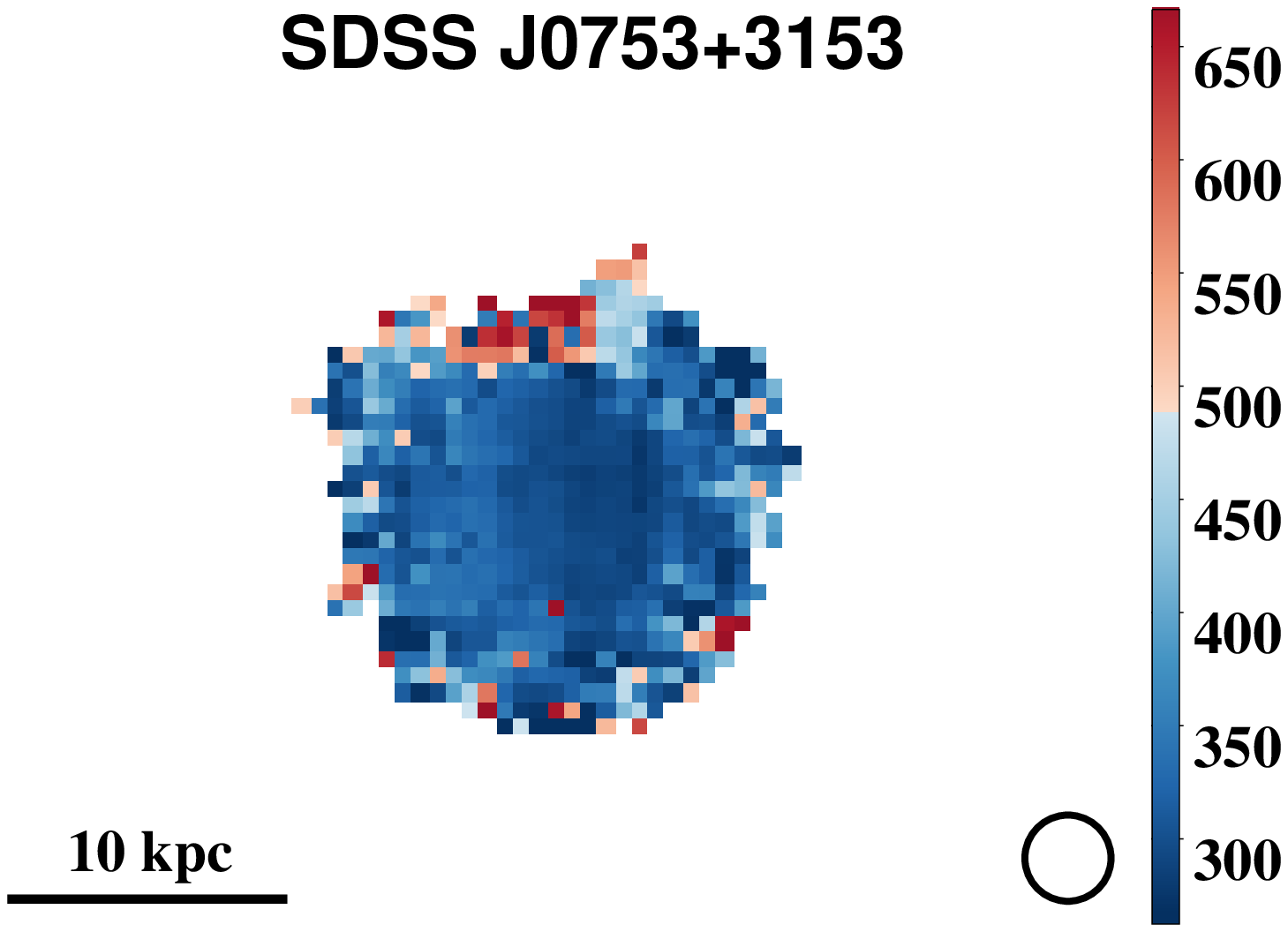}%
\includegraphics[scale=0.33,trim=0cm 0mm 10mm 0mm]{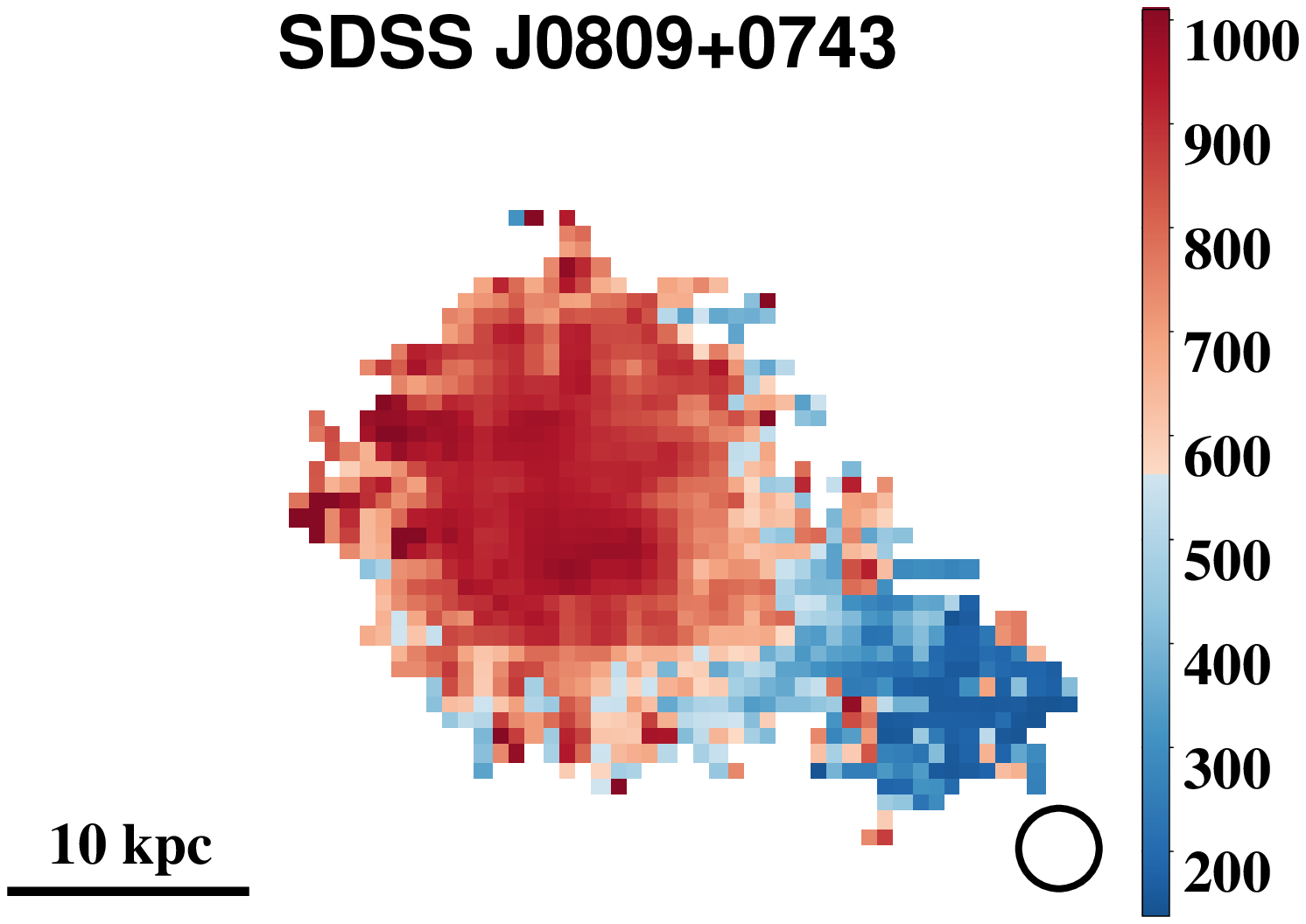}\\
\includegraphics[scale=0.33,trim=0cm 0mm 10mm 0mm]{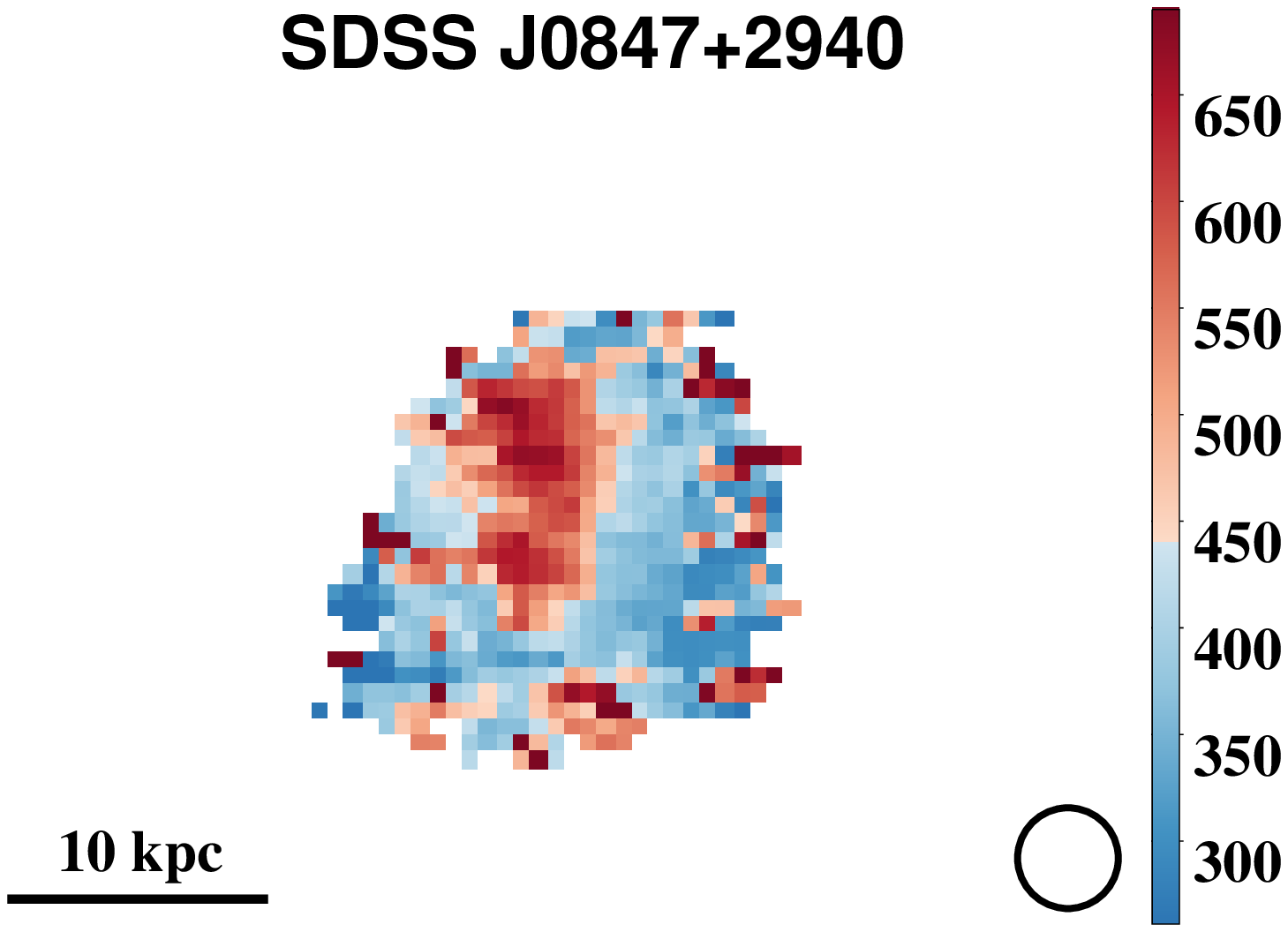}%
\includegraphics[scale=0.33,trim=0cm 0mm 10mm 0mm]{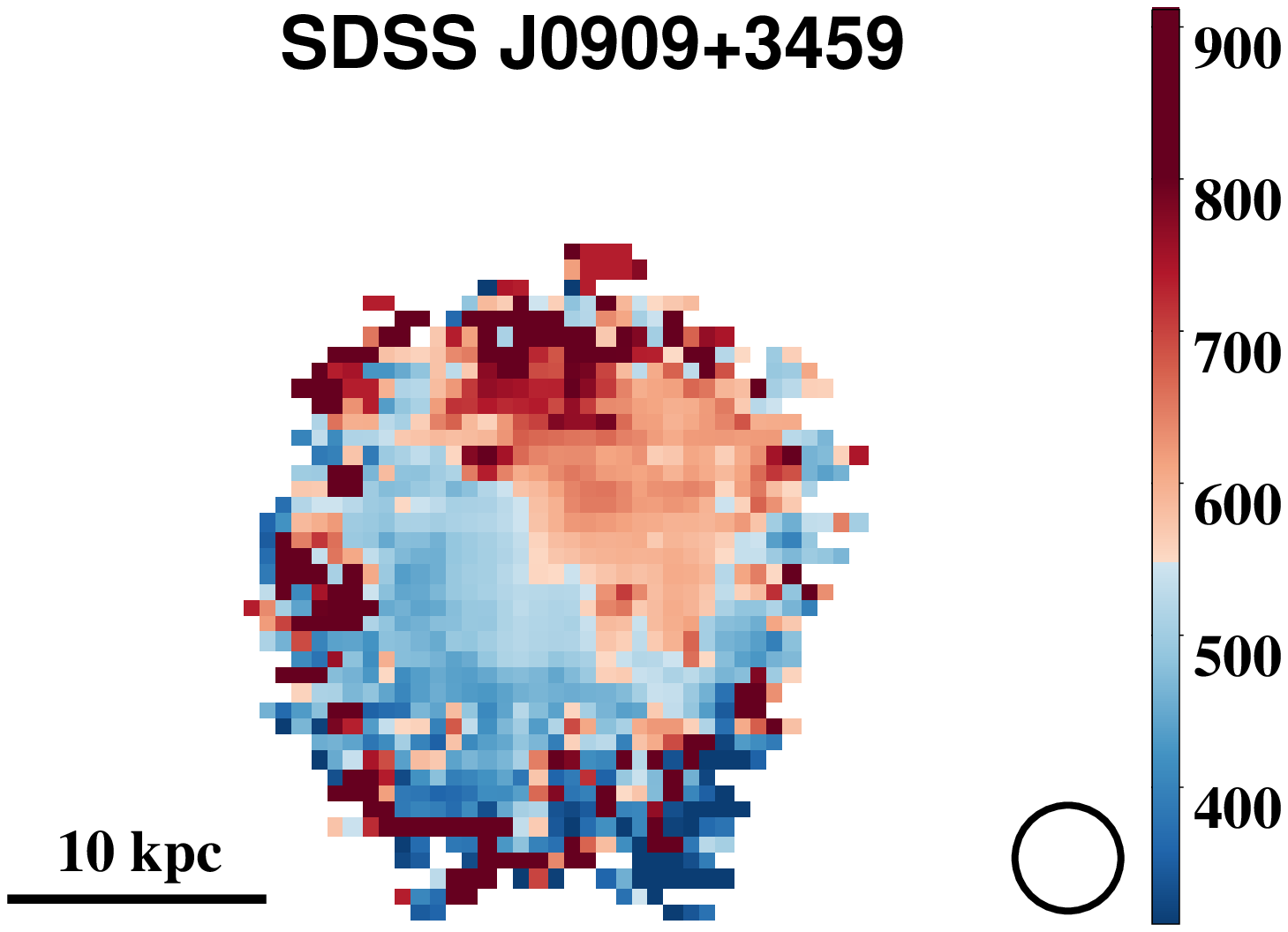}%
\includegraphics[scale=0.33,trim=0cm 0mm 10mm 0mm]{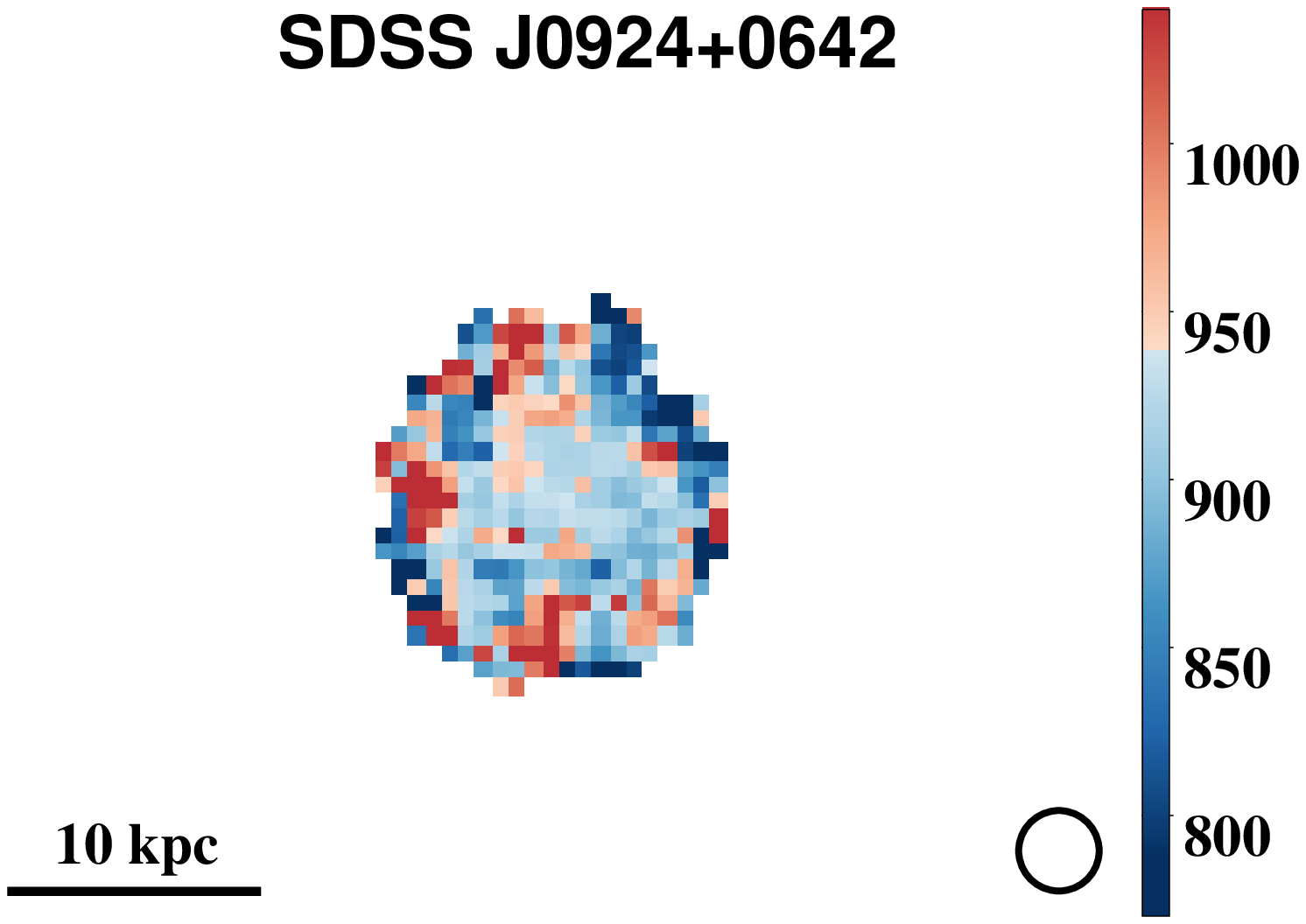}\\
\includegraphics[scale=0.33,trim=0cm 0mm 10mm 0mm]{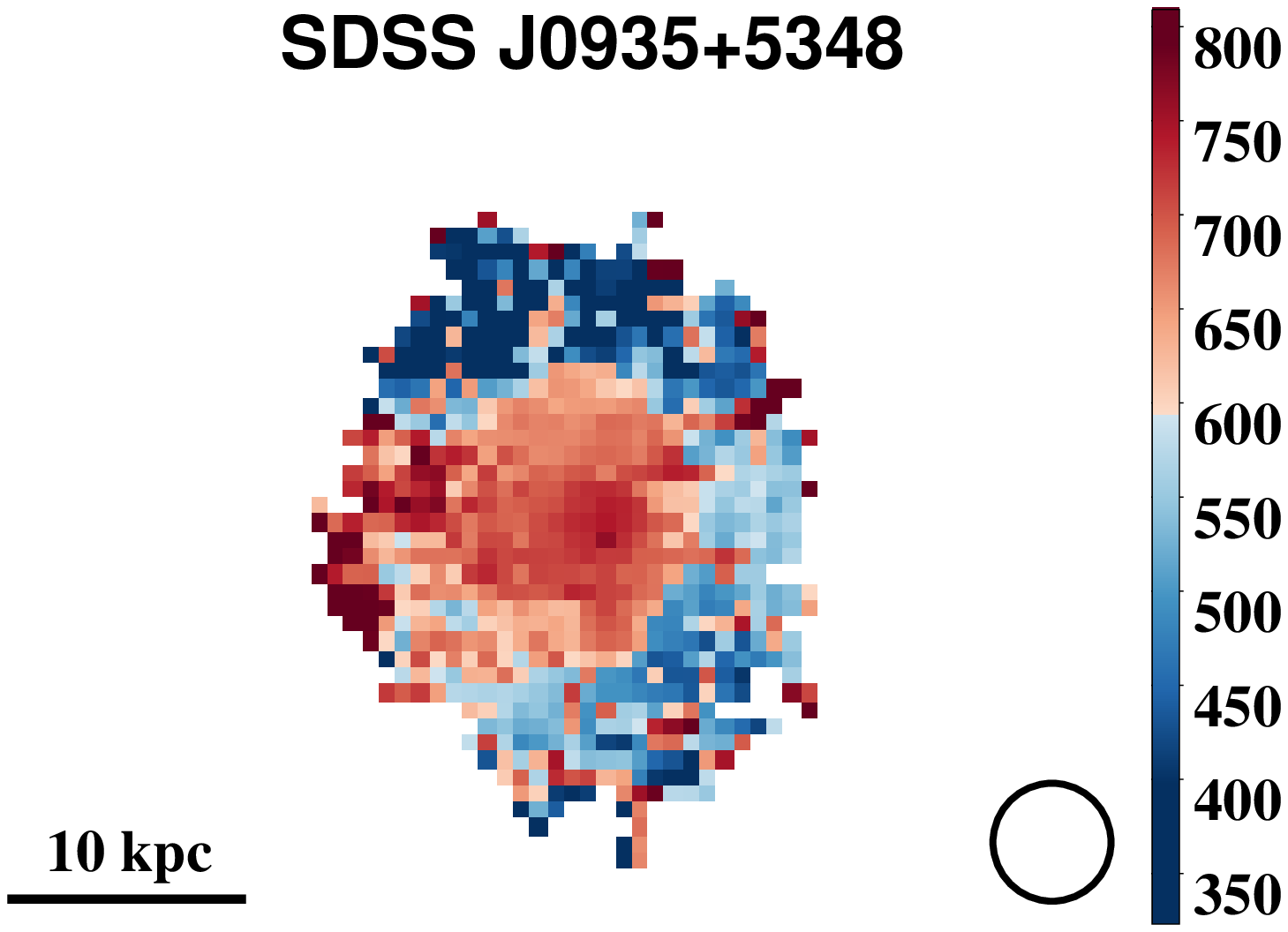}%
\includegraphics[scale=0.33,trim=0cm 0mm 10mm 0mm]{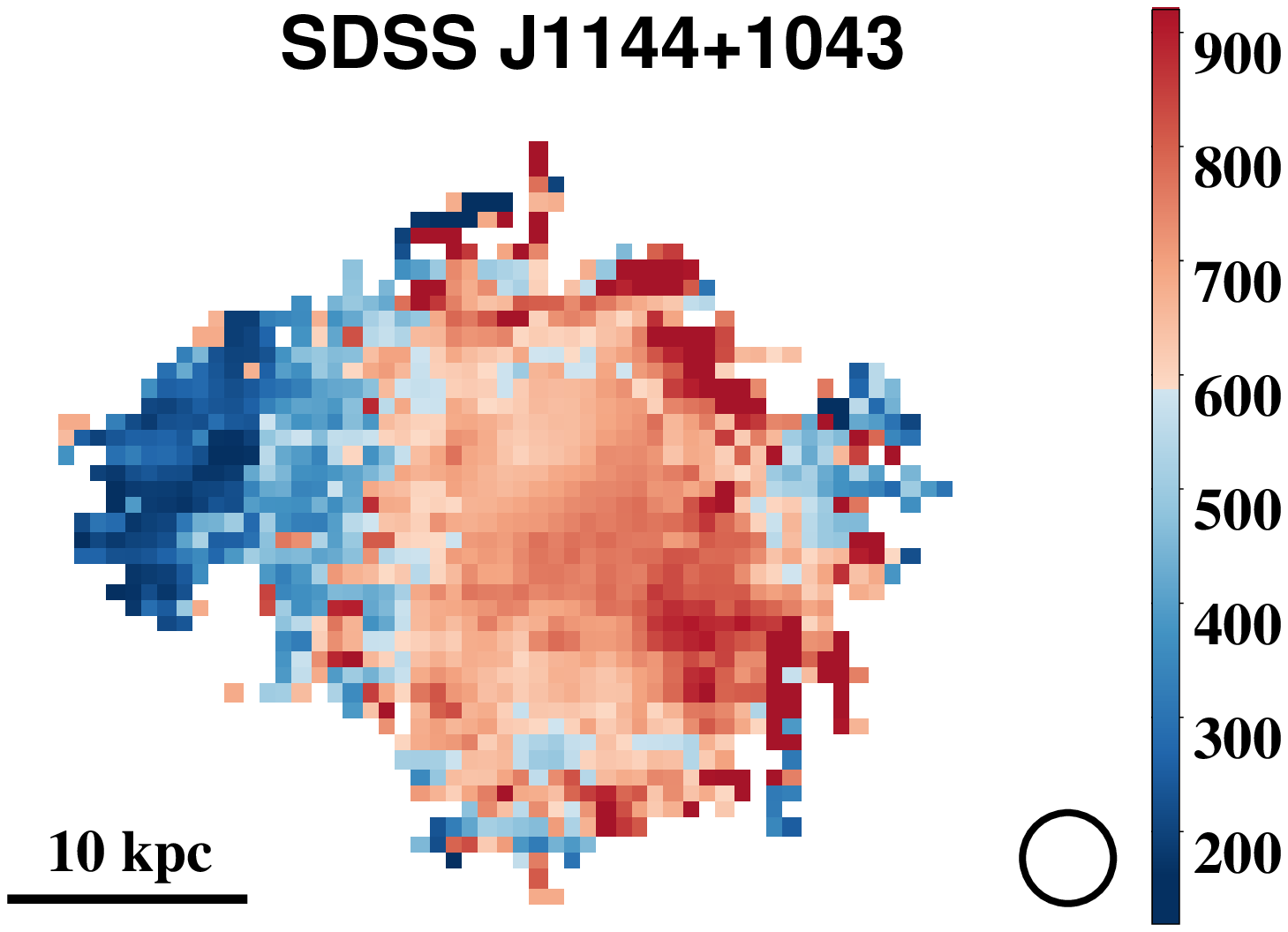}%
\includegraphics[scale=0.33,trim=0cm 0mm 10mm 0mm]{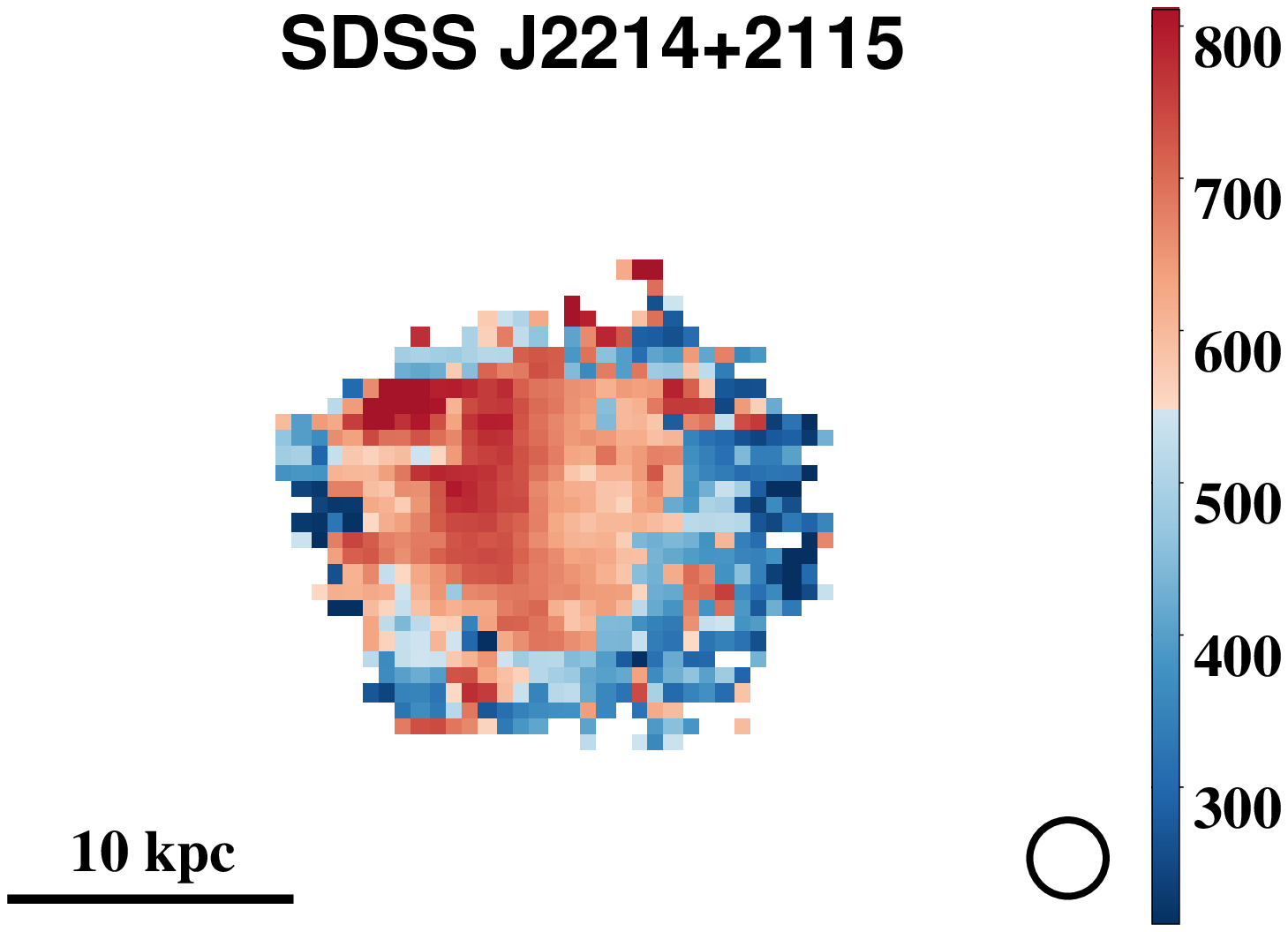}
\caption{The maps of the $W_{80}$ parameter (measured in km s$^{-1}$) for the type 1 quasars in our sample. $W_{80}$ is a measure of the line-of-sight velocity dispersion of the emission-line gas. For a Gaussian profile, $W_{80}=1.088\times$FWHM, and for the heavy-winged line profiles characteristic of our objects, $W_{80}>1.088\times$FWHM. Only spaxels where the peak of the \oiii\ $\lambda$5007\AA\ line is detected with S/N $>5$ are plotted.}
\label{fig:W80}
\end{figure*}

To characterize the line-of-sight velocity and velocity dispersion of the ionized gas, in every spaxel we use the multi-Gaussian fits to the \oiii\ line to  measure the median velocity ($v_{\rm med}$) and the velocity interval that contains 80\% of the total emission centered at the median velocity ($W_{80}$). These parameters, first introduced by \citet{whit85}, were also used to characterize the kinematics of the \oiii\ emission in the obscured quasar sample \citep{liu13b}. The spatial distributions of $v_{\rm med}$ and $W_{80}$ are shown in Figures \ref{fig:Vmed} and \ref{fig:W80}. 

Across every nebula, the median velocity of \oiii\ emission varies by up to a few hundred km s$^{-1}$. Although some noisy spaxels are present, most of the velocity variation proceeds in a smooth fashion from one spaxel to the next, which implies that the velocity profiles of \oiii\ vary among the spaxels. The well-resolved velocity structure confirms that the \oiii\ emission is spatially resolved in our targets. Several of the objects show well-organized velocity structures, with blue-shifted emission predominantly on one side of the nebula and redshifted emission predominantly on the other and with the line of nodes centered at the brightness peak (e.g., SDSS~J0304+0022, SDSS~J0311$-$0707, and SDSS~J0935+5348). 

In Table \ref{tab2} we report the maximum difference in $v_{\rm med}$ between the redshifted and the blueshifted regions $\Delta v_{\rm max}$ after excluding the 5\% highest and 5\% lowest $v_{\rm med}$ values to minimize the effect of noise. The maximum projected velocity difference ranges between 83 and 576 km s$^{-1}$ among the twelve unobscured quasars. The same measurement performed in the obscured sample yielded values of 89--522 km s$^{-1}$ \citep{liu13b}. 

The \oiii\ velocity widths are just as high as those seen in type 2 quasars. Both the luminosity-weigthed overall $\langle W_{80} \rangle$ (measured from the SDSS fiber spectra which capture the integrated emission from the nebulae) and the maxima of $W_{80}$ in the spatially resolved maps (excluding the 5\% highest values) are listed in Table \ref{tab2}. The maximal values are $\sim$1000 km s$^{-1}$ or greater in most of the sample, reaching 2000 km s$^{-1}$. The object with the smallest line-of-sight velocity dispersion appears to be SDSS~J0753+3153 which has a nearly constant $W_{80}\sim300$ km s$^{-1}$ across the nebula. 

\begin{figure}
\includegraphics[scale=0.75,trim=0mm 0mm 0mm 5mm]{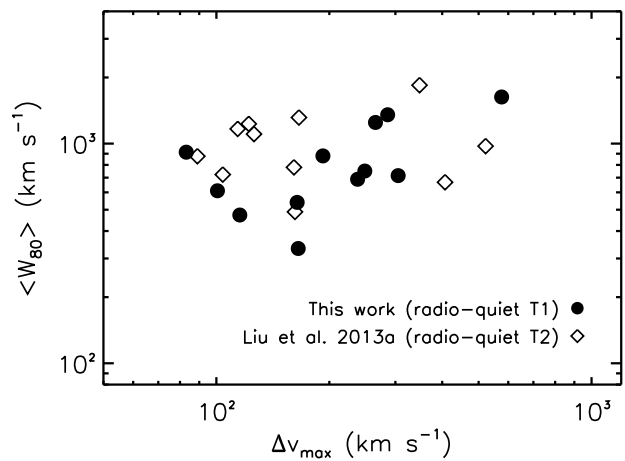}
\caption{The relationship between the luminosity-weighted line width $\langle W_{80}\rangle$ (measured from the SDSS fiber spectrum) and the maximal range of line-of-sight velocities $\Delta v_{\rm max}$ for our unobscured quasar sample, compared to the obscured quasars from \citet{liu13a,liu13b}. Each data point represents an entire quasar. }
\label{fig:w80dv}
\end{figure}

In Figure \ref{fig:w80dv}, we show the distributions of the overall line widths $\langle W_{80}\rangle$ and the maximum velocity difference $\Delta v_{\rm max}$ for both type 1 and type 2 quasars. Excluding the only likely unresolved target, SDSS J0924$+$0642, we do not find a significant difference between $\langle W_{80}\rangle$ values in type 1 and type 2 objects, with the Kolmogorov-Smirnov (K-S) test giving a $p=0.37$ probability that they are drawn from the same distribution. The median $\Delta v_{\rm max}$ is slightly larger in the unobscured sample ($\langle \Delta v_{\rm max}^{\rm T1} \rangle=238\pm138$ km s$^{-1}$) than in the obscured quasars ($\langle \Delta v_{\rm max}^{\rm T2} \rangle=161\pm146$ km s$^{-1}$), although the difference is not statistically significant (the K-S test yields probability $p=0.15$ that the two samples follow the same distribution).

We find no correlation between the two quantities for the obscured sample \citep{liu13b}. A weak correlation might exist among the unobscured quasars (the Kendall rank correlation coefficient $\tau=0.60$ with probability $p=0.01$ that no correlation is present).

The radial profiles of $W_{80}$ are almost flat at projected distances $R\lesssim5$ kpc. At larger radii, the scatter increases with $R$, while $W_{80}$ appears to decrease in 10/12 of the sample (by $\sim$5 \% per kiloparsec from the brightness center) and remains roughly constant in the two remaining objects (J0753$+$3153 and J0924$+$0642). This is very similar to the profiles seen in type 2 quasars. As we discussed previously \citep{liu13b}, at least some of this decline is due to the decrease in the signal-to-noise ratio which prevents us from fitting multiple Gaussian components to the \oiii\ in the outer parts of the nebulae and from reliably measuring weak broad bases to the \oiii\ emission. However, we concluded that this effect was not sufficient to explain the entirety of the $W_{80}$ decline and that it was likely that the line-of-sight velocity width was in fact slightly decreasing with projected distance. Given the similarity of $W_{80}$ profiles between the two samples, the same arguments apply for the unobscured quasars as well. A declining $W_{80}$ radial profile becomes more plausible when the PSF smearing effect is taken into account, which makes the high $W_{80}$ values in the center spill into outer regions and thus flattens the overall radial profile.

We measure line asymmetry $A$ for the integrated \oiii\ emission (tabulated in Table \ref{tab2}) as well as for every spaxel. Asymmetry is defined as $A=\left((v_{90}-v_{\rm med})-(v_{\rm med}-v_{10})\right)/W_{80}$, based on velocities at 10\% and 90\% of the cumulative line flux. For lines with blue-shifted excess, this value is negative; this is what is predominantly seen in our sample. Negative line asymmetries strongly suggest that at least some of the narrow-line-emitting clouds form an outflow whose redshifted part (directed away from the observer) is partly extincted by the dust in the quasar host galaxy \citep{heck81,whit85}. The distribution of asymmetries in unobscured quasars is statistically indistinguishable from that in obscured ones (the K-S test gives a probability $p=0.62$ that they are drawn from the same distribution). Similarly, the measurements of the shape parameter $K=W_{90}/(1.397\times {\rm FWHM})$ (defined via the velocity width enclosing 90\% of the flux) indicate that \oiii\ profiles in type 1 quasars have heavier wings than a Gaussian and that the typical shapes of the \oiii\ lines in type 1 and type 2 quasars are similar ($p=0.20$ as given by the K-S test).

\section{Discussion}
\label{sec:discuss}

\subsection{Size-luminosity relation}

It would seem reasonable that quasars with higher luminosity would have larger photo-ionized nebulae around them. The number of photons available for photo-ionization varies with distance from the quasar $r$ as $\propto L_{\rm bol}/4 \pi r^2$. Therefore, in the simplest possible model -- in which the density of particles within the line-emitting clouds is the same everywhere -- the size of a region with a given ionization parameter is $\propto {L_{\rm bol}^{1/2}}$. If the narrow-line luminosity is an accurate indicator of the bolometric luminosity with $L_{\soiii}\propto L_{\rm bol}$, then the \oiii\ sizes and luminosities of nebulae should be related via $R \propto L_{\soiii}^{0.5}$. Somewhat shallower slopes (down to 0.33) can be obtained under other assumptions about the $L_{\soiii}$--$L_{\rm bol}$ relation \citep{hain13,benn06,schm03}.

\begin{figure*}
\includegraphics[scale=0.65,trim=0cm 0mm 0mm 0mm]{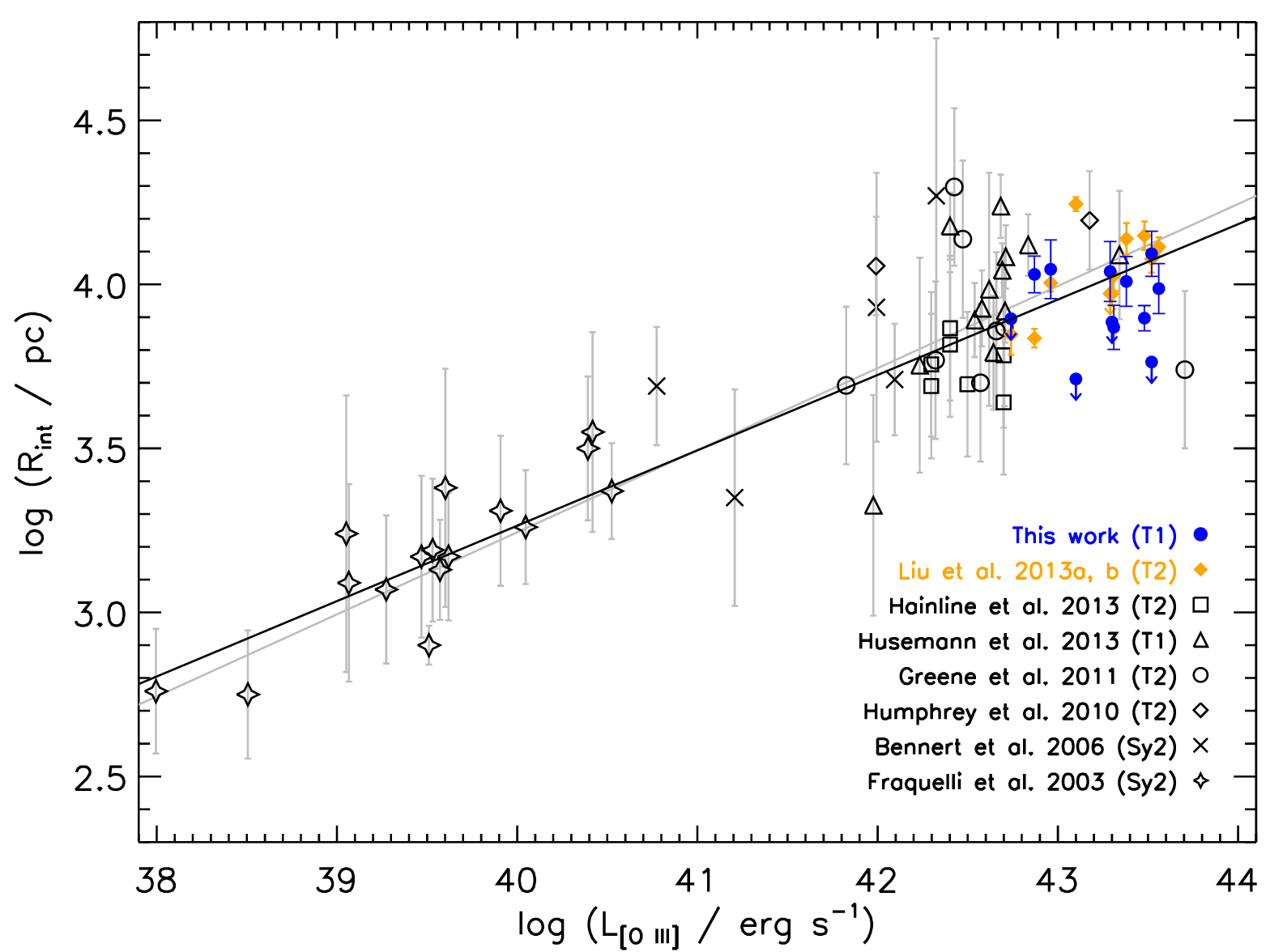}
\includegraphics[scale=0.65,trim=0cm 0mm 0mm -5mm]{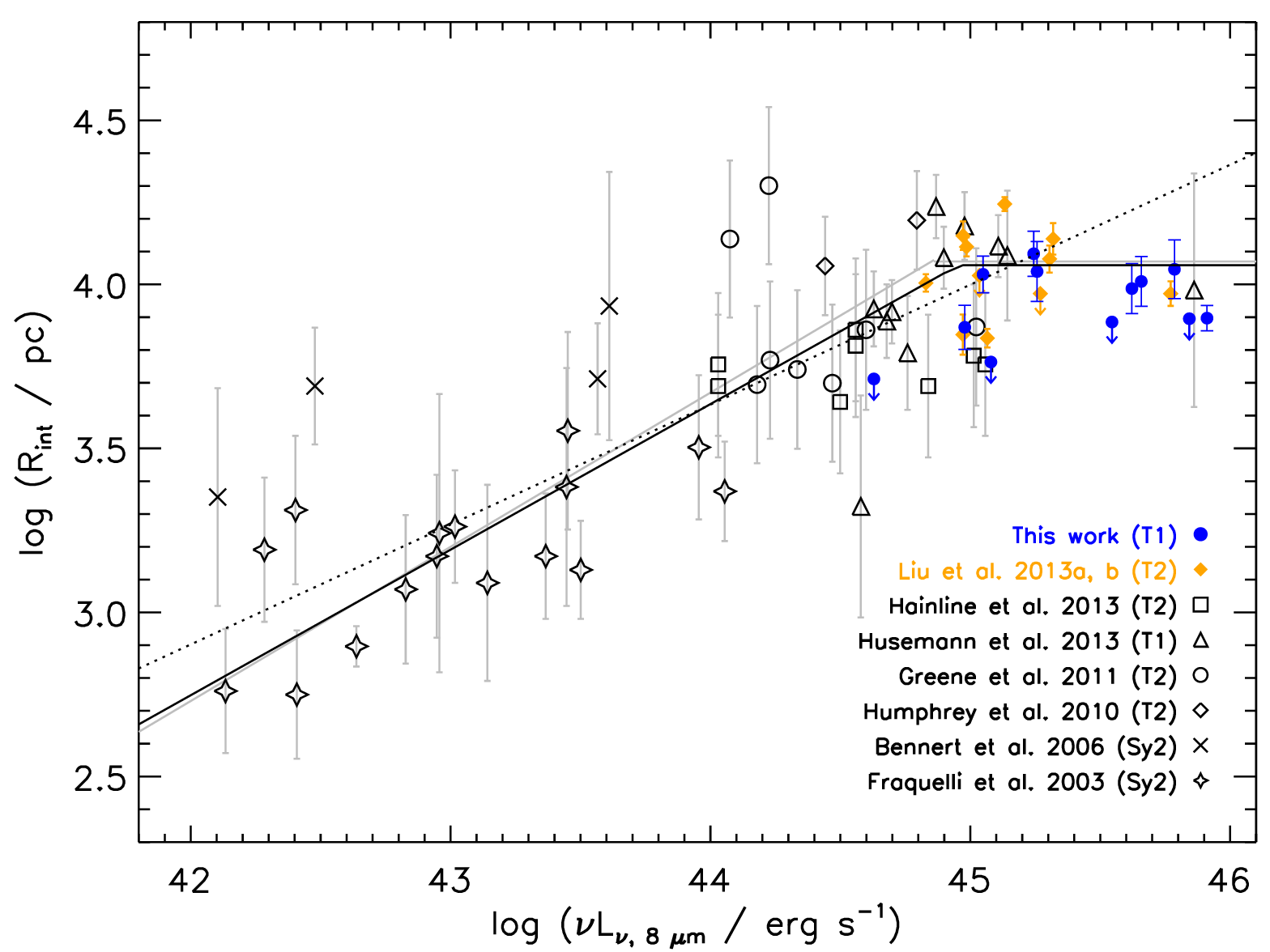}
\caption{Size-luminosity relation of the radio-quiet/weak unobscured quasars from our sample (filled circles) 
in terms of \oiii\ (top) and mid-IR (bottom) luminosities. Also included are previous observations 
of type 2 quasars \citep{liu13a,hain13,gree11,hump10}, type 1 quasars \citep{huse13} and Seyfert 2 galaxies 
\citep{benn06,fraq03}. $R_{\rm int}$ is the isophotal radius at the surface brightness of 
$10^{-15}/(1+z)^4$ erg s$^{-1}$ cm$^{-2}$ arcsec$^{-2}$ \citep{liu13a}, and $\nu L_{\rm \nu,8\;\mu m}$ is 
the rest-frame 8 \micron\ luminosity obtained by interpolating between WISE bands. 
Our linear fit to the $R_{\rm int}$ vs. $L_{\rm [O\;{\scriptscriptstyle III}]}$ relation results in a 
slope of $0.23\pm0.02$ (solid black line), consistent with \citet[][$0.25\pm0.02$, grey line]{liu13a}.
Our best-fit piecewise linear fit to the $R_{\rm int}$ vs. $\nu L_{\rm \nu,\;8\;\mu m}$ relation (solid black
line) has a slope of $0.44\pm0.02$, a break at $\log \nu L_{\rm \nu,\;8\;\mu m}^{\rm br}=44.95\pm0.06$, a 
maximum radius of $\log R_{\rm max}/({\rm pc})=4.06\pm0.01$, well consistent with the result of 
\citet[][gray line]{hain13}. For the objects with $\nu L_{\rm \nu,\;8\;\mu m} < 10^{45.5}$ erg s$^{-1}$, 
we find a best-fit slope of $0.37\pm0.03$.}
\label{fig:sl}
\end{figure*}

The observed slope of the size-luminosity relation is significantly shallower than these simple estimates. By combining measurements of \oiii\ sizes in quasars with those in Seyfert 2 galaxies, \citet{gree11} found a slope of 0.22$\pm$0.04. \citet{liu13a} supplemented these observations with 11 objects at the very luminous end and used the same uniform distance-independent definition of nebular size $R_{\rm int}$ to find an exponent of 0.25$\pm$0.02. 
Incorporating our data points and observations from \citet{hain13} and \citet{huse13} does not lead to significant changes to the slope. In fact, taking into account the marginally or unresolved sources as upper limits of $R_{\rm int}$, we perform the Bayesian linear fit employed by \citet{liu13a}, which uses Markov Chain Monte Carlo to calculate the posterior probabilities \citep{kell07}, and find the following best-fit relation (Figure \ref{fig:sl}):

\begin{equation}
{\log\left(\frac{R_{\rm int}}{\rm pc}\right)}={(0.23\pm0.02)\log\left(\frac{L_{\rm [O\;{\scriptscriptstyle III}]}}{\rm 10^{42}~erg~s^{-1}}\right)+(3.72\pm0.02)}.
\end{equation}

The observed shallow slope of the size-luminosity relationship likely indicates that the clouds located in the outer part of the gas nebulae around luminous quasars are matter-bounded (as implied by the line ratio measurements), i.e. the gas is fully ionized, and the recombination rate cannot keep up with the ionization rate \citep{liu13a}.

Because the conversion from $L_{\rm bol}$ to $L_{\soiii}$ involves a somewhat uncertain slope, \citet{hain13} suggest using a more direct measure of the bolometric luminosity, for example the luminosity at rest-frame 8 \micron\, which traces emission from the warm to hot dust near the supermassive black hole and thus should be a good measure of the power of the central engine. We obtain the rest-frame $\nu L_{\rm \nu,\;8\;\mu m}$ by spline interpolating the photometry from the Wide-field Infrared Survey Explorer \citep[WISE;][]{wrig10} which provides all-sky catalogs at 3.4, 4.6, 12 and 22 \micron. 

Based on two objects from \citet{huse13} and \citet{liu13a},
\citet{hain13} report a tentative flattening of the relationship at
the high luminosity end, suggesting the existence of an upper limit on the size of the narrow-line region beyond which the increase in quasar luminosity does not result in an increase in the size of the narrow-line region. As seen from Figure \ref{fig:sl}, the addition of our sample strengthens this finding: our data yield 6 quasars (half of the sample) that have 
$\nu L_{\rm \nu,\;8\;\mu m}>10^{45.5}$ erg s$^{-1}$ but $R_{\rm int}\lesssim11$ kpc, so that the total number of quasars in this regime reaches eight. 
Excluding these objects, we perform the Bayesian linear fit employing Markov Chain Monte Carlo used by \citet{liu13a} again and find a best-fitting linear relation for the targets with $\nu L_{\rm \nu,\;8\;\mu m} < 10^{45.5}$ erg s$^{-1}$:

\begin{equation}
{\log\left(\frac{R_{\rm int}}{\rm pc}\right)}={(0.37\pm0.03)\log\left(\frac{\nu L_{\rm \nu,\;8\;\mu m}}{\rm 10^{44}~erg~s^{-1}}\right)+(3.63\pm0.03)}.
\end{equation}

This result is consistent with that of \citet[][slope = $0.41\pm0.02$, 
intercept = $3.65\pm0.02$]{hain13} and is depicted by a dotted line
in Figure \ref{fig:sl}. If the flattening is characterized by a plateau,
we find the following best-fitting piecewise linear fit for all data 
points but the upper limits:

\begin{equation}\label{eq:piece}
 \log{\left(\frac{R_{\rm int}}{\rm pc}\right)} = \left\{ 
  \begin{array}{l r l l r}
     (0.44\pm0.02)\log\left(\frac{\nu L_{\rm \nu,\,8\,\mu m}}{\rm 10^{44}\,erg\,s^{-1}}\right) + (3.64\pm0.02) \\
     ~~~~~~~~~~~~~~~~~~~~~~~~{\rm for~\nu L_{\nu,\,8\,\mu m} < 10^{44.95\pm0.06}~erg~s^{-1}}; \\
     ~\\
     4.06\pm0.01 \\  
     ~~~~~~~~~~~~~~~~~~~~~~~~{\rm for~\nu L_{\nu,\,8\,\mu m} \geqslant 10^{44.95\pm0.06}~erg~s^{-1}}. \\
  \end{array} \right.
\end{equation}

This piecewise linear equation is shown by a solid black line in the same figure, 
and is also in good agreement with \citet[][slope = $0.47\pm0.02$, 
intercept = $3.67\pm0.02$, break luminosity = $10^{44.86\pm0.05}$ erg s$^{-1}$, 
and the limiting radius = $10^{4.07\pm0.03}$ pc]{hain13}.

We must be cautious in interpreting the flattening seen in Figure \ref{fig:sl}. In particular, our type 1 quasars have been intentionally selected to be matched in \oiii\ luminosity to the type 2 quasars from
\citet{liu13a,liu13b}, and yet the former appear to be more luminous by a factor
of $\sim3$ than the latter in mid-infrared (mid-IR). The significant IR-to-\oiii\
ratio difference points to the intrinsic difference in the spectral
energy distribution (SED) of the two populations. The mid-IR continuum of type 2 quasars is much redder than that of type 1s \citep{liu13b}, which likely means that even the mid-IR emission is not an entirely isotropic luminosity indicator. Specifically, the slopes of the IR SEDs (defined as $\nu L_{\nu}\propto \nu^{\alpha}$) measured between rest-frame 5 and 12 \micron\ are $\alpha=-0.33\pm0.27$ for the type 1 sample (median and standard deviation) and $\alpha=-1.13\pm0.55$ for the type 2 sample. In type 2 objects the thermal emission from hot dust seen at mid-IR wavelengths can be affected by obscuration when propagating through the surrounding much colder material in order to reach the observer, which would result in both its reddening and overall suppression. This hypothesis is further strengthened by the observation that many type 2 quasars show silicate absorption which is a bi-product of this process \citep{knac73,zaka08,hao05}. If the \oiii\ line is a more or less isotropic indicator of the bolometric luminosity of the quasar, then type 1's would appear more luminous in the mid-IR than their \oiii-matched type 2 counterparts, as we see in Figure \ref{fig:sl}. Ryan C. Hickox et al. (private communication) find a similar (0.3--0.4 dex) difference at $L_{\soiii}>10^{43}$ erg s$^{-1}$ between the type 1 and 2 spectral energy distributions when averaged over large SDSS samples. Our knowledge of the details of the mid-IR opacity curve of dust suffers from large systematic uncertainties \citep{smit07}. To crudely estimate the amount of extinction necessary to produce such difference in color between type 1 and type 2 quasars, we assume that dust opacity is $\propto 1/\lambda$ over optical-to-IR range and compute reddening in the simplistic ``screen of cold dust'' approximation. To produce a $>0.3$ dex reddening of the unobscured quasar continuum in the 5--12 \micron\ range, $A_V>13$ mag worth of dust obscuration would be required, in line with typical limits on the amount of obscuration in type 2 objects (\citealt{zaka04}; also see \citealt{clea07}, \citealt{haas08} and \citealt{lacy13} for extinction studies in mid-IR).

Thus there are two possible interpretations of Figure \ref{fig:sl}. The first is that the 8 \micron\ monochromatic luminosity is an accurate probe of the bolometric luminosity of quasars regardless of the type; then we must conclude that our new sample of unobscured quasars is 0.3 dex more intrinsically luminous than the sample of the obscured ones. In this case, since $R_{\soiii}$ and $L_{\soiii}$ are so similar between the two, this implies that there exists an upper limit on both the size (10 kpc) and the luminosity of the narrow-line region and neither of these values further increases with bolometric luminosity.  

The second interpretation is that $L_{\soiii}$ is a reasonably isotropic measure of quasar luminosity, unbiased with respect to quasar type. Then the difference in the 8 \micron\ luminosity is due to circumnuclear obscuration. As for the sizes, $R_{\soiii}$ are statistically indistinguishable in type 1 and type 2 samples, and we are unable to comment on the existence of the upper limit to the size, since our new data do not in fact lead to an increase in the range of intrinsic luminosities probed. We are inclined to accept this latter interpretation because the mid-IR spectral energy distributions of the two samples are undeniably different. In order to reliably probe the flattening of the size-luminosity relationship \citep{netz04}, objects with higher \oiii\ and IR luminosities must be observed and quasars of different types need to be considered separately.

The derived $B$-band absolute magnitude of our sample is $-25.6\pm0.7$ mag on average (Table \ref{tab1}), higher than that of the 19 radio-quiet type 1 quasars studied in \citet{huse08} and \citet{huse13} located at $z$ = 0.06--0.24 ($-23.5\pm0.08$ mag). The detection rate of extended nebulae around our radio-quiet/weak quasars (11/12) is significantly higher than that of theirs (6/19), but the physical extents are comparable (Figure \ref{fig:sl}). Among their 6 detected radio-quiet nebulae, 3 show one-sided or biconical morphology and the rest are round, in contrast to our quasar nebulae that are morphologically smoother and rounder in every case (ellipticity $\sim0.1$).

\subsection{\oiii-to-H$\beta$ ratio}

The \oiii-to-H$\beta$ ratio is a powerful diagnostic of ionization conditions in the nebula. In particular, in type 2 quasars this ratio remains nearly constant over most of the nebula and then rapidly declines with the distance from the central source, which we interpret as the signature that the line-emitting clouds become fully ionized and matter-bounded in the outer parts of the nebulae \citep{liu13a}. In type 1 quasars, this is a difficult measurement to replicate because of the contribution to the H$\beta$ profile from the broad-line region of the quasar. 

To perform this measurement, in every spaxel we subtract the overall quasar continuum and \feii, and then we fit the H$\beta$ line using a combination of Gaussians to isolate the narrow component. In eight cases, only two Gaussian components are sufficient -- one for the broad component and one to the narrow one. The resulting radial profiles of the \oiii/H$\beta_{\rm narrow}$ ratio are similar to the ones we found in type 2 quasars, that is, the ratio persists at a constant value ($\sim10$) within a radius of $R_{\rm br}\simeq5$--9 kpc (7 kpc on average) from the center, and beyond this distance we detect a marginal decline in \oiii/H$\beta_{\rm narrow}$. In the remaining four objects, the spectral decomposition of H$\beta$ into a narrow and a broad component is highly uncertain, even using the spatially integrated spectra. 

\begin{figure}
\includegraphics[scale=0.65,trim=0cm 0mm 0mm 0mm]{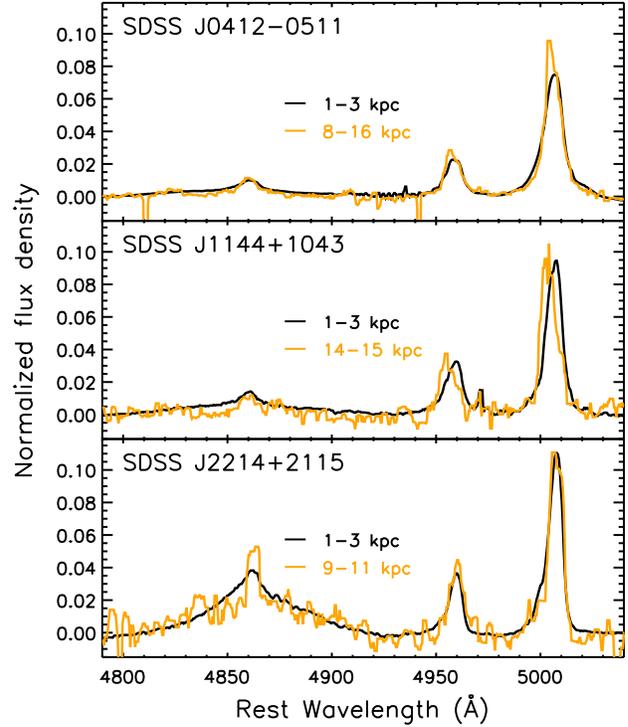}
\caption{Spectra of three quasars, collapsed within an annulus inside the break radius of
the \oiii/H$\beta_{\rm narrow}$ profile, and another annulus beyond this radius. 
All spectra are normalized so that the integrated \oiii\ line intensity is unity.
}
\label{fig:o3hbspec}
\end{figure}

For three objects, we show in Figure \ref{fig:o3hbspec} the spectra collapsed within two circular annuli that are within and beyond the break radius, respectively. The narrow H$\beta$ component becomes stronger relative to the broad H$\beta$ component in the outer parts, indicating that the narrow H$\beta$, like the \oiii, is spatially resolved while the broad emission is not. Furthermore, the ratio \oiii/H$\beta_{\rm narrow}$ is marginally smaller, which is especially visible in the bottom panel for SDSS J2214+2115. 


In summary, despite the difficulties associated with the emission from the broad-line region of the quasar, we are able to measure the \oiii/H$\beta_{\rm narrow}$ ratios in the extended nebulae around type 1 quasars. We tentatively detect a decline of this ratio in the outer parts of the nebulae, beyond the break radius, which we interpret as due to the clouds becoming over-ionized (or matter-bounded), by analogy to our findings in type 2 quasars. In that case, our hypothesis was supported by the increase in the He {\sc ii} $\lambda$4686\AA-to-H$\beta$ ratio in the outer parts of the nebulae \citep{liu13a}, which we cannot determine in type 1 quasars because of the contamination from the broad-line region, but even without this additional diagnostic it appears that the ionization conditions are very similar in type 1 and type 2 samples. Therefore, the uniform size of the nebulae is likely set by the pressure profile (which in turn determines the ionization conditions in the gas) and not necessarily by the presence or absence of gas.

\subsection{Origin of gas}

The main result of this paper is the striking similarity of the ionized nebulae around obscured and unobscured radio-quiet/weak quasars at a matched \oiii\ luminosity. The physical extents, shapes, morphologies, and ionization conditions of the nebulae in the two samples are very similar. Neither sample is dominated by illuminated merger debris which would appear as spatially and kinematically separated clumps with relatively small velocity disperion \citep{fu09}. The roundness of nebulae in both samples and the similarity of nebular sizes between type 1 and type 2 samples point to wide-angle illumination patterns, which would make the appearance of the nebulae relatively insensitive to the position of the observer.  

We previously demonstrated that the large line widths combined with relatively small velocity differences across the nebulae point to quasi-spherical outflows with typical velocities of $\sim$800 km s$^{-1}$ \citep{liu13b}. Alternative explanations, such as host galaxy rotation, fall short of explaining the combination of morphological and kinematic data. In particular, the velocity widths of the emission lines are too high to be comfortably accommodated by the rotation of a disk galaxy, and the ubiquitous blue-shifted asymmetry of the lines in both samples also points to gas in outflow. Furthermore, the nebulae are too large and too round to be produced in inclined galaxy disks, not to mention that luminous type 2 quasars predominantly reside in elliptical hosts \citep{zaka06} and that the presence of large galactic disks would have likely produced systematic differences in the type 1 and type 2 samples. 

Thus, the outflow hypothesis still provides the most natural explanation for our data for both type 1 and type 2 samples. At typical velocities of $\sim$800 km s$^{-1}$, the lifetimes of the nebulae is set by the gas travel time to 13--14 kpc, roughly $10^7$ years, comfortably close to the typical lifetime of luminous quasars \citep{mart01, mart04}. 

However, given the close similarity of nebulae in type 1 and type 2 sources we run into a potential problem. How can all quasars, both obscured and unobscured, show these large ionized regions with long inferred lifetimes? For instance, in a classic ``blow-out'' picture \citep{sand96, hopk06}, we might expect that the type 1 objects have already expelled the majority of their gas, leaving themselves bare. Instead, we see no discernible differences between the obscured and unobscured sources. 

One possible solution to this problem is that instead of pushing the gas out of the galaxy, the quasars are simply lighting up pre-existing gas, but then we still have the puzzle of what brought the gas out to 13--14 kpc and moving with velocities that are inconsistent with the dynamical equilibrium of the gas within the host galaxy. Perhaps the gas was brought there by a previous episode of quasar activity \citep{ciot01} or even by starburst-driven winds \citep{heck90}, although the typical velocities of the gas seen in our sample are more consistent with quasar-driven feedback than with starburst-driven feedback \citep{rupk13, hill13}. 
Nevertheless, H {\sc i} halo gas has been found to commonly surround all galaxies, being either early-type or
star-forming \citep{thom12}. The bimodal metallicity distribution of this $\sim$10$^{4\mhyphen5}$ K halo gas indicates
a mixture of metal-rich gas originating from galactic outflows and tidally stripping and pristine metal-poor gas freshly 
transported through cold accretion streams \citep{lehn13}. The discovery of the probably pristine halo gas offers an 
alternative possibility that our quasars are actually lighting up circumgalactic gas in a non-equilibrium but almost 
steady state.

Another solution -- one that does not involve appealing to a previous episode of activity -- is to apply the purely geometric unification model, which is to say that type 1 and type 2 quasars in the two samples are intrinsically very similar and are at the same evolutionary stage. Then it would be unsurprising to find that the large-scale distribution of ionized gas, relatively unaffected by the circumnuclear obscuration, is similar in the two samples. In fact, for a bi-conical model of quasar illumination \citep[as commonly seen in Seyfert galaxies;][]{mulc96a,mulc96b} we expect to see somewhat smaller nebular sizes in type 1 objects (which are viewed closer to on-axis) than in type 2s according to the standard geometric unification model, and in fact we do see a small (although not a statistically significant) difference in nebular sizes between the two samples, $R_{\rm int}=10.7\pm 1.7$ kpc for type 1s versus $12.9\pm 3.4$ kpc for type 2s. The purely geometric unification model is indirectly supported by HST images and spectropolarimetric observations of type 2 quasars \citep{zaka05, zaka06}. These observations show that our type 2s are definitely seen as type 1s along some directions, although this does not guarantee that the type 1 and type 2 samples we observed with Gemini IFU are drawn from the same intrinsic distribution of obscuration covering factors. In fact, this difference is also in line with the theoretical evolutionary scenarios that anticipate unobscured quasars to be relatively gas-poor, because they are in the post-``blow-out'' phase when the gas that fuels both quasar activity and star formation has been expelled from the host galaxy.

If we postulate that the type 1 and type 2 samples are intrinsically very similar and the gas seen in the halos of their host galaxies got there as a direct result of the ongoing quasar activity, then we find that both type 1 and type 2 quasars we are observing are at the same --- and fairly advanced --- evolutionary stage. It is quite possible that these findings are strongly predicated on the exact method of target selection. Indeed, in this paper we study type 1 and type 2 objects of very similar --- and very high --- \oiii\ luminosities. It is likely that such luminosities are not characteristic of earlier or later stages of quasar feedback. In particular, in later stages of feedback one can expect to see diffuse gas at large distances from the quasar which does not necessarily manifest itself in line emission at optical wavelengths, so the \oiii\ luminosity expected in such object would be low and it would be missed by our survey. At the other extreme, the initial acceleration and propagation of the outflow --- the ``blow-out'' phase \citep{sand96, hopk06} --- could be so dust-enshrouded that the photo-ionization of the clouds by the quasar is suppressed, so that again the narrow-line region emission is weak and such object would not be observed in our study. Therefore in order to identify quasars at varying stages of the quasar feedback process it is important to diversify target selection methods. Perhaps red quasars \citep{glik07, geor09} or FeLoBAL quasars \citep{farr07, fauc12} represent a young population and are therefore worthwhile targets to examine in the search of the younger phase of quasar-driven winds.

\section{Conclusions}
\label{sec:summary}

In this paper we examine the morphology and kinematics of the ionized gas around twelve unobscured (type 1) quasars using data from Gemini GMOS IFU. These objects are well matched in luminosity, redshift and observational setup to the sample of eleven obscured (type 2) quasars we studied previously using the same method \citep{liu13a, liu13b}. Observations of ionized gas around type 1 quasars are notoriously difficult because of the emission from the bright central point-like source. Continuum, \feii\ and broad H$\beta$, all of which arise close to the supermassive black hole, contaminate the \oiii\ and H$\beta$ emission that arises over the entire host galaxy. We mitigate these difficulties by modeling the spectrum in every spatial element of the field of view and removing the contaminating components. 

Extended ionized gas emission is detected in most cases; in a couple of objects, the \oiii\ distribution is almost as compact as the PSF, but even in these objects some velocity gradients are seen across the nebula, indicating that faint extended emission is present on top of a bright compact source of \oiii. 

The shapes, the morphologies and the surface brightness profiles of the \oiii\ emission in type 1 quasars are statistically indistinguishable from those in type 2 objects. The median sizes and the standard deviations are $R_{\rm int}=10.7\pm 1.7$ kpc for type 1s vs $12.9\pm 3.4$ kpc for type 2s, and thus the nebulae in type 1 objects are slightly smaller than those in type 2s, but given our sample sizes the values are still consistent with being drawn from the same distributions. Similarly, we find no statistical differences between any of the kinematic measures (velocity gradients across the nebulae, line widths, line shape parameters) in the two samples. The \oiii/H$\beta$ ratio is much harder to measure in type 1s than in type 2s because of the contamination by the broad-line region of the unobscured quasar, but within the measurement uncertainties this ratio follows the same plateau + decline trend in type 1s as we see in type 2s. The only significant differences between the two samples are in their mid-IR luminosities (higher in type 1s) and mid-IR colors (redder in type 2s) which both suggest that quasar emission is anisotropic even at mid-IR wavelengths, while the \oiii\ luminosity is a more isotropic indicator.  

The similarity of morphological and kinematic properties between type 1 and type 2 samples suggests that they are intrinsically very similar and that the standard geometry-based unification model of active nuclei \citep{anto93} likely applies to these sources. The ionized gas seen around quasars of both types is not in dynamical equilibrium with the quasar host galaxies and is likely in the form of an outflow. If this gas ultimately originated in a compact distribution close to the quasar, then our observations suggest that we are observing both samples at a fairly late evolutionary stage, when most of the gas has already been removed from the galaxy. Our samples of type 1 and type 2 objects were carefully selected to have very similar -- and very high -- \oiii\ luminosities, and therefore it is perhaps unsurprising that they probe the same evolutionary phase. The search for the earlier (``blow-out'') and later stages of quasar feedback should continue among other populations which are not necessarily characterized by high narrow-line luminosities. 

\section*{Acknowledgments}

N.L.Z. and J.E.G. are supported in part by the Alfred P. Sloan fellowship. G.L. and N.L.Z. acknowledge support from the Theodore Dunham, Jr. Grant of the Fund for Astrophysical Research. We acknowledge the use of Edward L. Wright's online cosmology calculator \citep{wrig06}. This publication makes use of data products from the Wide-field Infrared Survey Explorer, which is a joint project of the University of California, Los Angeles, and the Jet Propulsion Laboratory/California Institute of Technology, funded by the National Aeronautics and Space Administration.

Funding for SDSS-III has been provided by the Alfred P. Sloan Foundation, the Participating Institutions, the National Science Foundation, and the U.S. Department of Energy Office of Science. The SDSS-III web site is http://www.sdss3.org/.

SDSS-III is managed by the Astrophysical Research Consortium for the Participating Institutions of the SDSS-III Collaboration including the University of Arizona, the Brazilian Participation Group, Brookhaven National Laboratory, Carnegie Mellon University, University of Florida, the French Participation Group, the German Participation Group, Harvard University, the Instituto de Astrofisica de Canarias, the Michigan State/Notre Dame/JINA Participation Group, Johns Hopkins University, Lawrence Berkeley National Laboratory, Max Planck Institute for Astrophysics, Max Planck Institute for Extraterrestrial Physics, New Mexico State University, New York University, Ohio State University, Pennsylvania State University, University of Portsmouth, Princeton University, the Spanish Participation Group, University of Tokyo, University of Utah, Vanderbilt University, University of Virginia, University of Washington, and Yale University.

\bibliographystyle{mn2e}
\bibliography{master}




\label{lastpage}

\end{document}